\documentclass[aps,pre,showpacs,floats,twocolumn,floats,superscriptaddress,floatfix]{revtex4}
\usepackage{bm}
\usepackage{url}
\usepackage{times}
\usepackage{verbatim}
\usepackage{theorem}
\usepackage{makeidx}
\usepackage{amsmath}
\usepackage{bbm}
\usepackage{xy}
\usepackage{amssymb}
\usepackage{graphicx}
\usepackage{color}
\usepackage{epsfig}
\newif
\iffigs
\figstrue     
\makeatletter \def\fps@figure{tp} \makeatother
\iffigs
\fi
\def\drawing #1 #2 #3 {
\begin{center}
\setlength{\unitlength}{1mm}
\begin{picture}(#1,#2)(0,0)
\put(0,0){\framebox(#1,#2){#3}}
\end{picture}\end{center} }

\newcommand{\kg}{{K_{\scriptscriptstyle \mathrm{G}}}}
\newcommand{\tg}{{t_{\scriptscriptstyle \mathrm{G}}}}
\newcommand{\lambdag}{{\lambda_{\scriptscriptstyle \mathrm{G}}}}
\newcommand{\pkg}{\mathrm{P}_{\!\!\! _{K_{\scriptscriptstyle
        \mathrm{G}}}}}
\newcommand{\gkg}{g_{ _{K_{\scriptscriptstyle         \mathrm{G}}}}}

\newcommand{\ue}{\mathrm{e}}
\newcommand{\ui}{\mathrm{i}}
\newcommand{\ts}{t_{\star}}
\newcommand{\taus}{\tau_{\star}}
\newcommand{\us}{u_{\star}}
\newcommand{\kb}{\bm k}
\newcommand{\ut}{\tilde u}
\newcommand{\up}{u ^{\prime}}
\newcommand{\ul}{u ^{\scriptscriptstyle <}}
\newcommand{\ug}{u ^{\scriptscriptstyle >}}

\newcommand{\uhp}{{\hat u} ^{\prime}}
\newcommand{\uhk}{{\hat u}_k}
\newcommand{\uhpk}{{\hat u} ^{\prime}_k}
\newcommand{\uhlkmkp}{\hat u ^{\scriptscriptstyle <}_{\star,\, k-k'}}
\newcommand{\uhlp}{\hat u ^{\scriptscriptstyle <}_{\star p}}
\newcommand{\uhgp}{\hat u ^{\scriptscriptstyle >}_{\star p}}
\newcommand{\uhgq}{\hat u ^{\scriptscriptstyle >}_{\star q}}
\newcommand{\uhpkp}{{\hat u} ^{\prime}_{k'}}

\newcommand{\uvk}{{\check u}_k}

\newcommand{\uvkmkp}{\check u _{k-k'}}
\newcommand{\uvkpkp}{\check u _{k+k'}}

\newcommand{\vhk}{{\hat v}_k}
\newcommand{\fhk}{{\hat f}_k}
\newcommand{\fhkg}{{\hat f}_{_{K_{\scriptscriptstyle
        \mathrm{G}}}}}
\newcommand{\psij}{\psi ^{(j)}}
\newcommand{\psijk}{\psi_k ^{(j)}}
\newcommand{\upj}{u ^{\prime(j)}}
\begin{document}
\title{Resonance phenomenon for the Galerkin-truncated Burgers and Euler equations}
\author{Samriddhi Sankar Ray}
\affiliation{UNS,~CNRS,~OCA,~Lab.~Cassiop\'ee,~B.P.~4229,~06304~Nice~Cedex~4,~France}
\altaffiliation{SSR: also at Centre for Condensed Matter Theory, Department of
Physics, Indian Institute of Science, Bangalore, India}
\author{Uriel Frisch}
\affiliation{UNS,~CNRS,~OCA,~Lab.~Cassiop\'ee,~B.P.~4229,~06304~Nice~Cedex~4,~France}
\author{Sergei Nazarenko}
\affiliation{University of Warwick, Mathematics Institute, Coventry
    CV4 7AL, UK}
\author{Takeshi Matsumoto}
\affiliation{Dept. of Physics, Kyoto University, Kitashirakawa
  Oiwakecho Sakyoku, Kyoto 606-8502, Japan}
\begin{abstract}
It is shown that the solutions of inviscid hydrodynamical equations with
suppression of all spatial Fourier modes having wavenumbers in excess of a
threshold $\kg$ exhibit unexpected features. The study is
carried out for both the one-dimensional Burgers equation and the
two-dimensional 
incompressible Euler equation.  For large $\kg$ and smooth initial conditions, the first
symptom of truncation, a localized short-wavelength oscillation which we call
a ``tyger'', is caused by a resonant interaction between fluid particle motion
and truncation waves generated by small-scale features (shocks, layers with
strong vorticity gradients, etc).  These tygers appear when
complex-space singularities come within one Galerkin wavelength
$\lambdag = 2\pi/\kg$  from the real
domain and typically arise far away from preexisting small-scale structures at
locations whose velocities match that of such structures. Tygers are weak
and strongly localized at first---in the Burgers case at the time of
appearance of the first shock their amplitudes and widths are proportional to
$\kg ^{-2/3}$ and $\kg ^{-1/3}$ respectively---but grow and eventually invade
the whole flow. They are thus the first manifestations of the thermalization
predicted by T.D.~Lee in 1952.

The sudden dissipative anomaly---the presence of a finite dissipation in the
limit of vanishing viscosity after a finite time $\ts$--, which is well known
for the Burgers equation and sometimes conjectured for the 3D Euler equation,
has as  counterpart, in the truncated case, the ability of tygers to store a
finite amount of energy in the limit $\kg\to\infty$.  This leads to Reynolds
stresses acting on scales larger than the Galerkin wavelength and thus
prevents the flow from converging to the inviscid-limit solution. There
are indications that it may eventually be  possible to purge the tygers and thereby to
recover the correct inviscid-limit behaviour.


\end{abstract}
\pacs{05.20.Jj, 05.45.-a, 47.27.Jv}
\maketitle
\section{Introduction and formulation}
\label{s:intro}
 When the motion of a fluid is described at the microscopic level, a
 \textit{conservative} Hamiltonian formulation is appropriate and statistical
 steady states can be described using Gibbs ensembles. At the macroscopic
 level, however, one obtains a \textit{dissipative} hydrodynamical description
 because macroscopic motion can be irreversibly degraded into thermal
 molecular motion.  Curiously, T.D.~Lee observed that, starting from the
 hydrodynamical or magnetohydrodynamical equations for an \textit{ideal
   fluid}, one can obtain a conservative dynamical system to which Gibbsian
 statistical mechanics becomes applicable \cite{lee52}. For this he used a
 Galerkin truncation of the equations, a procedure that keeps only a finite
 number of spatial Fourier harmonics. For the case of the Galerkin-truncated
 3D incompressible Euler equation, Lee obtained thermalized equilibrium
 statistical states having an equipartition of kinetic energy among all the
 Fourier harmonics and thus a $k^2$ energy spectrum. This is very far from the
 spectrum of fully developed turbulence as observed experimentally, which
 could lead one to believe that Galerkin truncation applied to the Euler
 equation cannot tell us anything about the dissipative states of turbulence.

Kraichnan was the first to think otherwise. Considerations of the
Galerkin-truncated equilibria of the 2D Euler equation played an
important role in his conjecture about an inverse energy cascade
\cite{rhk67,eyinkfrisch}. In 1989 he and S.~Chen  went much further
and wrote (\cite{rhkchen}, p.~162):
\begin{quote}
the truncated Euler system can imitate
NS [Navier-Stokes] fluid: the high-wavenumber degrees of freedom act like a thermal sink
into which the energy of low-wave-number modes excited above equilibrium is
dissipated. In the limit where the sink wavenumbers are very large compared
with the anomalously excited wavenumbers, this dynamical damping acts
precisely like a molecular viscosity. 
\end{quote}

Supporting evidence was found in 2005 with very-high-resolution spectral
simulations of the 3D Galerkin-truncated Euler equation that showed the
following: when initial conditions are used that have mostly low-wavenumber
modes, the solutions have long-lasting transients in which only the
high-wavenumber modes are thermalized, while the lower-wavenumber modes behave
in a way similar to that for viscous high-Reynolds-number flow
\cite{cichowlasetal}. This seems to hold not only when the
low-wavenumber modes are weak (as implicitly assumed by Kraichnan who
invoked
the fluctuation-dissipation relations) but also  in the strong
turbulence regime  that displays a 
K41-type inertial range.  One possible interpretation is
that the thermalized modes act as a kind of artificial molecular world,
thereby allowing dissipative (Navier--Stokes) dynamics for the
lower-wavenumber modes.

We understand far too little 
about the mathematics of the 3D Euler and Navier--Stokes equations to
start a serious analytical investigation of what happens to solutions
of the Galerkin-truncated 3D Euler equation when $\kg \to \infty$. However,
such matters may be within reach for the one-dimensional inviscid Burgers
equation, a well-understood problem in the absence of 
truncation. Even in that ``simple'' case, the behavior at large $\kg$ is far
from obvious. Indeed,  there are known
instances where an energy-conserving modification of the inviscid
Burgers equation with a small parameter is found not to
converge \footnote{In a weak or distributional sense, that is after
multiplication by suitable smooth test functions and integrations by
parts.} to the inviscid limit \cite{laxlevermore,goodmanlax}
(see also \cite{houlax}). Hence caution is needed and we shall discuss
this issue further.

There is also an important practical  reason to be interested in
Galerkin-truncated hydrodynamics. Spectral methods (and their
pseudo-spectral variants) are among the most precise methods for the
numerical integration of hydrodynamical equations \cite{orszag}. By
necessity, a finite resolution must then be used. In other words one
integrates not the full hydrodynamical equations but their
Galerkin-truncated modifications. If the high-wavenumber modes are 
sufficiently damped by viscous dissipation the difference may 
be extremely small. Yet the desire to push the Reynolds number
can lead to serious truncation errors. Furthermore in investigations
of the blow-up problem for the 3D Euler equation (cf., e.g.,
Refs.~\cite{blue,gibbon} and references therein) it is important to be able to
distinguish genuine blow-up from truncation effects.

We had one additional reason to investigate what exactly are the
consequences of Galerkin truncation. M.E.~Brachet (private
communication 2007) informed us about a  strange phenomenon  
observed when Galerkin truncation is used in conjunction with
the  one-dimensional inviscid Burgers equation
\begin{equation}  
  \partial_tu+u\partial_x u =0; \qquad u(x,0) =u_0(x).
\label{burgers}
\end{equation}
The initial condition $u_0(x)$ has just a few Fourier
harmonics and the number of retained Fourier modes is large. The first symptom
of Galerkin truncation found by Brachet was  a spurious oscillation in physical space,
seemingly born not at all where one would expect it, namely
in the neighborhood of genuine small-scale structures such as shocks
and their precursors called preshocks, but completely ``out of the
blue'' at a place having no particular small-scale activity, as
illustrated in Fig.~\ref{f:smtygerappearance}.

In this paper we shall understand why this happens, both for the Burgers
equation and for the Euler equation (so far mostly in 2D). The
phenomenon is here called \textit{tyger} after William Blake's poem
for reasons given later.
 Before proceeding to explain the organization of the paper, it is useful
to define our Galerkin-truncated problem more precisely for the case
of the Burgers equation (the 2D Euler case will be formulated in
Sec.~\ref{ss:twod}).  We restrict ourselves to $2\pi$ periodic solutions
which can be expanded in a Fourier series
\begin{equation}  
u(x) = \sum_{k= 0, \pm 1, \pm 2 \ldots} \ue ^{\ui kx}\hat u_k .
\label{fourier}
\end{equation}
Let $\kg$ be a positive integer, here called the Galerkin truncation
wavenumber. We define the Galerkin projector $\pkg$ as the low-pass filter
which sets to zero  all  Fourier components with wavenumbers
$|k|>\kg$.
In other words
\begin{equation}  
\pkg u(x) =   \sum_{|k| \le \kg} \ue ^{\ui kx}\hat u_k.
\label{defpkg}
\end{equation}
The (untruncated) inviscid Burgers equation, written in conservation form, is
\begin{equation}  
  \partial_tu+\partial_x (u ^2/2) =0; \qquad u(x,0) =u_0(x).
\label{inviscidburgers}
\end{equation}
The associated Galerkin-truncated (inviscid) Burgers equation whose solution is 
denoted $v(x,t)$ is obtained by applying the low-pass filter to both
the initial condition and to the nonlinear term \cite{majdatimofeyev2000}. It reads
\begin{equation}  
  \partial_tv+ \pkg\partial_x (v ^2/2) =0; \qquad v_0 =\pkg u_0.
\label{gtburgers}
\end{equation}
As is well known, the gradient $\partial_x u$ of the solution to the inviscid Burgers equation with
smooth initial data  typically
blows up after a finite time $t_\star$. At $t_\star$ the solution
$u$ has a cubic-root singularity, called a preshock
\cite{fournierfrisch,frischbecleshouches}. Beyond $t_\star$ the solution can be
continued by introducing a small viscous term into the r.h.s. of 
\eqref{burgers}; in the
limit of  vanishing viscosity one obtains what we here call the
inviscid-limit solution, which  has one or several shocks
\cite{hopf}. This is a generalized solution which
satisfies the inviscid Burgers equation only in a weak sense. The inviscid-limit solution has a \textit{dissipative
  anomaly}, i.e., it dissipates energy even in the limit of vanishing
viscosity. In contrast, the solution to the Galerkin-truncated
equation \eqref{gtburgers} stays smooth and conserves energy forever.

This paper has two main parts: Sec.~\ref{s:simulpheno} deals with the numerical exploration of the
tyger phenomenon and includes soft phenomenological
interpretations of our various
findings. 
More specifically, in Sec.~\ref{ss:tygerresonance} we identify the
resonant particle-wave interaction mechanism responsible for the birth of tygers.
In Sec.~\ref{ss:temporal} we present
the whole temporal scenario from the birth of tygers to full
thermalization.
In Sec.\ref{ss:twod} we show that the 2D incompressible Euler
equation
also gives rise to tygers by a mechanism similar to what we find for
the Burgers equation. In Sec.~\ref{ss:energy} we investigate
the energetics (dissipative anomaly) and the issue of the (weak) limit
of the truncated solutions when $\kg \to \infty$. Sec.~\ref{s:birth} is restricted  to the birth
of tygers for the Burgers equation; it involves state-of-the-art
simulations of the scaling properties with $\kg$ up to $40,000$ and
also a fair amount of analytic theory. Open problems and conclusions are
presented in Sec.~\ref{s:conclusion}. There are four technical
appendices.

\section{Simulations and phenomenology}
\label{s:simulpheno}

\subsection{Tygers and resonance}
\label{ss:tygerresonance}

Henceforth, when we write about a/the ``untruncated solution of the Burgers
equation'',
without specifying more, it is the inviscid
limit of the untrucated Burgers equation which is understood. 
 A particularly
simple $2\pi$-periodic solution of the Burgers equation is
obtained with the initial condition
\begin{equation}  
u_0(x) = \sin x,  
\label{sinx}
\end{equation}
which has two stagnation (zero-velocity) points, $x=0$ with positive
strain (gradient) and $x=\pi$ with negative strain. The latter gives rise
to a cubic-root preshock singularity at the time $t=\ts
=1$ \footnote{Elementary facts about the solution of the Burgers
  equation and its singularities are recalled in
  Appendix~\ref{a:sing}.}.
Hereafter the initial condition \eqref{sinx} and its space-translates will
be referred to as ``single-mode initial condition''. 

\begin{figure}[htbp]
\iffigs
\begin{center}
\includegraphics[height=5cm]{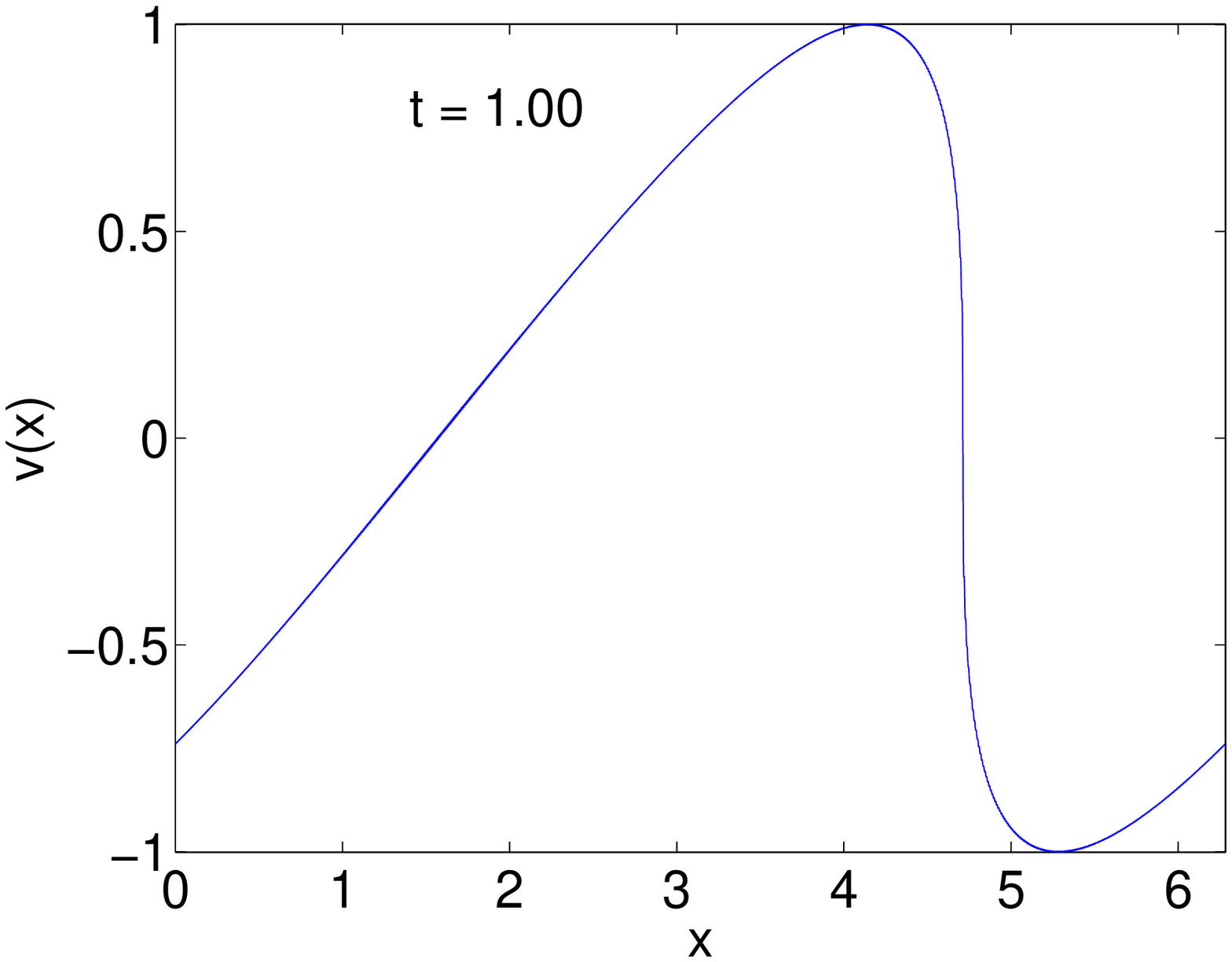}
\includegraphics[height=5cm]{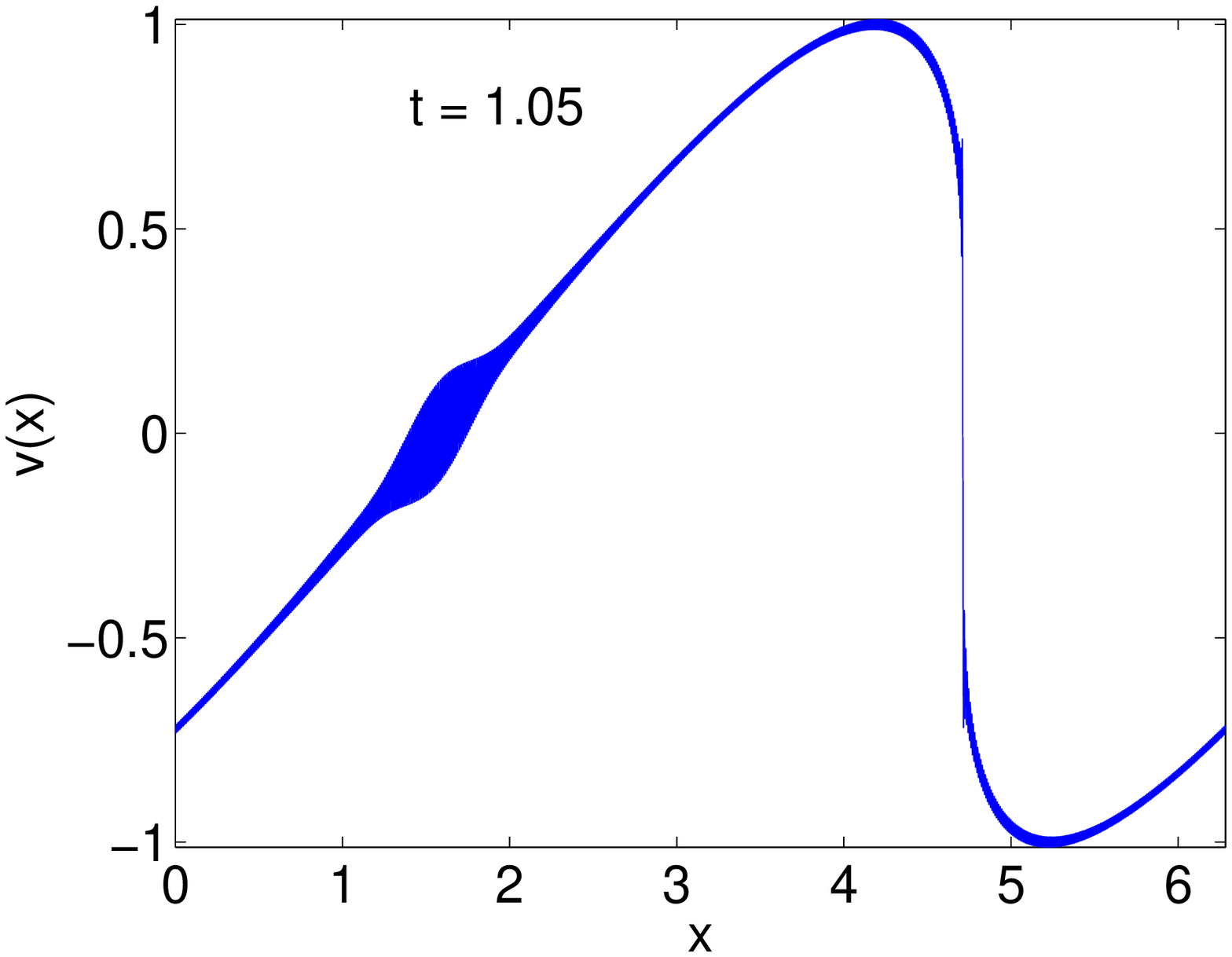}
\includegraphics[height=5cm]{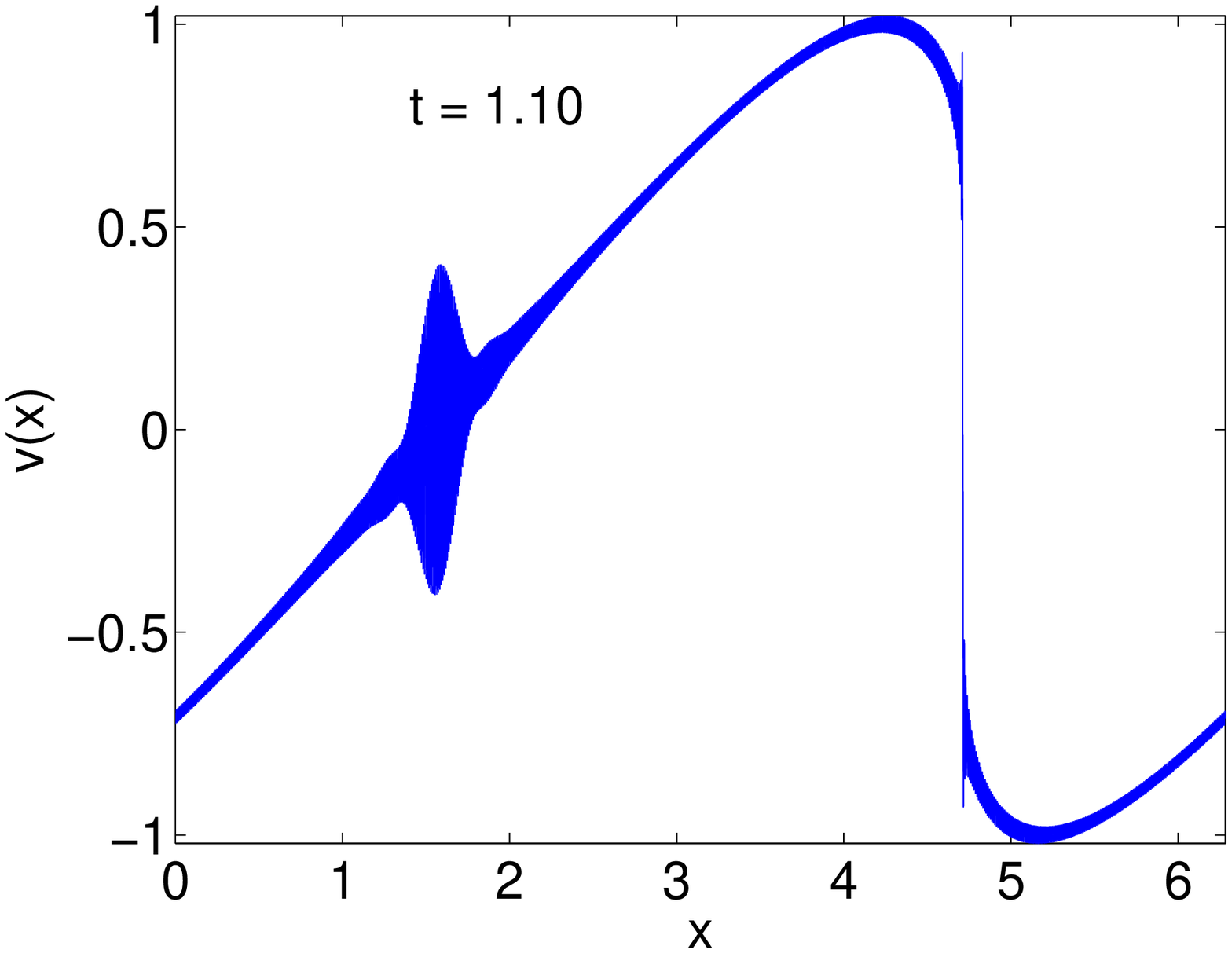}
\end{center}
\fi
\caption{(Color online) Growth of a tyger in the solution of the inviscid  Burgers
  equation
with initial condition $v_0(x) = \sin(x - \pi/2)$ (to avoid graphical
edge effects). Galerkin truncation at $\kg =700$. Number of
collocation
points $N = 16,384$. Output time as labelled.  Observe
that the bulge appears far from the place of birth of
the shock. In the PDF online version of this paper all high-resolution
  figures are fully zoomable.}
\label{f:smtygerappearance}
\end{figure}

\begin{figure}[htbp]
\iffigs
\begin{center}
\includegraphics[height=3.3cm]{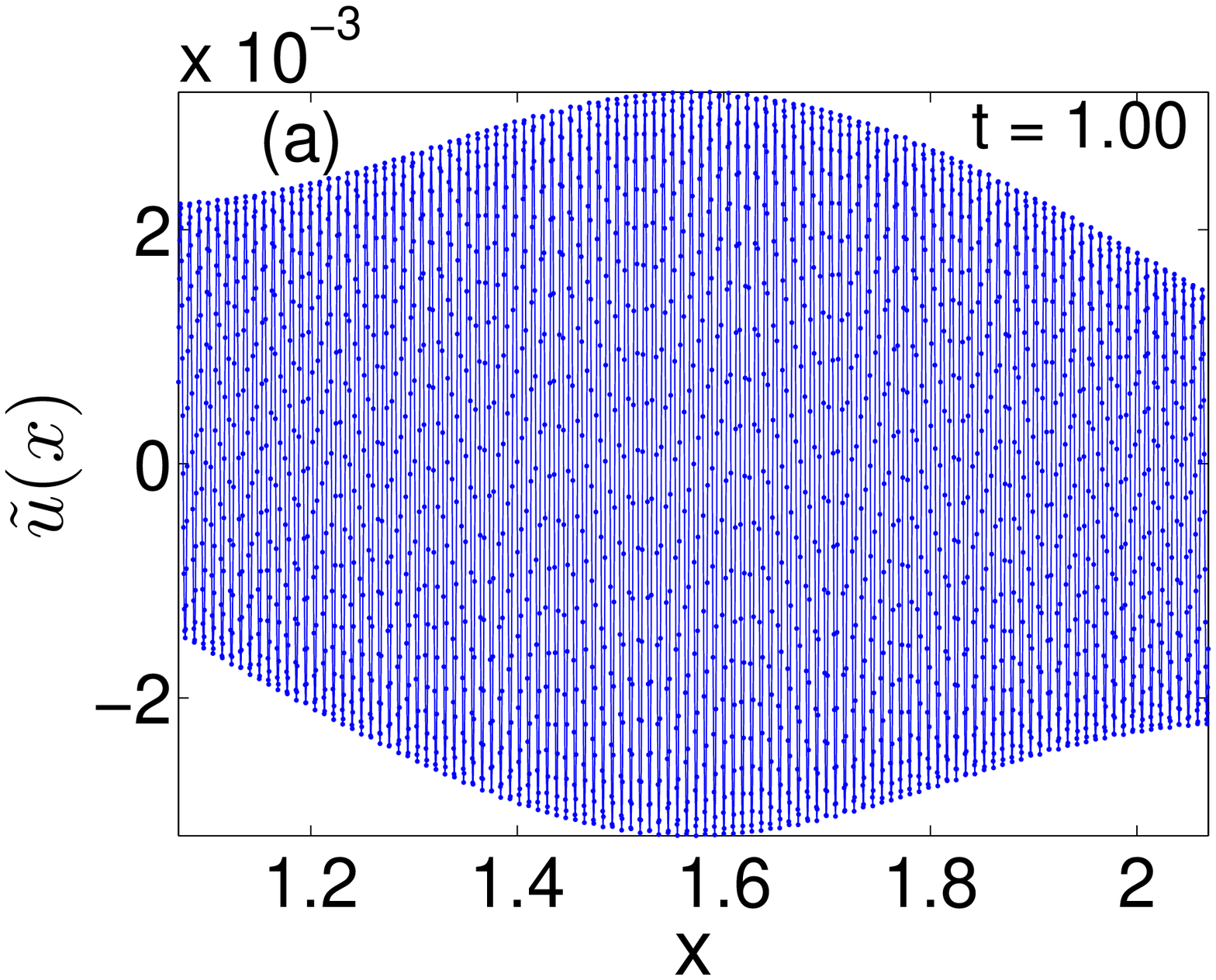}\,\,\,\,\,\,\,\,\,\,
\includegraphics[height=3.3cm]{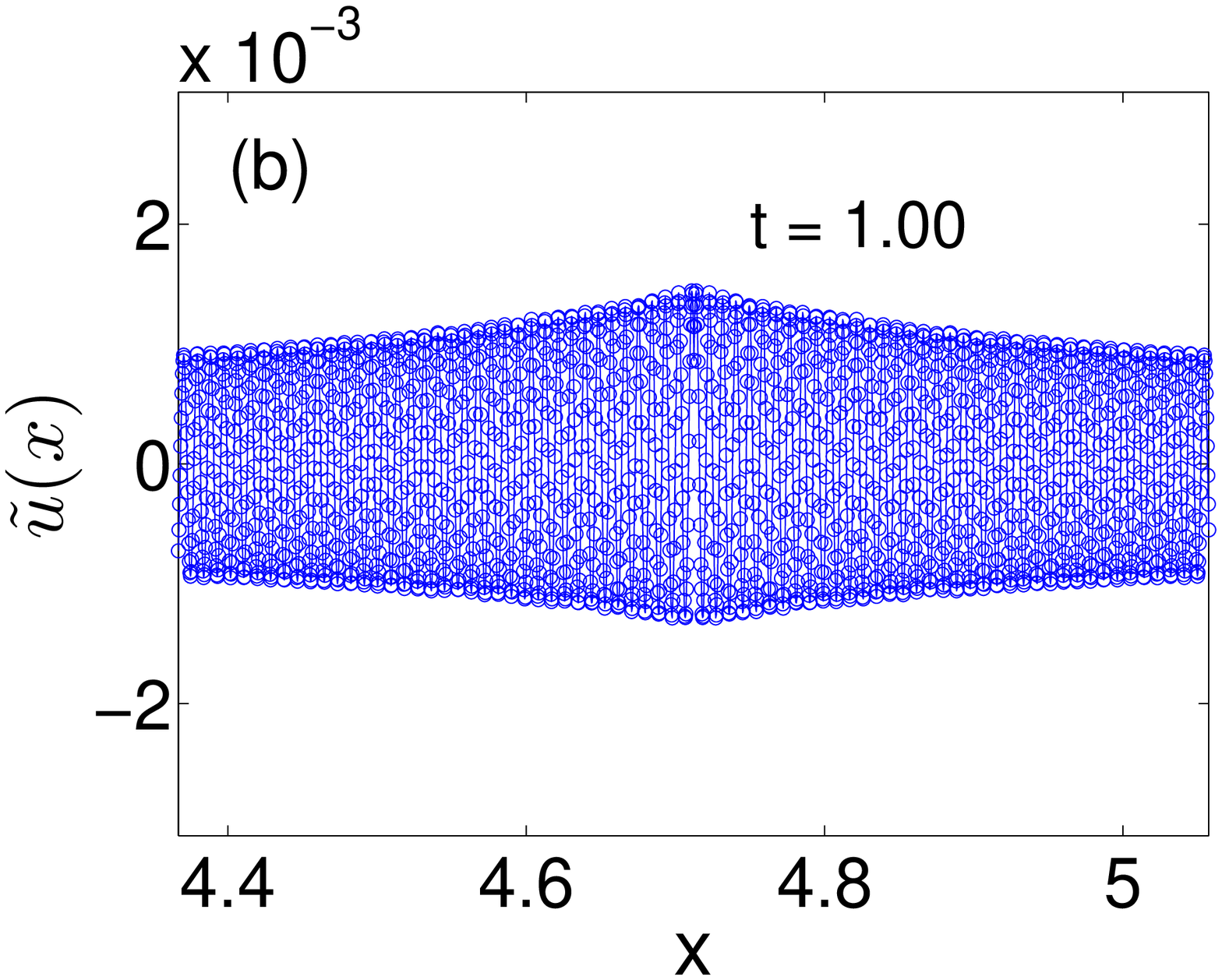}
\includegraphics[height=3.3cm]{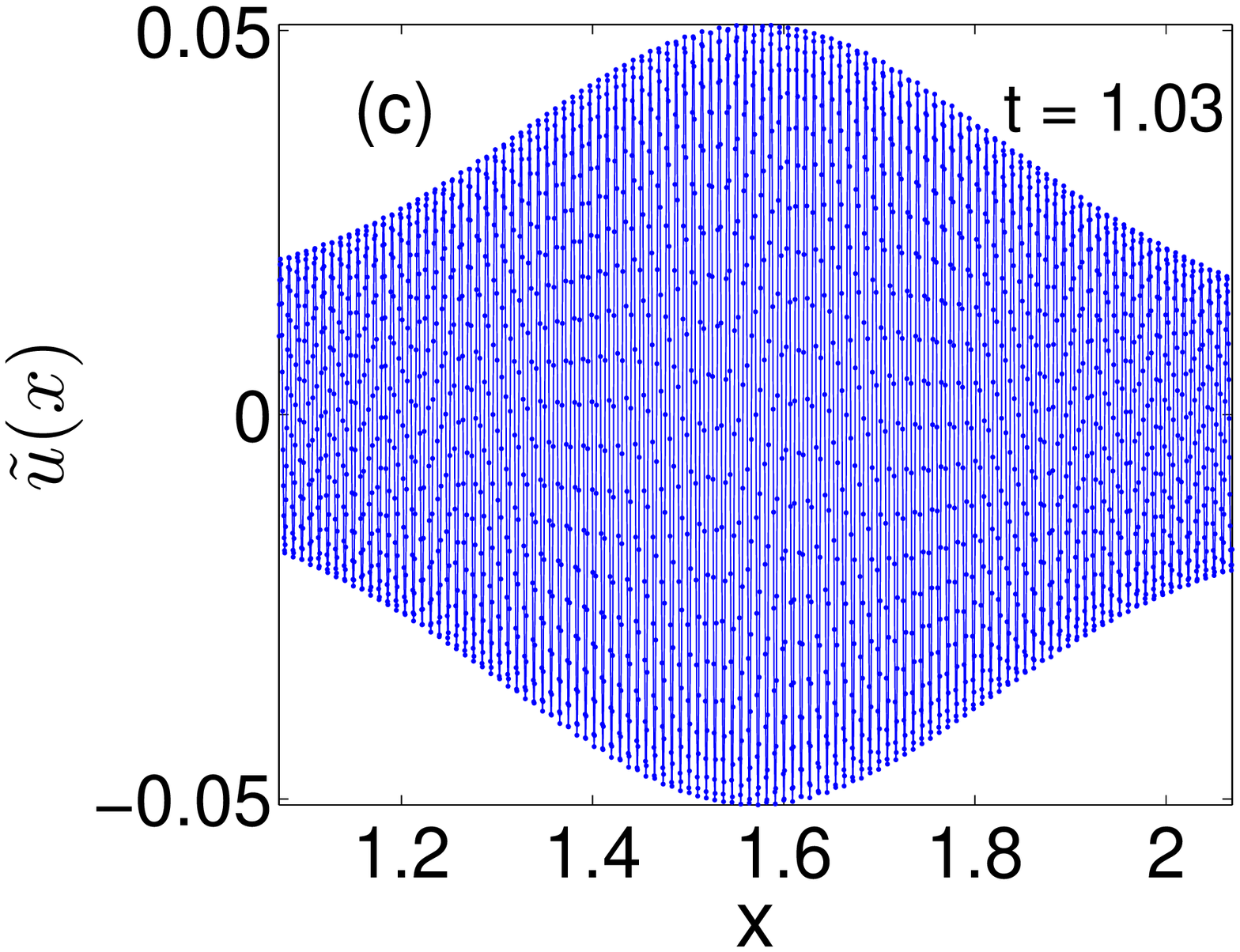}
\includegraphics[height=3.3cm]{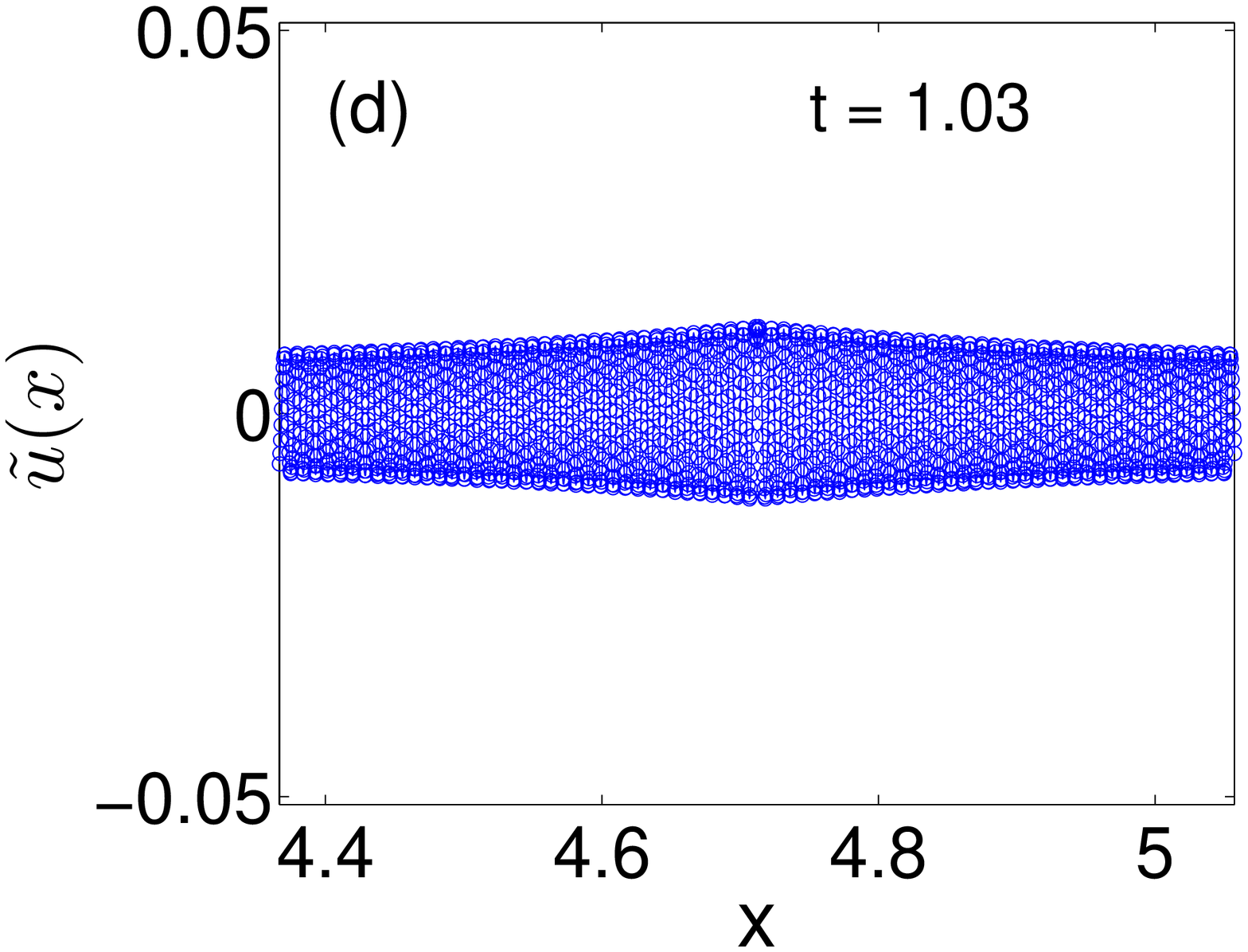}
\end{center}
\fi
\caption{(Color online) Close-up views of the discrepancy $\ut(x)$ at time $t=\ts =1$
and $t= 1.03$ in the tyger regions (panels (a) and (c), respectively) and in the 
preshock region (panels (b) and (d), respectively).  Same conditions as
Fig~\ref{f:smtygerappearance}. Note that the tyger bulge grows much faster
than the preshock bulge.}
\label{f:tygerpreshockbulge}
\end{figure}

With this initial condition, the tyger phenomenon is particularly
simple to observe. Fig.~\ref{f:smtygerappearance} shows the solution 
of the truncated Burgers equation with $\kg = 700$ at $t = \ts$ and at
slightly later times \footnote{The numerical method used to integrate
the Burgers equation is described in Appendix~\ref{a:num}.}. Near $x=
3\pi/2 \approx 4.712$ the
cubic-root preshock singularity at $\ts$ and the shock beyond that
time are standard features. The  new one is the ``tyger'', a growing   bulge
near the positive-strain stagnation point \footnote{A tyger is already
  seen but not commented upon in the last two panels of Fig. 6.2 of 
\cite{tadmor1989}. M.-E.~Brachet (Private communication, 2007) was
the first to draw our attention 
to this phenomenon for the case of the initial condition $\sin x$ but did not
propose an explanation.}.  A more detailed view is shown in
Fig.~\ref{f:tygerpreshockbulge} which shows the \textit{discrepancy} $\ut 
\equiv v-u$ between the truncated
solution and the untruncated one, and zooms into the tyger region and the
preshock regions at time $\ts$ and shortly thereafter.
The bulges seen consist basically of oscillations at the \textit{ Galerkin wavelength}
\begin{equation}  
  \lambdag \equiv \frac{2\pi}{\kg}.
\label{deflambdag}
\end{equation}
with a localized envelope function which is symmetrical around the center;
this symmetry will actually be destroyed later by nonlinearity, as we shall
see in Sec.~\ref{ss:temporal}. 

\begin{figure}[htbp]
\iffigs
\begin{center}
\includegraphics[height=5cm]{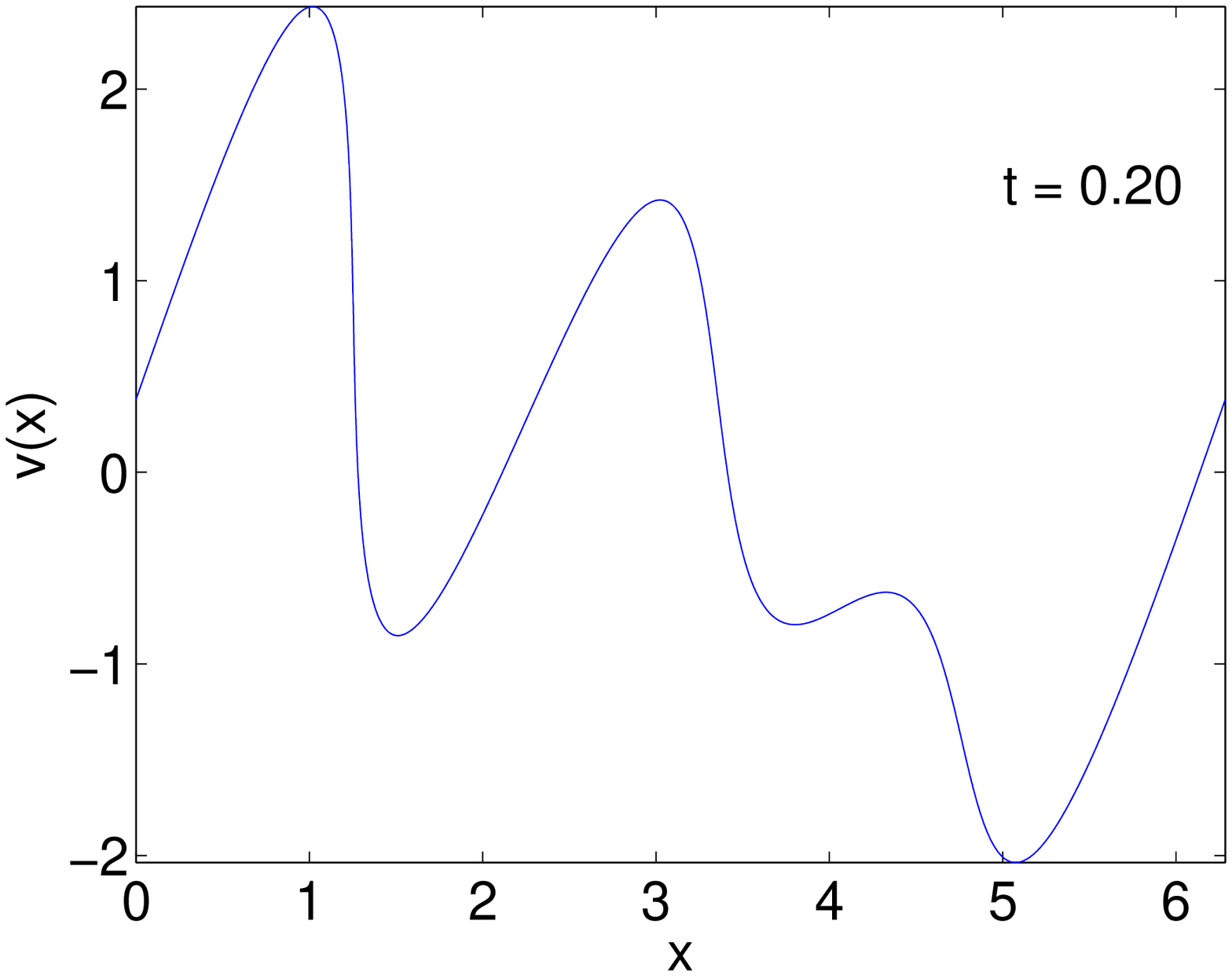}
\includegraphics[height=5cm]{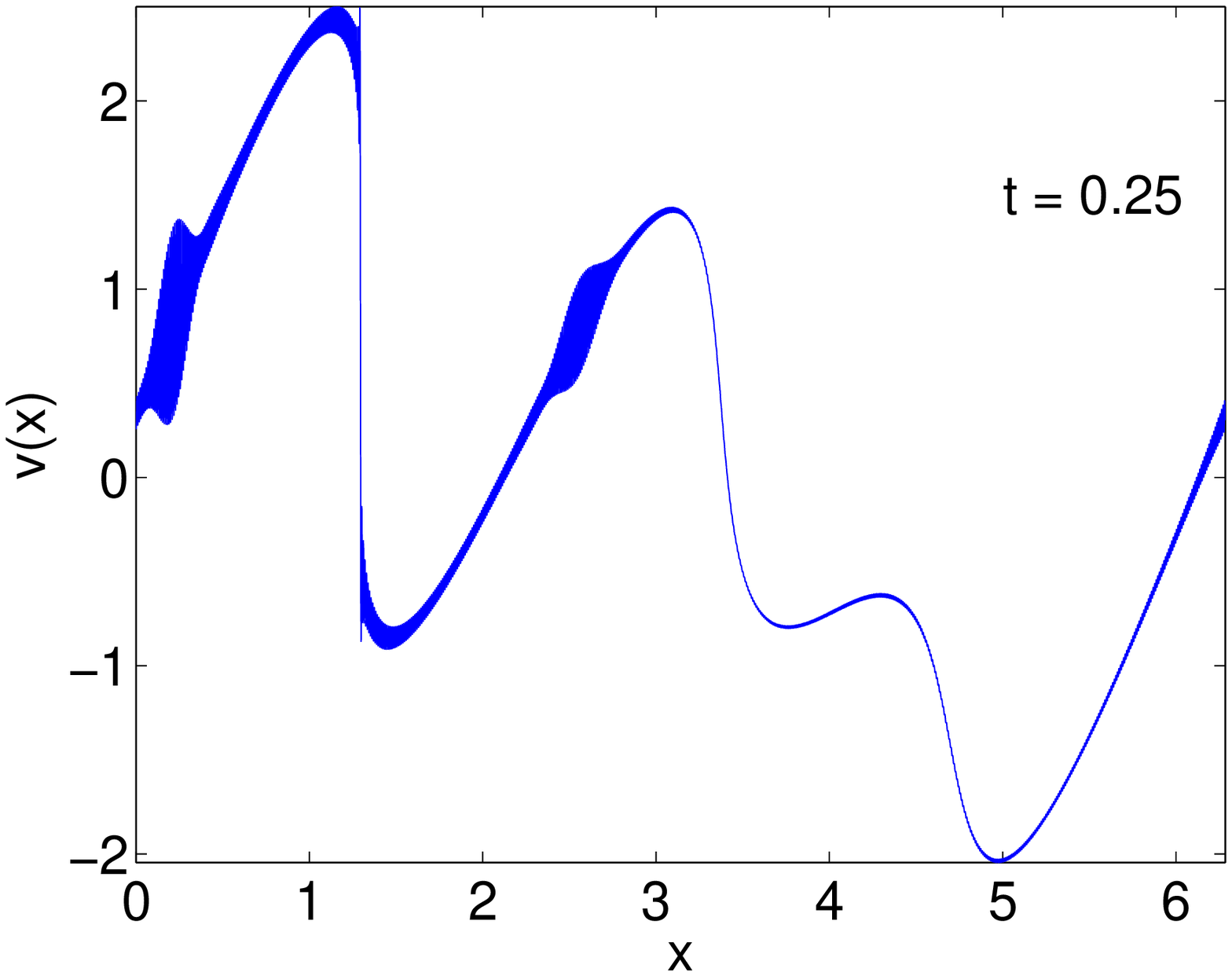}
\end{center}
\fi
\caption{(Color online) Three-mode initial condition. Other parameters as in
  Fig.~\ref{f:smtygerappearance}. Tygers appear at the points having
  the same velocity as the shock and positive strain.}
\label{f:3modeearly}
\end{figure}

We now use a multi-mode initial condition
\begin{equation}  
 u_0(x) =  \sin(x) + \sin(2x + 0.9) + \sin(3x), 
\label{3mode}
\end{equation} 
for which the first singularity is at $\ts =0.2218$. Simulating  the
truncated solution, again with $\kg =700$, we see (cf. Fig.~\ref{f:3modeearly}) that at 
$t=0.25$  there is a well-developed shock (near $x= 1.30$),
decorated by tygers on each side. 

The tygers are centered around points
where the velocity equals the half-sum of the limiting velocities
when approaching the shock from the left and the right. For the
Burgers equation this is precisely the velocity of the shock. Note that
there is yet another point (around $x=3.30$) which has the same
velocity
but no tyger; it has however a negative strain. Thus tygers appear
to be born at points of positive strain having the same velocity 
as a shock. 

\begin{figure}[htp]
\iffigs
\centerline{\includegraphics[height=5cm]{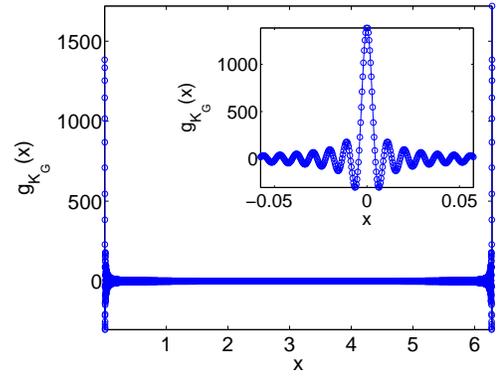}}
\fi
\caption{(Color online) Truncation wave with wavenumber $\kg = 700$.}
\label{f:truncationwave}
\end{figure}

We turn now to the phenomenological explanation of the tyger phenomenon,
leaving more systematic theory for Sec.~\ref{s:birth}. The presence of
the Galerkin truncation projector $\pkg$ makes \eqref{gtburgers}
nonlocal in physical ($x$) space. A localized strong nonlinearity, such as 
is present at a preshock or a shock, acts as a source of   a \textit{truncation 
wave}  whose spatial dependence is the Fourier transform of the
low-pass filter projector \footnote{The feeding mechanism producing
  truncation waves will be analyzed more systematically in
  Sec.~\ref{s:birth}.}. More precisely, in physical space the
nonlinear term involves a convolution with 
\begin{equation}  
  \gkg(x)\equiv\sum_{k=-\kg}^{k=\kg} \ue ^{\ui kx} =\frac{\sin \left(\kg +\frac{1}{2}\right)x}{\sin \left(\frac{x}{2}\right)}.
\label{gkg}
\end{equation} 
Away from the source (where it is close to a Dirac measure for large $\kg$), this ``truncation wave'' is mostly a plane wave with
wavenumber close to $\kg$ (see Fig.~\ref{f:truncationwave}) and thus
has a
wavelength close to the Galerkin wavelength,
\begin{equation}  
  \lambda_{\rm G} \equiv \frac{2\pi}{\kg}.
\label{redeflambdag}
\end{equation}
Observe that if a preshock/shock is moving with velocity $v_{\rm s}$, the
associated 
truncation wave becomes a progressive wave with phase  velocity $v_{\rm s}$.

Away from the shock region, the Burgers equation can be interpreted as
describing a particle dynamics: from a Lagrangian point of view, fluid particles
just move with their velocity unchanged. In the presence of
truncation,
those particles which happen to have a velocity equal to the phase
velocity of a truncation wave can \textit{resonantly interact} with
such waves \footnote{Nonlinear evolution equations for which there is
  no concept of fluid particle, such as the Constantin--Lax-Majda model
  \cite{CLM}, do not display the tyger phenomenon.}.

Resonant particle-wave interaction is a well known phenomenon. It is used, for
example, in plasma physics to explain Landau damping \cite{landau}:
the near coincidence of the velocity of particles and of the phase
velocity of Langmuir wave allows efficient interactions between the
two. 
This can lead to wave attenuation
(actual Landau damping) or enhancement (beam instability, bump on the tail
instability, \ldots). There are however substantial differences between the
Vlasov equation (governing the Landau instability) and the truncated Burgers
equation. For example the Langmuir wave evolves  through energy
transfers via the resonance, a problem which in the linear approximation can
be solved by use of the Laplace transform (leading to Landau's rule for a
pole-avoiding integration path); in contrast, our truncation waves are
completely prescribed by the singularities (preshocks or shocks) and undergo
no damping in the linear approximation. Furthermore the linear approximaton
does not have an easy analytic solution (cf. Sec.~\ref{s:birth}). In Landau
damping, resonant particles get trapped and a characteristic ``cat eyes''
phase-space distribution with progressively thinning filamentary structure is
obtained (cf. Fig.~II-1 of \cite{sagdeevgaleev}). In the truncated Burgers
dynamics thinning is arrested by truncation.

The radiation of truncation waves begins only at or close to the
time of formation of a preshock. After a time $\tau$ has elapsed,
those fluid particles having a velocity $v$ that does not match the
preshock velocity $v_{\rm s}$ may not feel much pull from the
truncation waves if phase mixing is present. More precisely, resonant
interactions are confined to particles such that 
\begin{equation}  
  \tau\Delta v \equiv \tau|v-v_{\rm s}|
\lesssim \lambda_{\rm G}.
\label{width}
\end{equation}
If $\tau$ is small, as in
Figs.~\ref{f:smtygerappearance} 
and \ref{f:3modeearly}, the region of resonance will be confined
to a small neighborhood of the point of resonance. In
Sec.~\ref{ss:3approx} we shall see that around the time of appearance
of a preshock the width of the associated resonance regions is
typically the order of $\kg ^{-1/3}$. Outside of such regions,
including near the preshock,  the
effect of truncation waves is just a small-amplitude oscillation at
the Galerkin wavelength, which on Figs.~\ref{f:smtygerappearance} and 
\ref{f:3modeearly} shows up as a thickening of the line with respect
to the inviscid-limit solution. 

Actually, only resonance points with positive strain produce
tygers. In a region of negative strain a wave of wavenumber
close to $\kg$ will be squeezed, potentially acquiring a larger
wavenumber, and thus disappearing beyond the \textit{truncation horizon} which
acts as a kind of black hole.

Observe that in the immediate neighborhood of a preshock or of a nascent shock
the strain is also negative and actually very large. Although the strongest
truncation waves are generated near such points (as one would infer by a
Gibbs-phenomenon argument), their growth is severely hampered by the negative
strain. Hence the bulge near the preshock grows in time much more slowly
than  that at the positive-strain resonance point, as illustrated
in Fig.~\ref{f:tygerpreshockbulge}.  Actually, the effect of
truncation near a shock will remain very small (and almost invisible without
zooming) until the tyger has fully spread out on the ramp
(cf. end of Sec.~\ref{ss:temporal}). This situation is in
contrast with what one observes with other energy-conserving semi-discrete
schemes \footnote{A semi-discrete scheme is continuous in time and discrete in
  space.}, such as the dispersive one studied by Goodman and Lax
\cite{goodmanlax}.


\subsection{From tygers to thermalization: the temporal evolution}
\label{ss:temporal}

It is well known that the Galerkin-truncated solution of the Burgers
equation, will eventually thermalize to a Gaussian state. The simplest
case is when the initial velocity $v_0$ integrates to zero over the
spatial period. An ergodicity argument, supported by numerical
simulations,
suggests that the thermalized state has 
equipartition of energy between all Fourier modes and thus
is just low-pass filtered white noise in the $x$ variable
\cite{majdatimofeyev2000,majdatimofeyev2002}. 

How does the highly 
organized and localized tyger structure seen in Figs.~\ref{f:smtygerappearance} 
and \ref{f:3modeearly} evolve into such a  totally random state?

\begin{figure*}[htbp]
\iffigs
\begin{center}
\includegraphics[height=4cm]{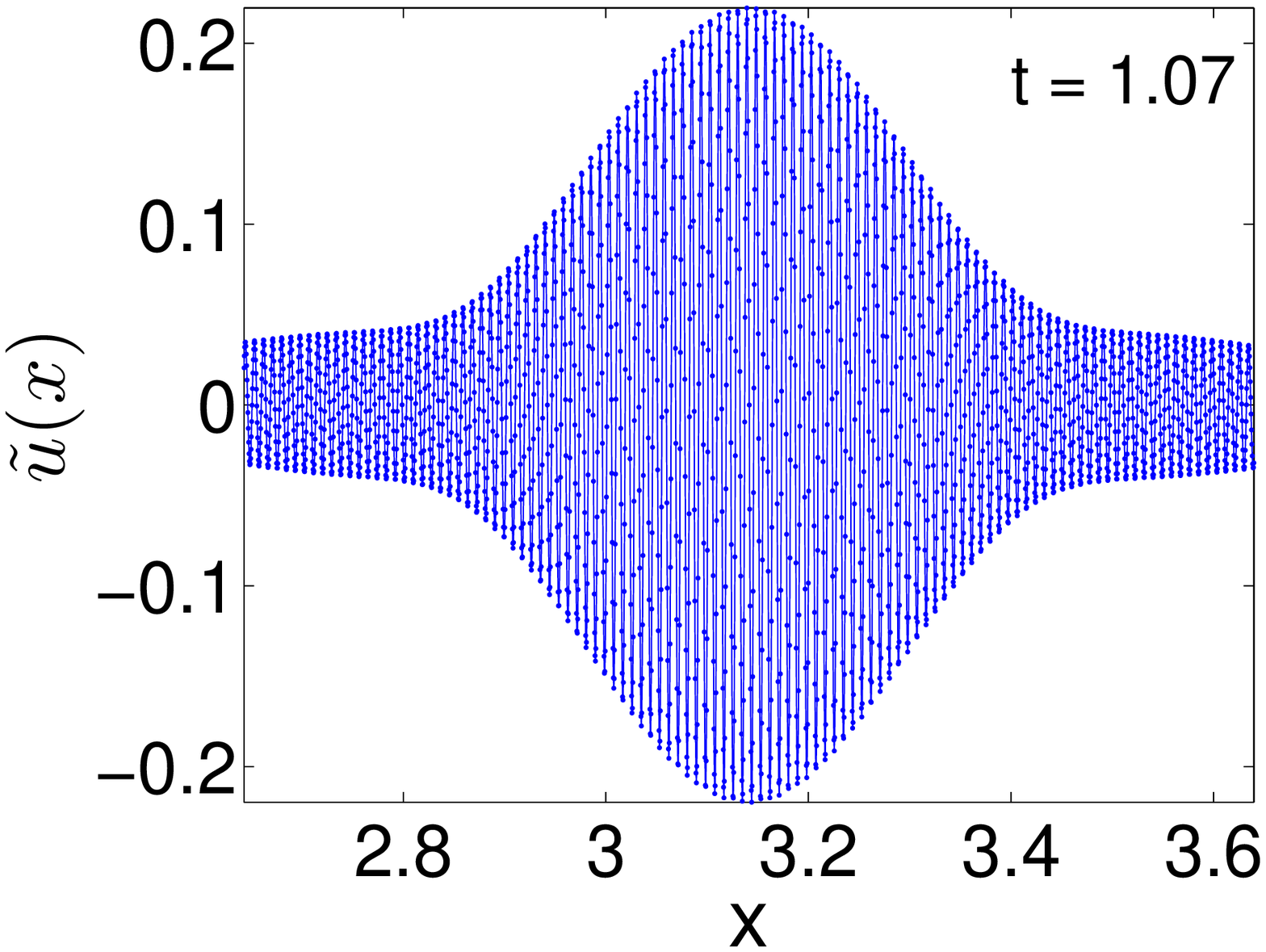}
\includegraphics[height=4cm]{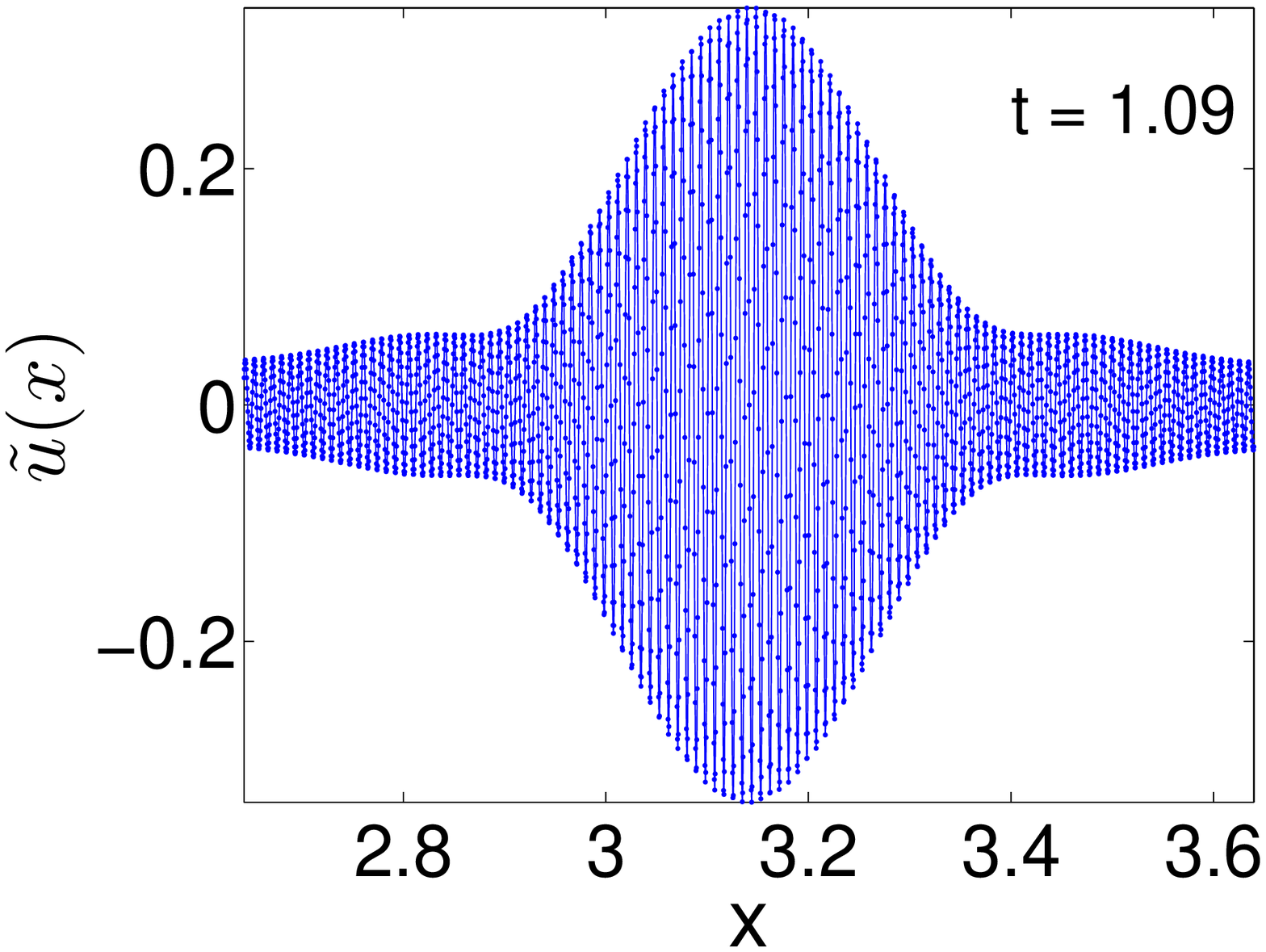}
\includegraphics[height=4cm]{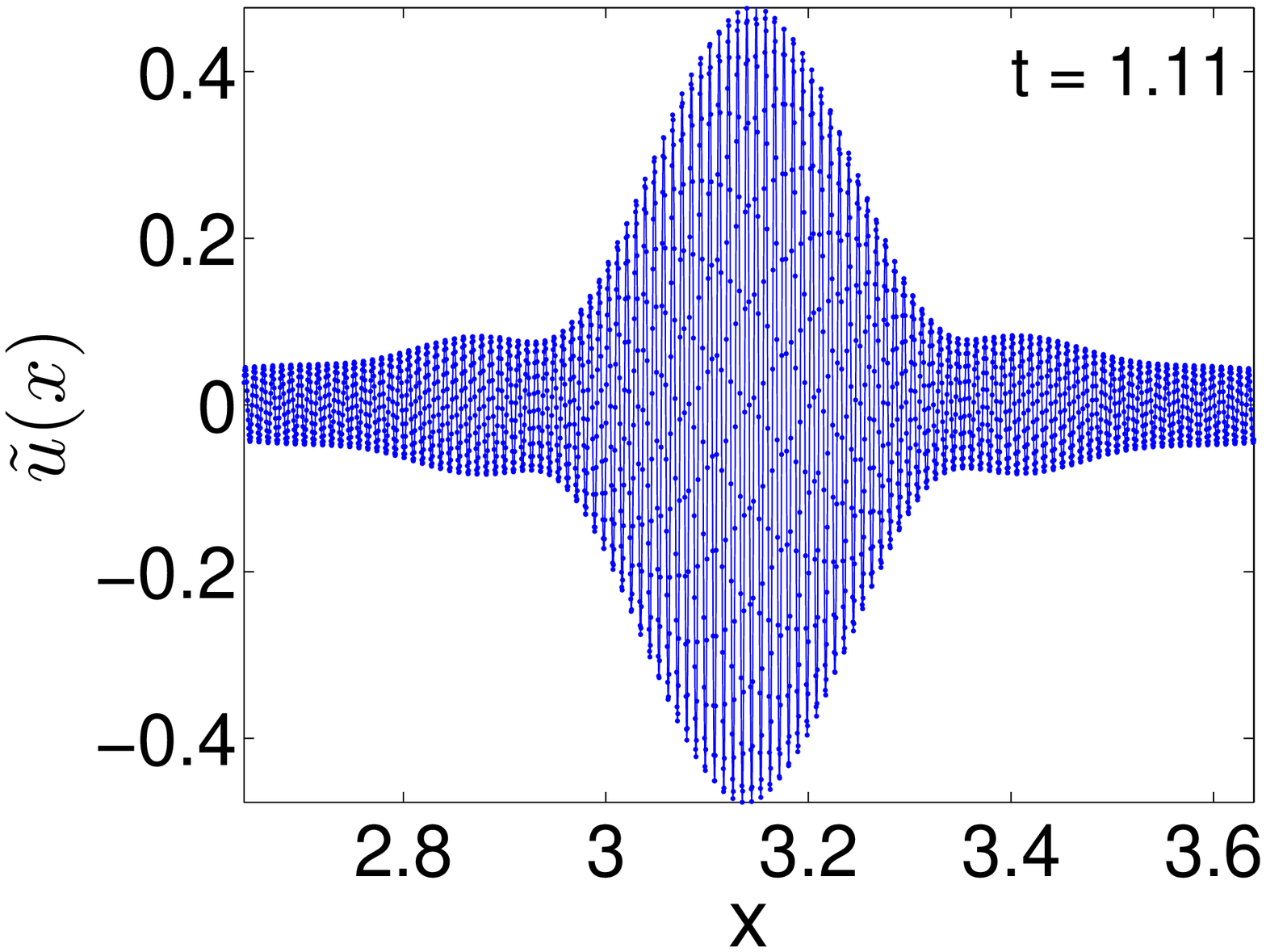}
\includegraphics[height=4cm]{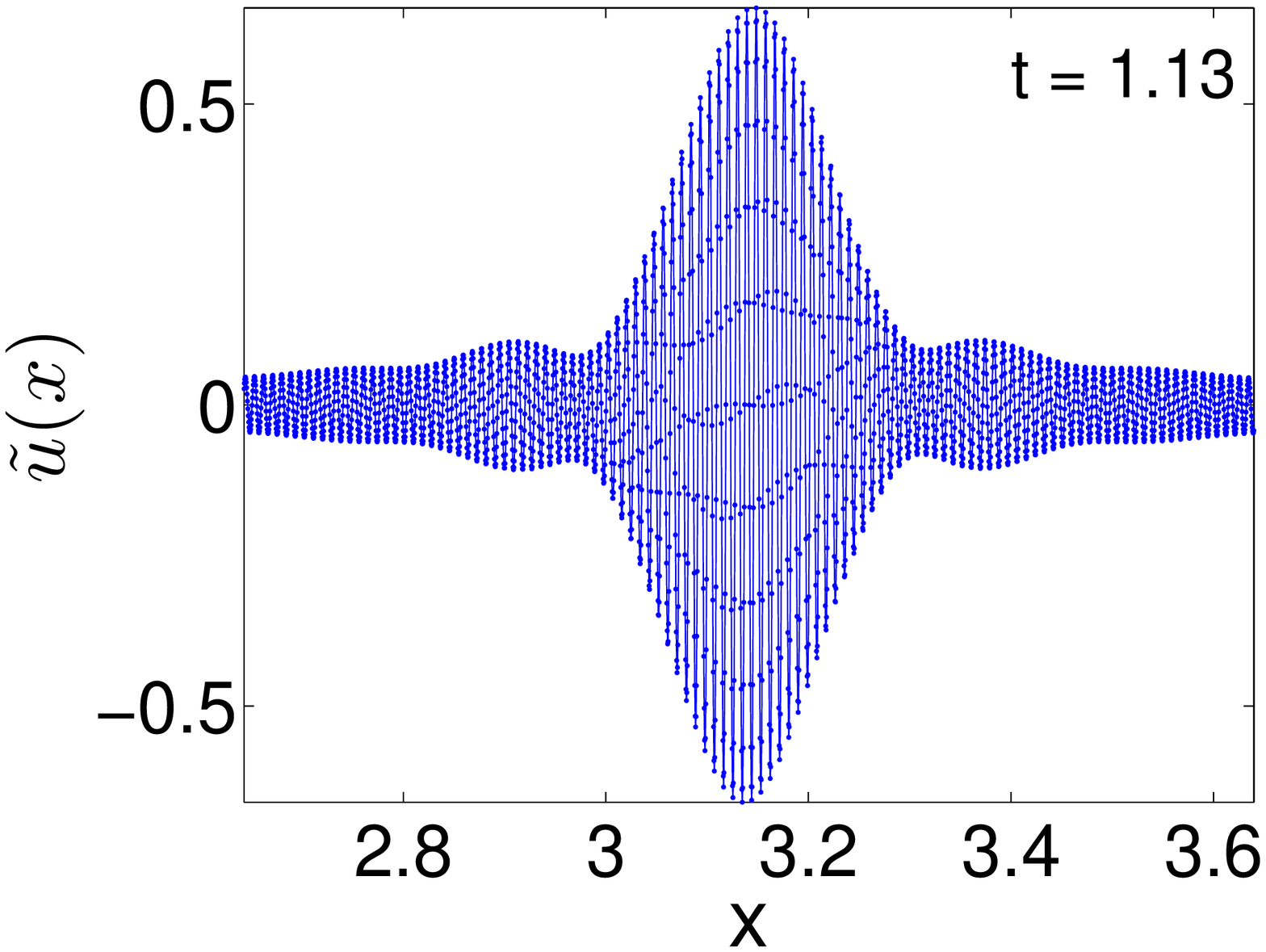}
\includegraphics[height=4cm]{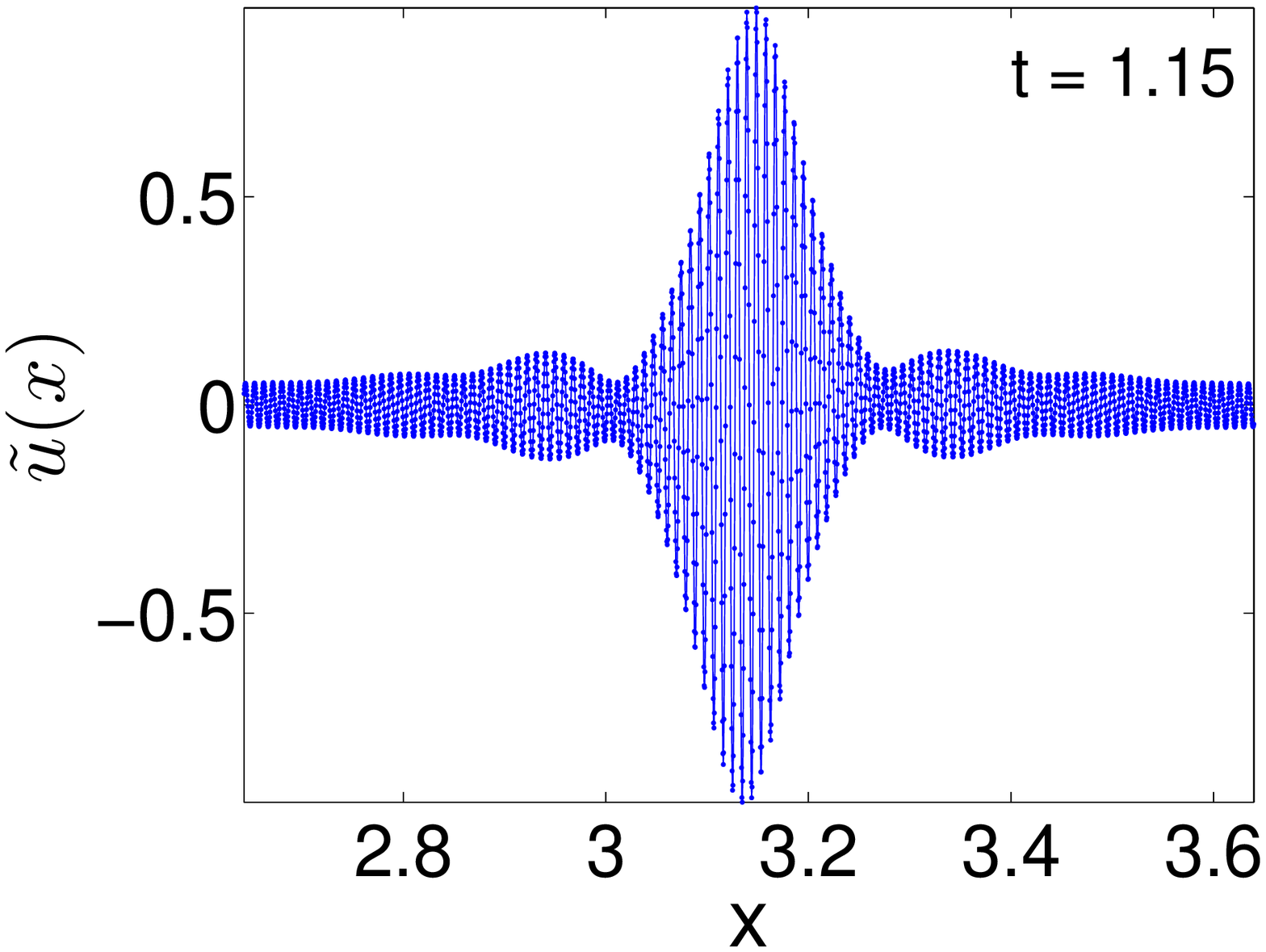}
\includegraphics[height=4cm]{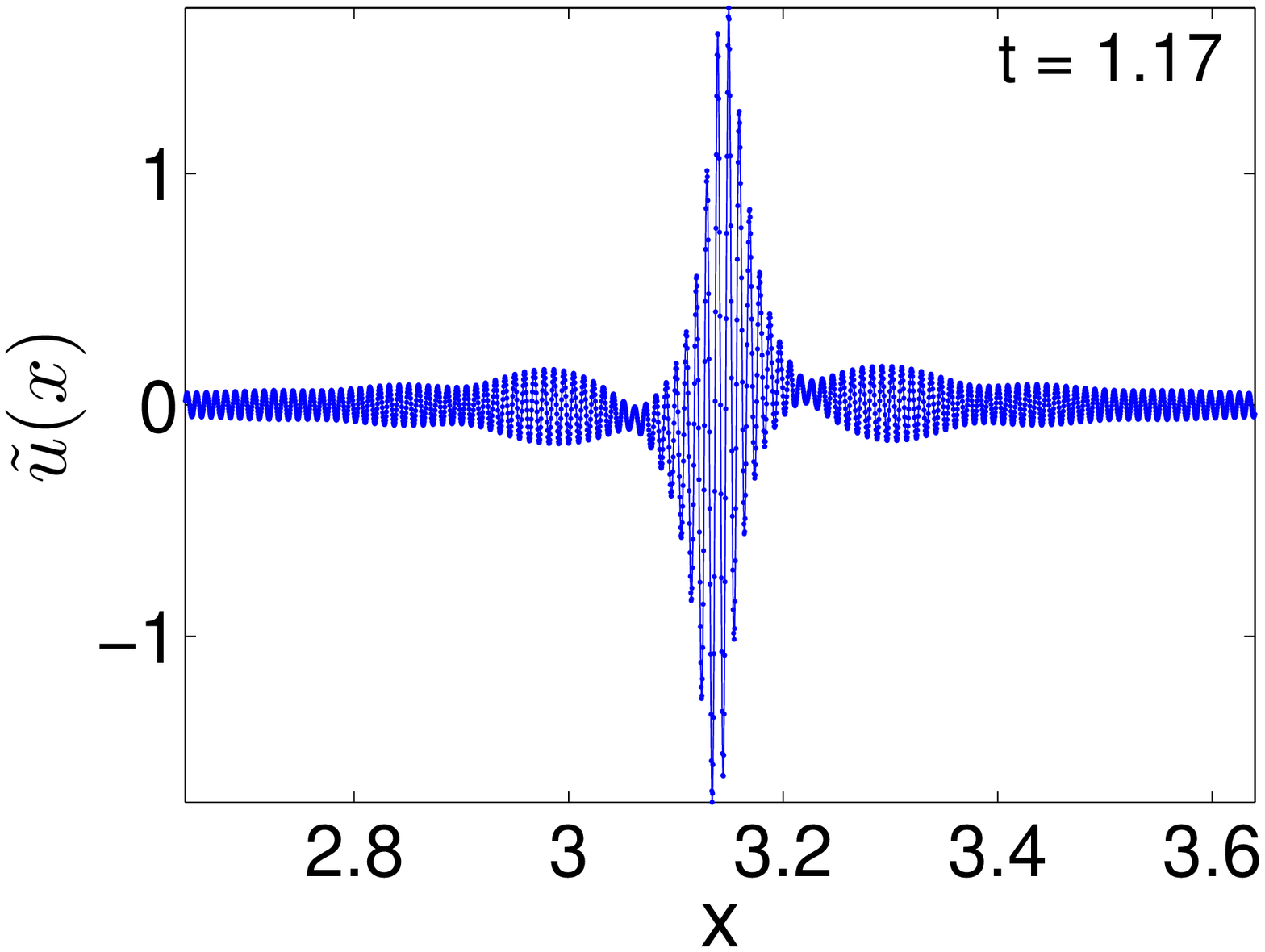}
\includegraphics[height=4cm]{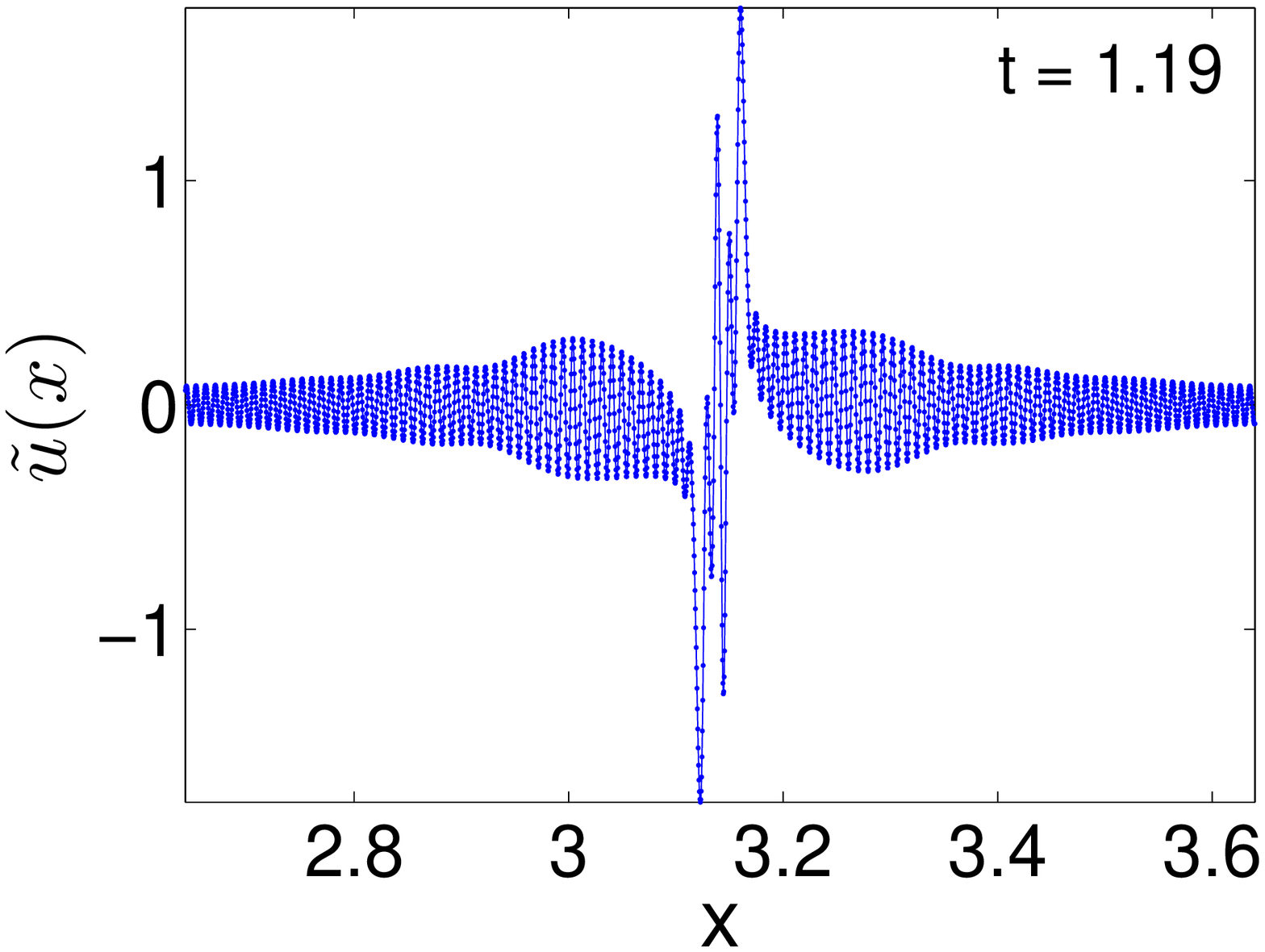}
\includegraphics[height=4cm]{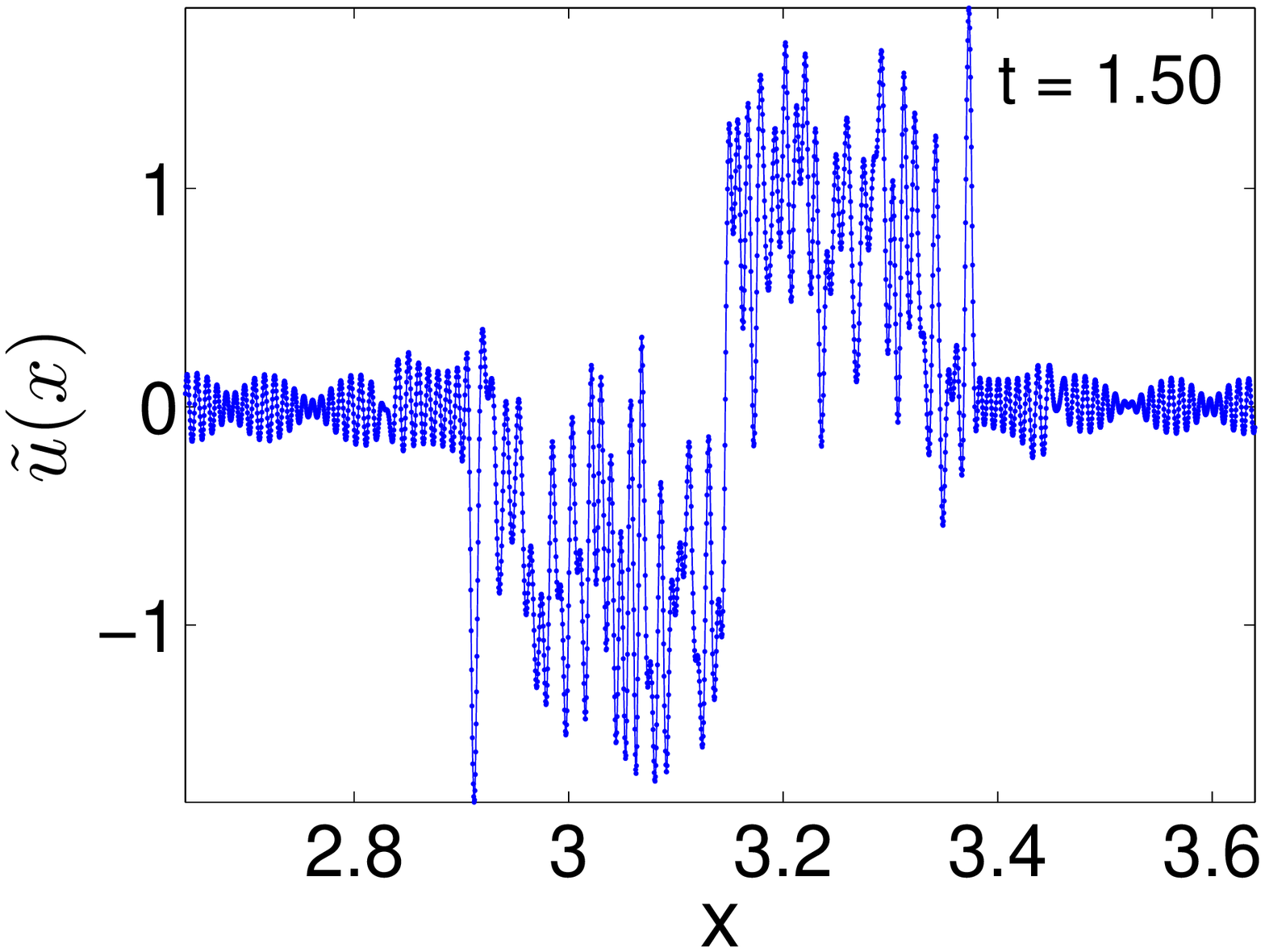}
\end{center}
\fi
\caption{(Color online) Evolution of the tyger (discrepancy) for same conditions as 
in Fig.~\ref{f:smtygerappearance}: growth, thinning, asymmetrization, collapse and chaotization.}
\label{f:collapse}
\end{figure*}

The birth of 
tygers around the time $\ts$ of the first preshock will be studied 
in Sec.~\ref{s:birth}. In particular, in Sec.~\ref{ss:tstardata}, we
shall present evidence for scaling  properties with $\kg$
of both tyger amplitude and width at $t=\ts$. Here we shall focus on
the temporal evolution at  later times.
Figs.~\ref{f:collapse}, corresponding to the single-mode initial
condition and $\kg =700$, show the
evolution of the tyger in terms of the discrepancy $\ut=v-u$ from $t=1.07$ to $t=1.50$, shortly after its
birth around $\ts =1$.  The first few panels display very symmetrical
(even)  bulges
\begin{figure*}[htbp]
\iffigs
\begin{center}
\includegraphics[height=4cm]{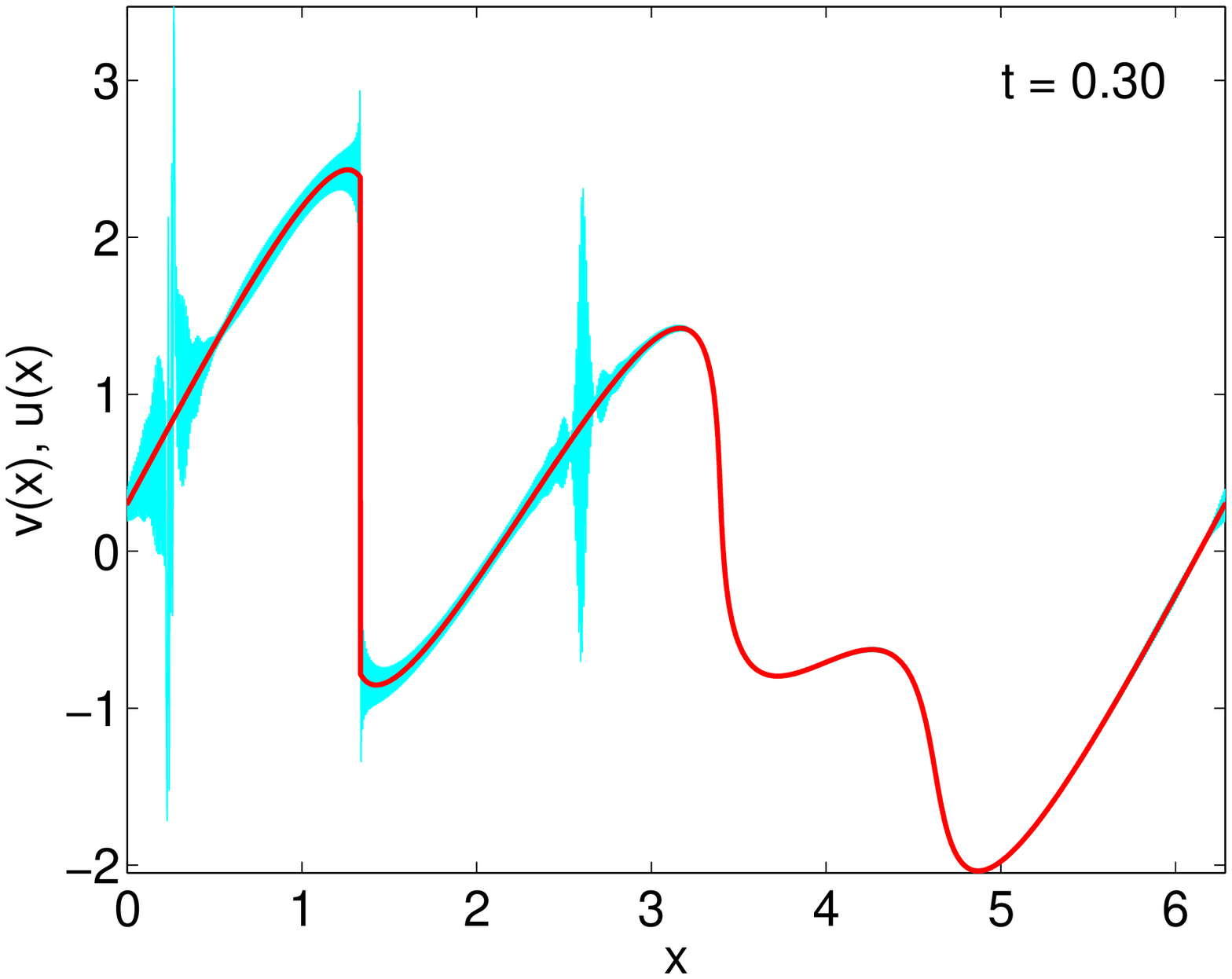}
\includegraphics[height=4cm]{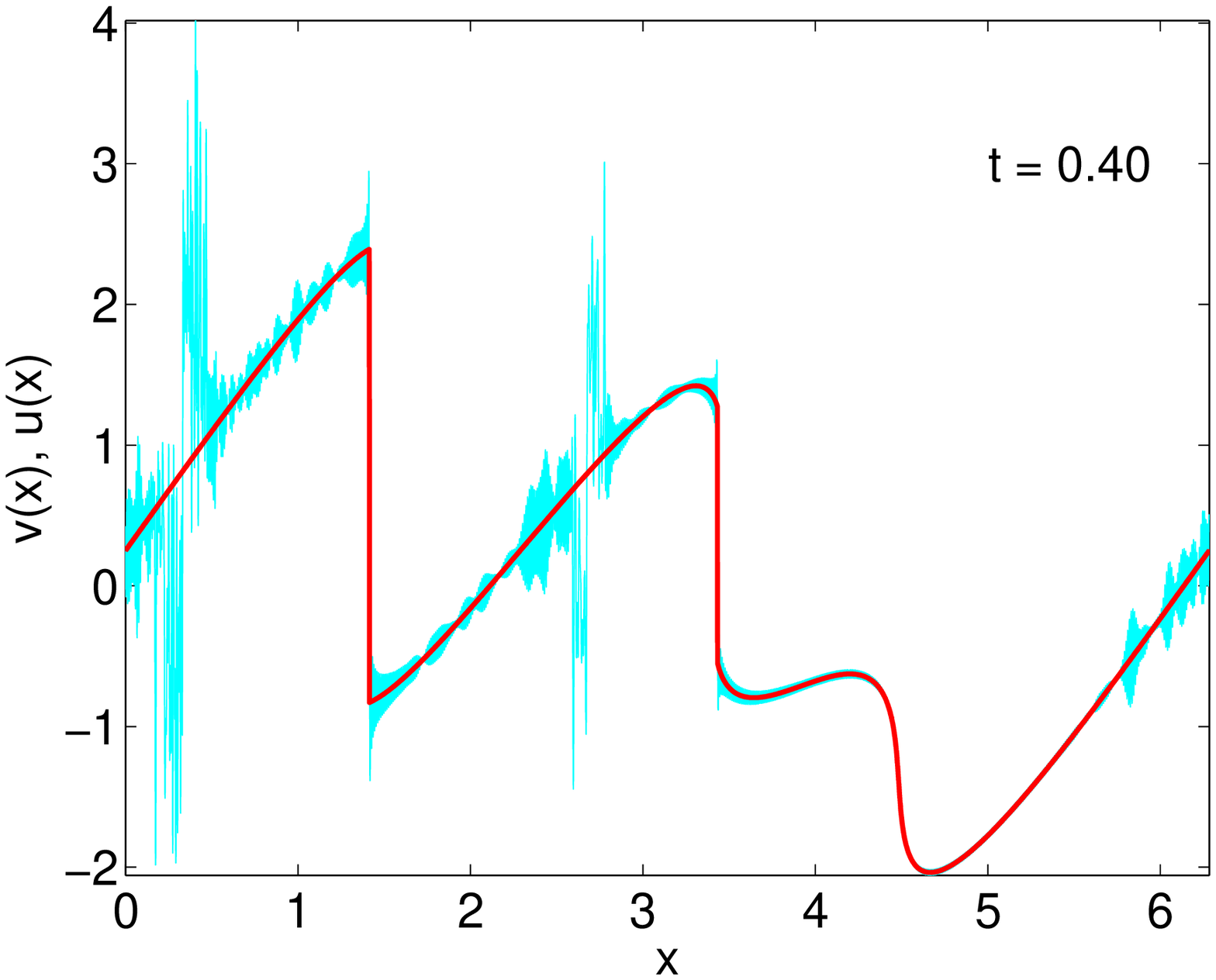}
\includegraphics[height=4cm]{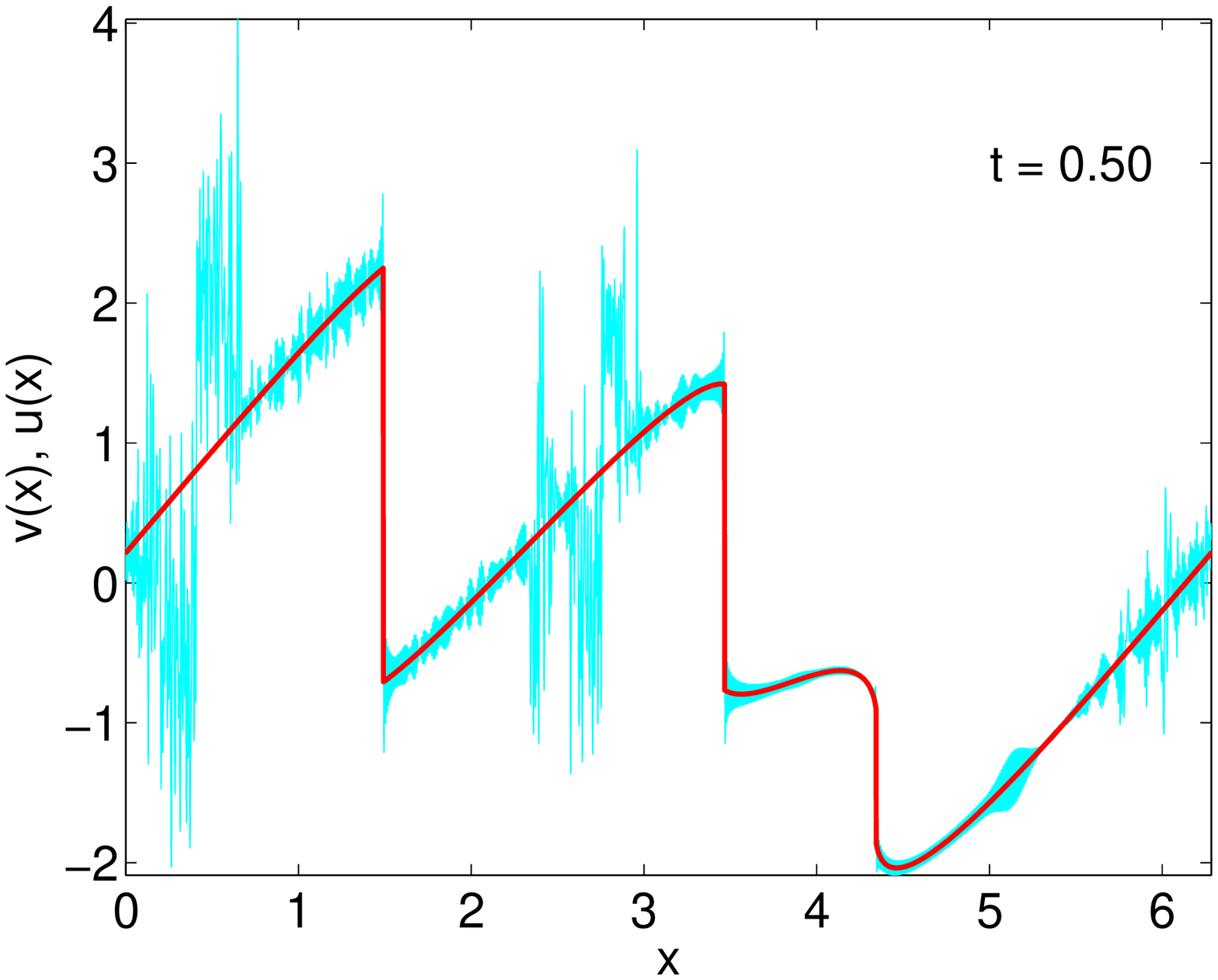}
\includegraphics[height=4cm]{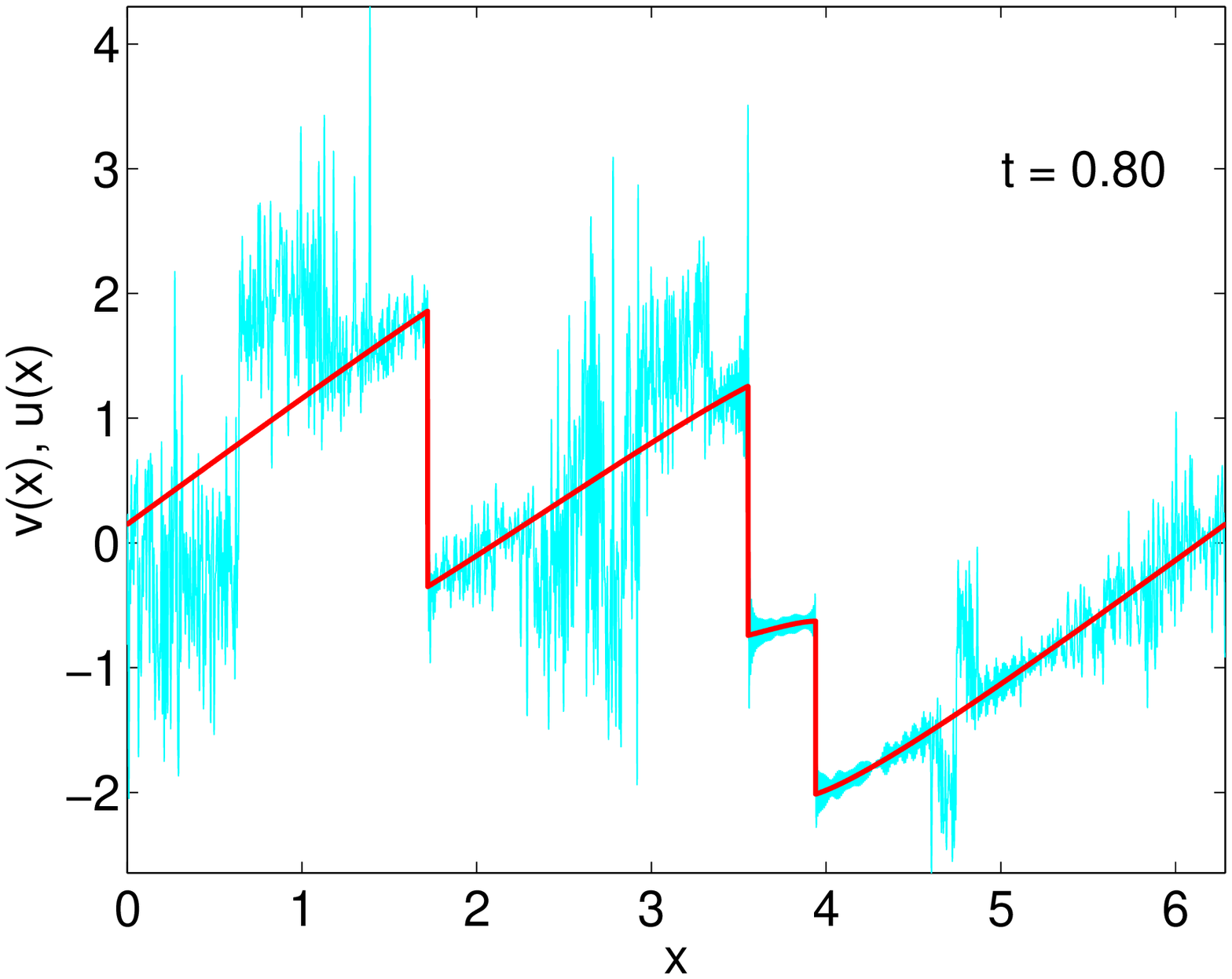}
\includegraphics[height=4cm]{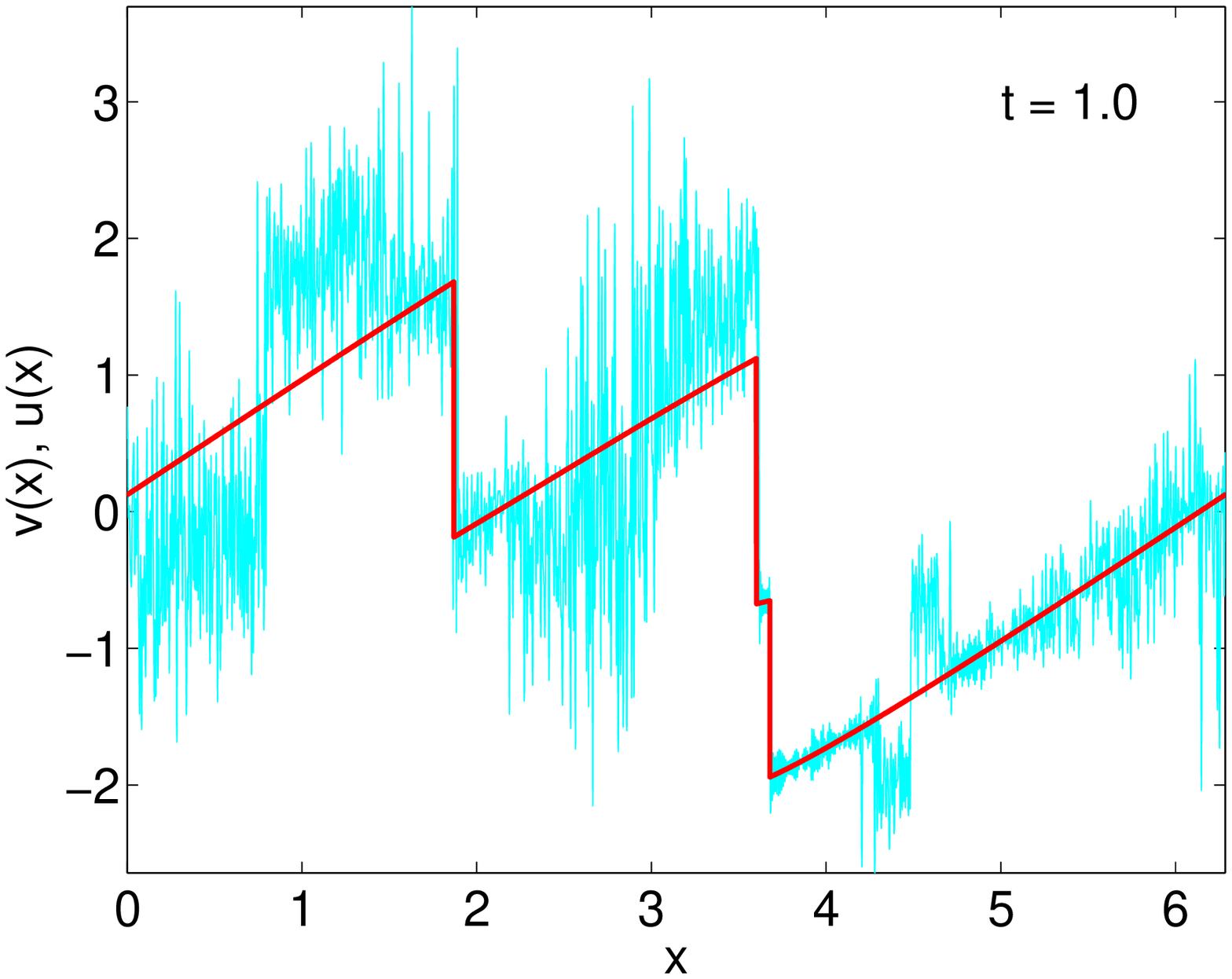}
\includegraphics[height=4cm]{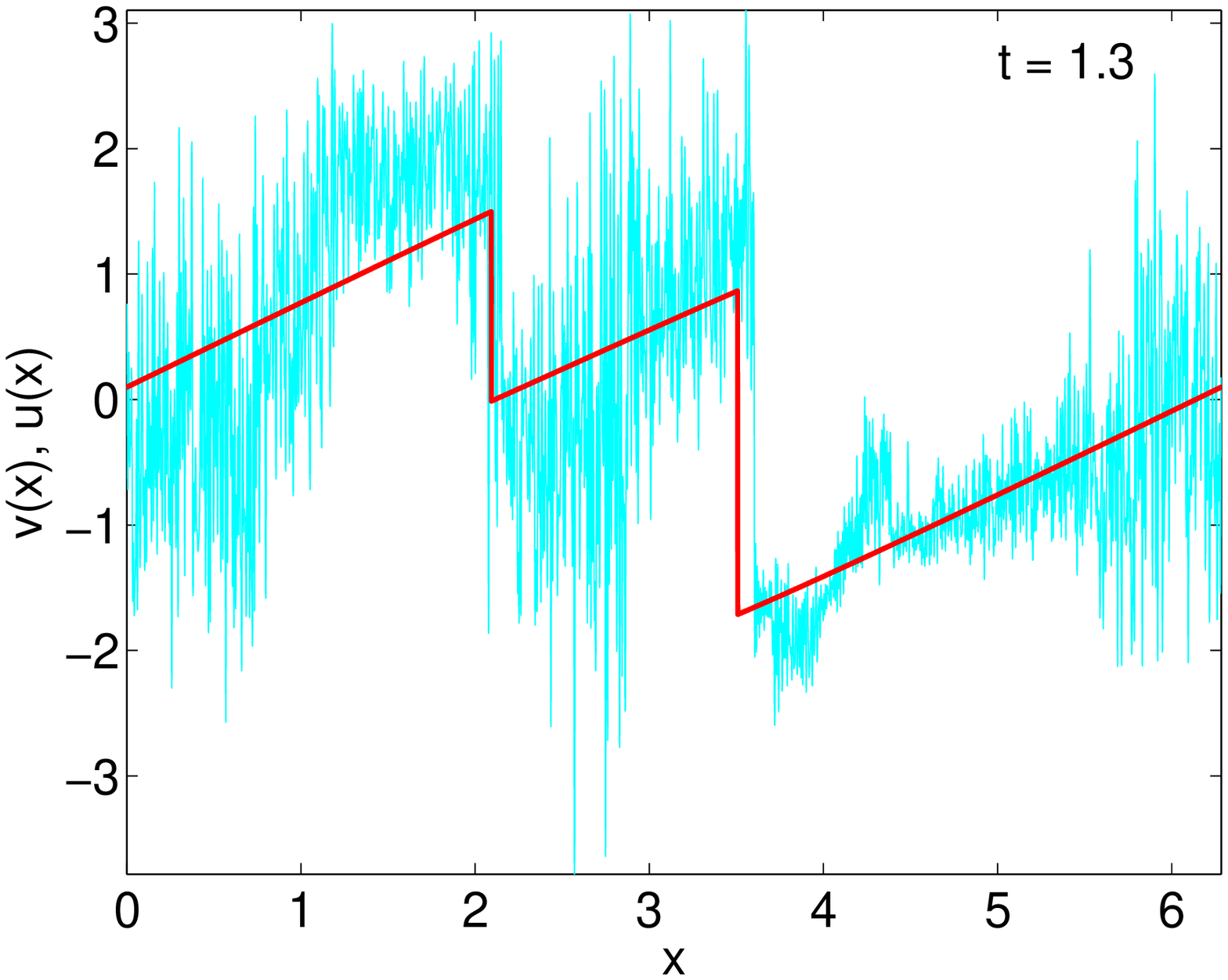}
\includegraphics[height=4cm]{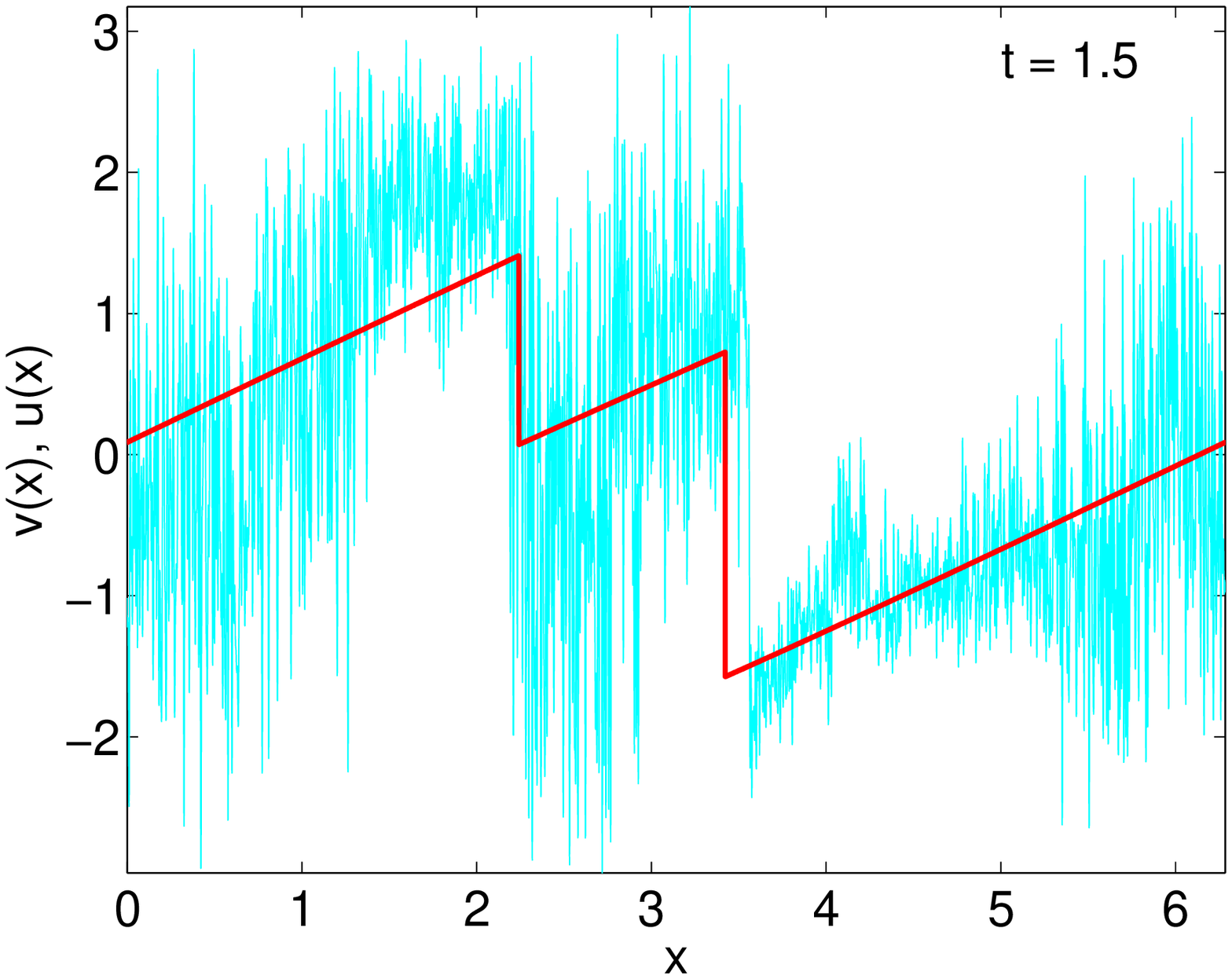}
\includegraphics[height=4cm]{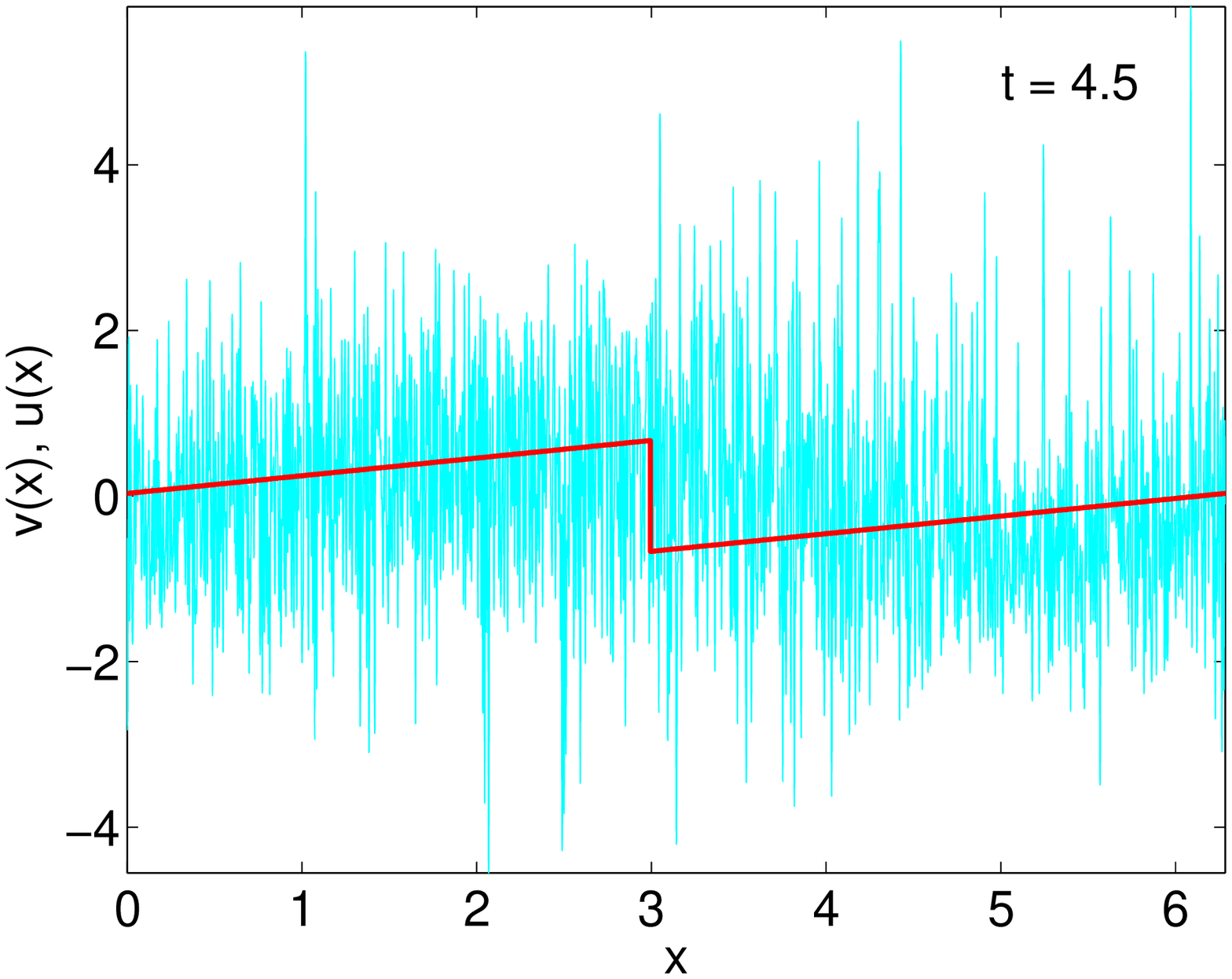}
\end{center}
\fi
\caption{(Color online) Long-time evolution of three-mode initial condition (other
  parameters as in Fig.~\ref{f:smtygerappearance}). The cyan curve (light grey) is
  the Galerkin-truncated solution and the red one (black) the untruncated
  inviscid-limit solution. Observe that tygers progressively
  invade the ramps between shocks but that shocks remain sharp and
  correctly
placed as long as the spreading out of the tygers on the ramps has not reached them. At very long
times,
the truncated solution is thermalized.}
\label{f:3modelate}
\end{figure*}
whose amplitudes grow in time, because truncation wave input
has accumulated, while their width decreases (thinning), as a consequence of
phase mixing.
Rather quickly, the decrease in width leads to a collapse of the tyger,
around $t=1.19$. This is preceded and accompanied by a growing
asymmetry of the tyger for which we offer the following
interpretation. The tyger contains kinetic energy in the form
of modulated oscillations at the Galerkin wavelength (we shall see
that this increasing energy compensates the loss of energy in shocks).
Over scales large compared to the Galerkin wavelength but small
compared
to the tyger width, this kinetic energy gives rise to an $x$-dependent
Reynolds stress, which pulls the tyger envelope up where 
the envelope has a negative slope and down where the slope is
positive. The resulting asymmetry becomes very conspicuous after the
collapse, as seen in the last panel of Fig.~\ref{f:collapse}.  This panel
has at least two other noteworthy features. To the right and left of
the central point $x=\pi$ we see two pieces that look like a portion 
of ($\kg$-truncated) white noise; this is the very beginning
of thermalization. Observe that the right piece is shifted vertically
with respect to the left one and that the transition looks almost like
an antishock (a shock which goes up rather than down, as prescribed
by the inviscid limit). 
\begin{figure*}[htp]
\iffigs
\begin{center}
\includegraphics[height=5.5cm]{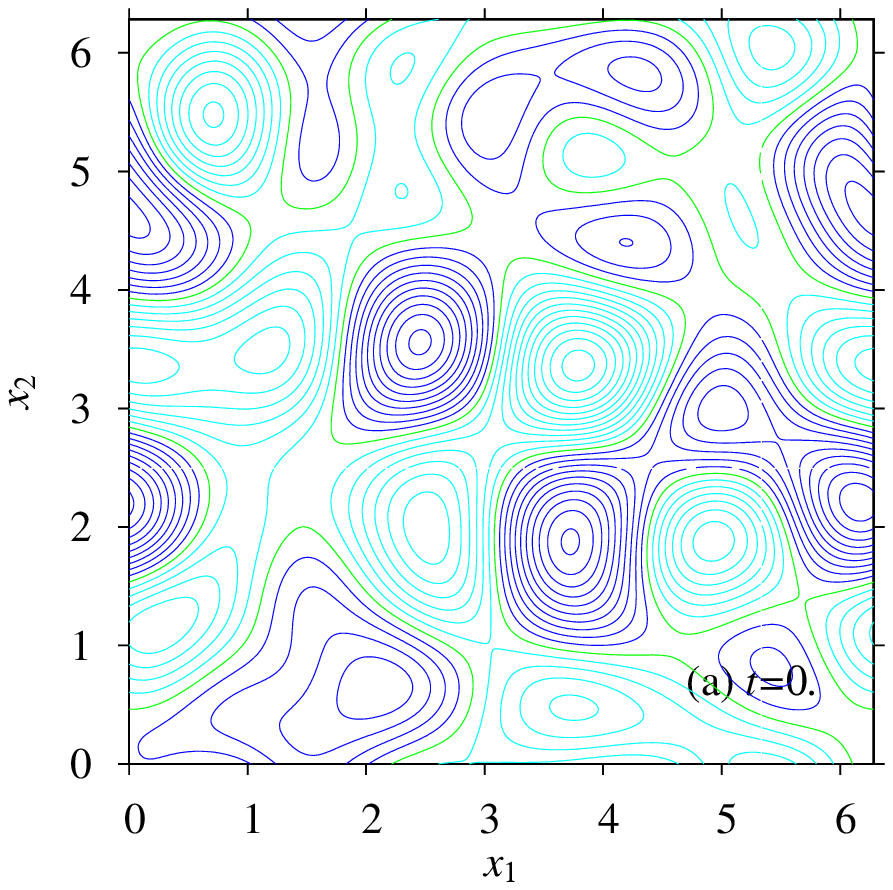}
\includegraphics[height=5.5cm]{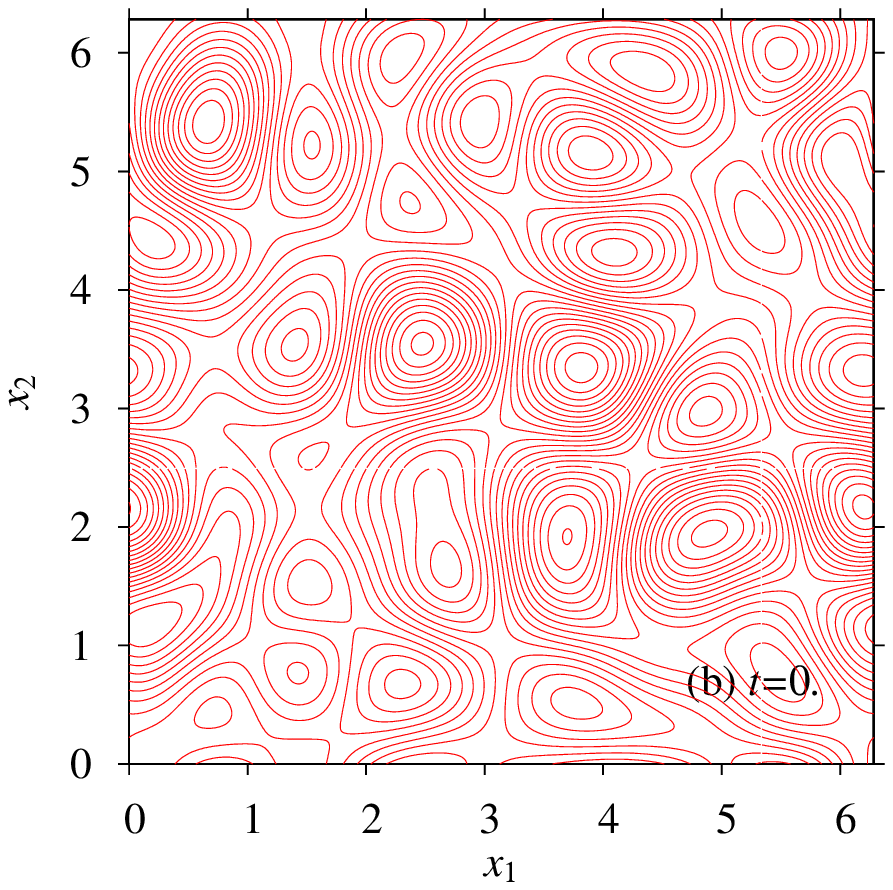}\\
\includegraphics[height=5.5cm]{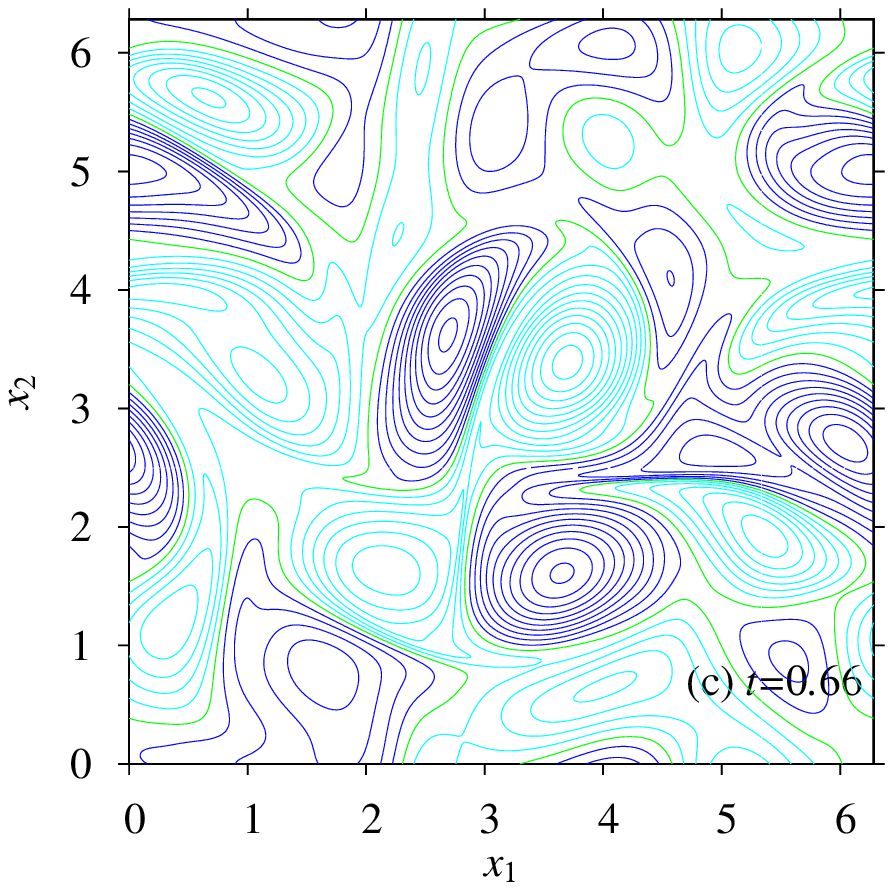}
\includegraphics[height=5.5cm]{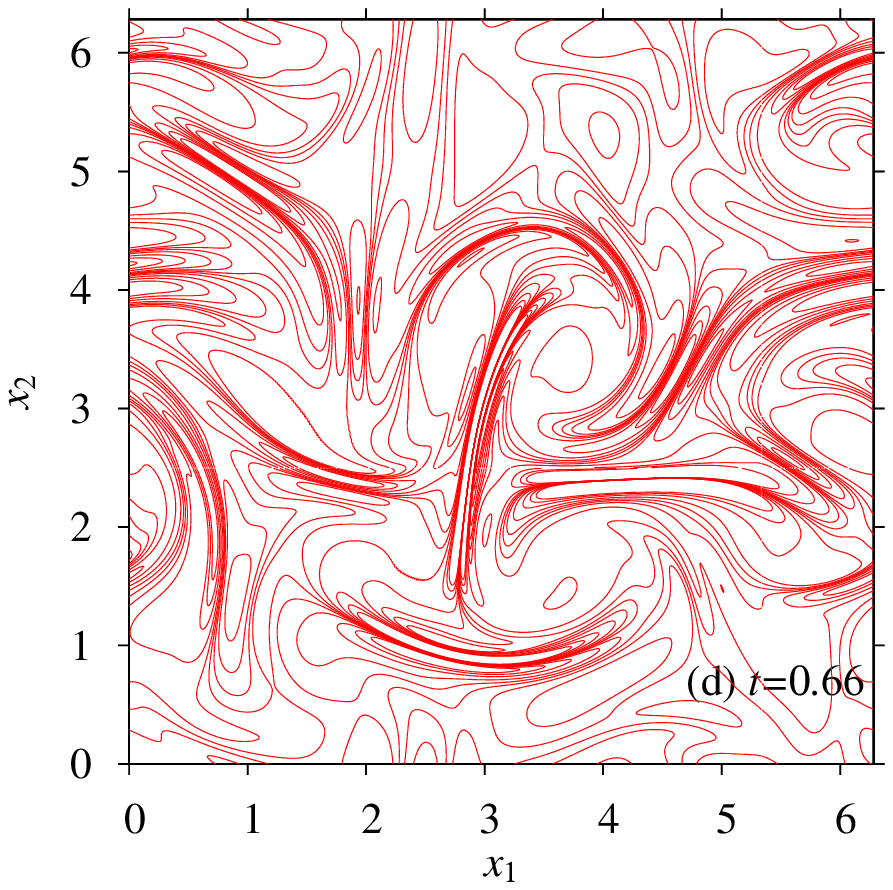}
\end{center}
\fi
\caption{(Color online) (a) shows contours of vorticity with random  initial conditions; positive values in dark blue (black), negative
  values in light blue (light gray); (c): same at time $t=0.66$ (before
  appearance of tygers); (b) and (d) are
  contours of the Laplacian of the vorticity at times $t=0$ and
  $t=0.66$. Notice the thin elongated red (black) ``cigars'' which play here
  the role of the preshocks/shocks in the Burgers case.}
\label{f:notyger}
\end{figure*}

Let us also observe that around the time
of collapse there is an apparent change of symmetry. Since the
single-mode initial condition is odd (after shifting the origin of the
$x$-axis to the center of the tyger) it stays odd at all later times.
This is however a statement about the full solution, down to the
Galerkin wavelength. If we concentrate on the larger-scale aspect
(the envelope of the bulge), we find that the discrepancy is even until somewhat
before collapse and odd after collapse.

For later phases of the tyger growth, it is better to show
simultaneously
the truncated solution and the untruncated one. Also, we switch to
the three-mode initial condition which has less symmetry and
is thus more generic. Fig.~\ref{f:3modelate}
shows the evolution from $t=0.3$, slightly after the first singularity
at $\ts =0.2218$, to $t=4.5$ when the solution of the truncated problem is
basically completely thermalized. 
From $t=0.3$ to $t=0.8$ we observe
that the chaotic-looking 
thermalized regions born after collapse are growing in extent and are
affecting more and more of the  ramp-like structure which are a well-known
feature of the  solution of the untruncated Burgers equation after the formation of
shocks. In short, we shall say that ``the tyger spreads out on the ramp''.

As long as significant tyger activity has not reached the shocks,
their positions, amplitudes and motions are correctly described
by the Burgers equation, down to the Galerkin wavelength. We have
checked that during this phase even shock merger is unaffected by truncation (an
instance
is seen around $t=1.0$).  Later, strong tyger activity near the
edges of shocks is able to shift them slightly (this is visible at $t=1.3$).
Once the shock amplitude has decayed to values much less than the
tyger fluctuations, the solution looks globally thermalized
($t=4.5$). It must be noted that the mechanism which prevents
thermalization
in the Fermi--Pasta--Ulam problem \cite{FPU} does not seem to be
present here.

\subsection{2D Euler}
\label{ss:twod}

So far we have worked with a very special hydrodynamical equation: the Burgers
equation is integrable, compressible and its solutions generically blow
up after a finite time. Is the tyger phenomenon also present when
none of these properties hold, as is the case for  the two-dimensional
incompressible Euler equation with smooth (analytic) initial data 
and space-periodic boundary conditions?

\begin{figure*}[htbp]
\iffigs
\begin{center}
\includegraphics[height=7.4cm]{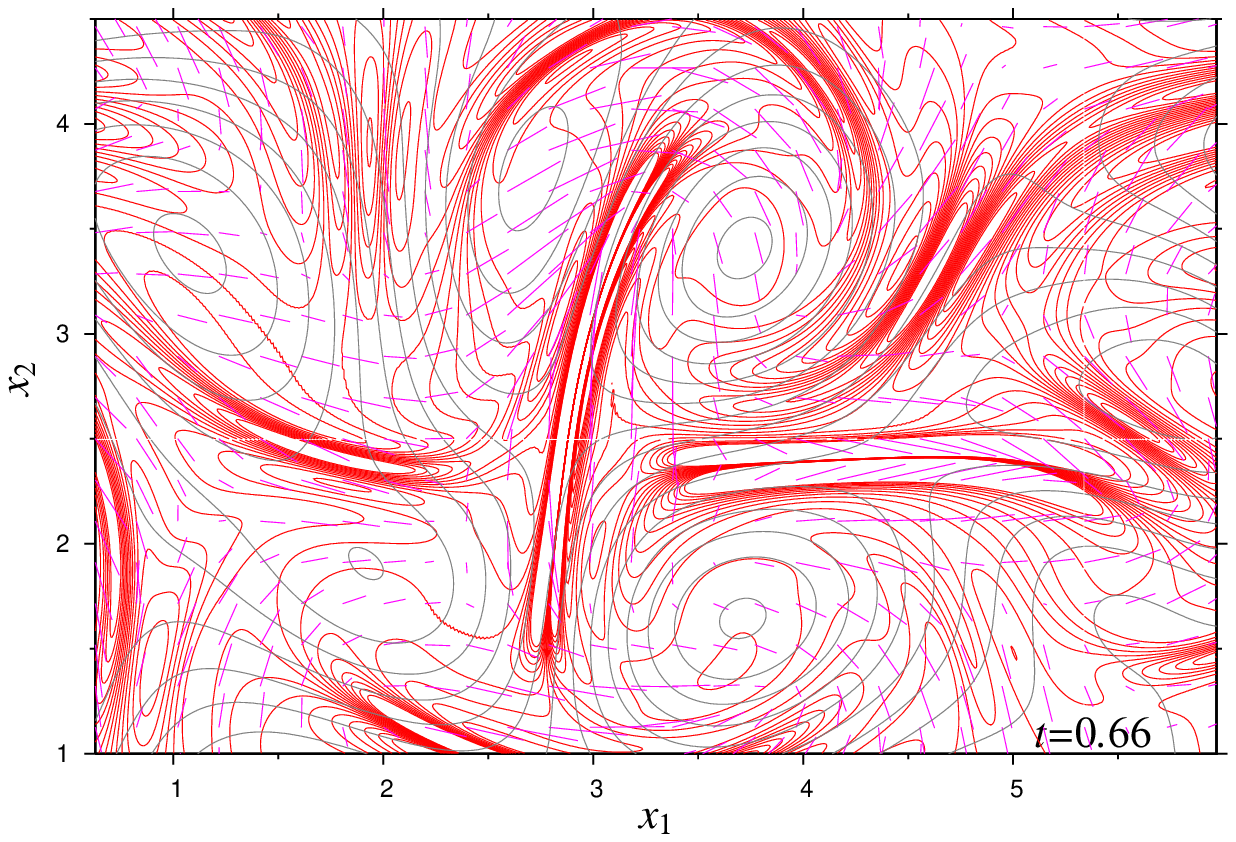}
\includegraphics[height=7cm]{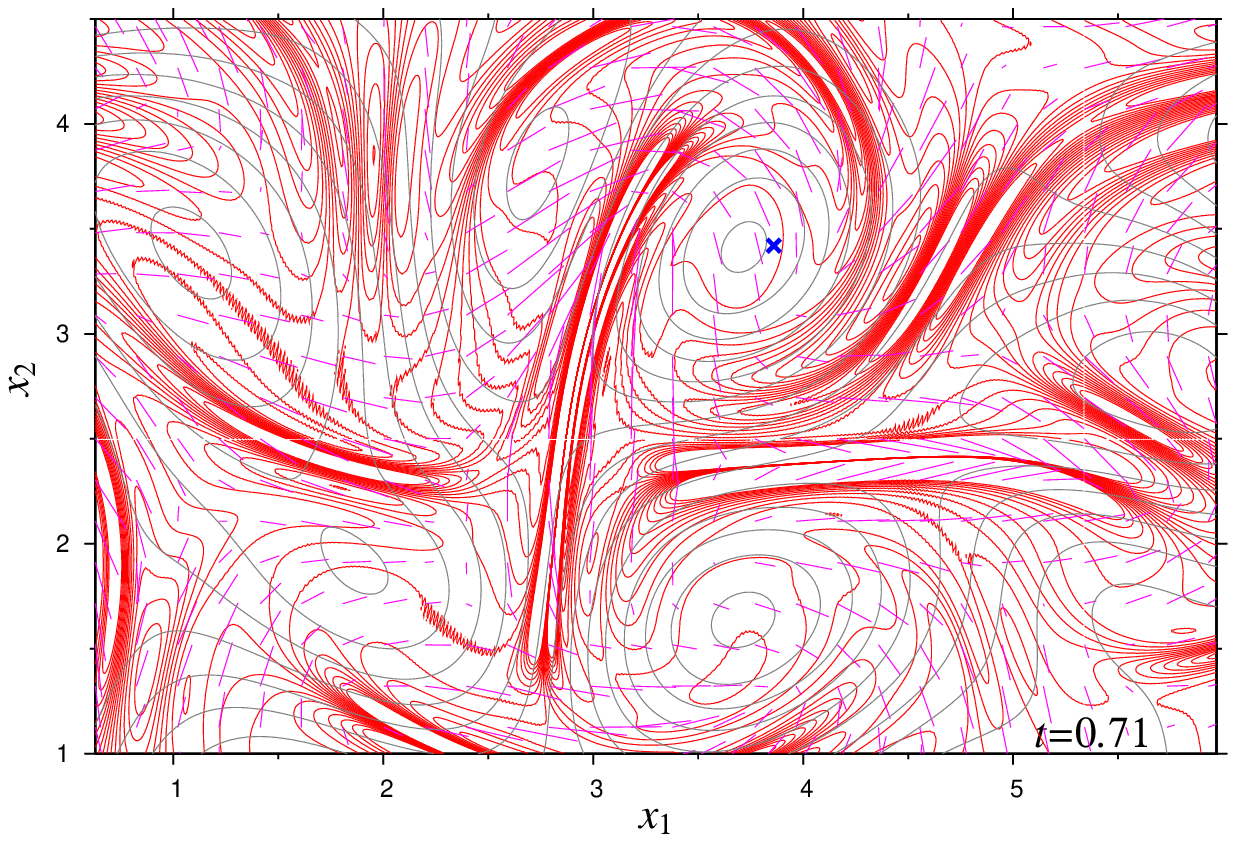}
\includegraphics[height=7cm]{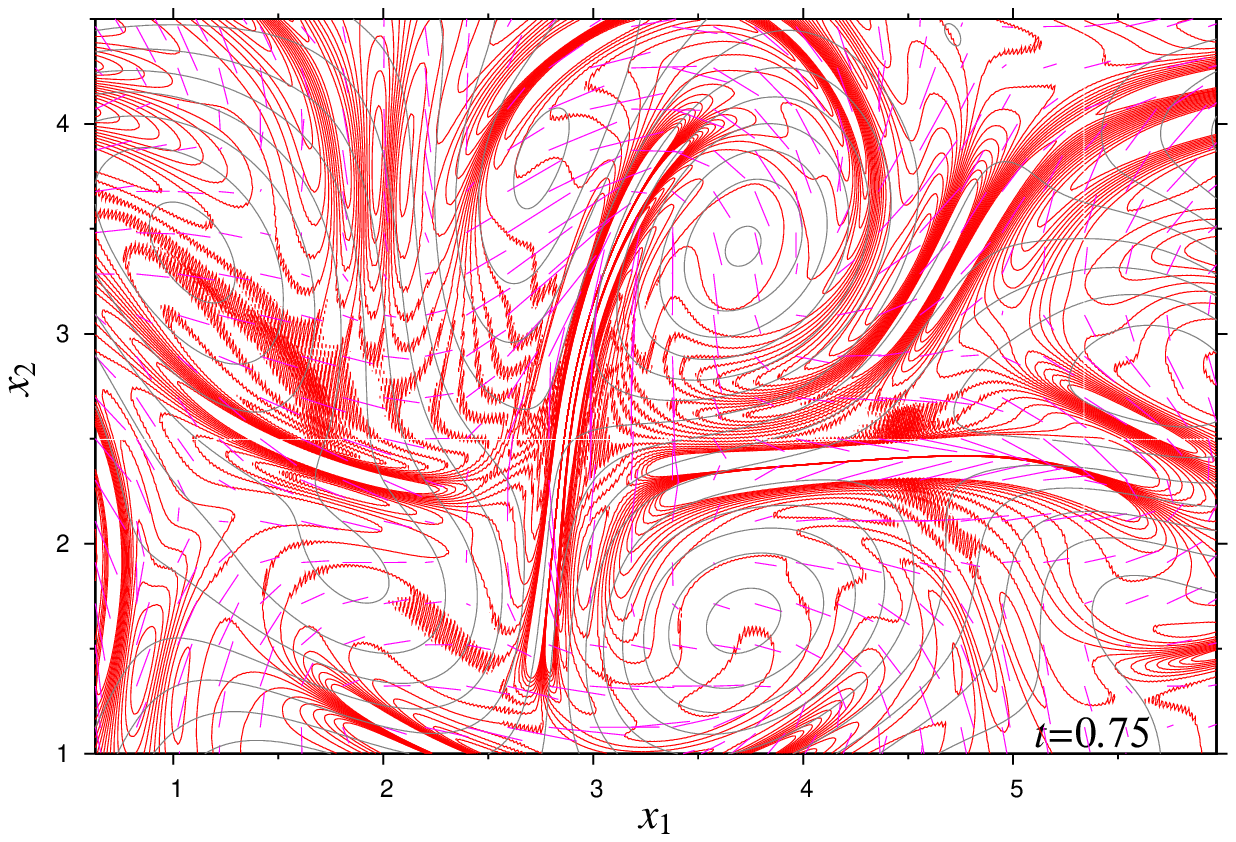}
\end{center}
\fi
\caption{(Color online) A 2D tyger: before ($t=0.66$), early ($t=0.71$) and
  later ($t=0.75$). Figures, moderately zoomed, centered on the main
  cigar. Contours of the Laplacian of vorticity in red (black), ranging from $-200$ to
  $200$ by increments of $25$, streamlines in gray, ranging from $-1.6$ to
  $1.6$ by increments of $2$ and positive strain eigendirections in pink (light
 gray) segments. Notice tygers appearing at $t=0.71$ in the form of wavy red (black)
  contours at places where the velocity is roughly parallel to the main cigar;
  later, the tygers grow in strength and extension. A blue  (black) x-mark is added at
  $t=0.71$ to underline that there is no tyger, in spite of resonance, because
  of a wrong strain direction. }
\label{f:2dtygerdevelopment}
\end{figure*}

The short answer is ``yes''. We have numerically investigated quite a number
of different initial conditions, including the two-mode Standard
Orthogonal Case (SOC) initial condition
used in \cite{paulsetal} and random initial conditions. The simulations were
done with resolutions from $512^2$ to $8192^2$.  Here we report the results
for the random  initial condition, which is $2\pi$-periodic in
$x_1$ and $x_2$. The Fourier space consists of couples of signed integers $\kb
\equiv (k_1,\, k_2)$. It is here decomposed for convenience into shells
corresponding to a $K\le |\kb|< K+1$, where $K$ is an integer. Each such shell
has $N(K)$ Fourier modes.  For $\kb$ in the $K$th shell, the Fourier
coefficients $\hat\omega_{\kb}$ of the initial vorticity are taken all with the same
modulus $2K ^{7/2} \exp(-K^2/4)/N(K)$ and with phases that are uniformly and
independently distributed in the interval $[0,2\pi[$, except that opposite
wavevectors are given opposite phases to preserve Hermitian
symmetry. The tyger calculations shown here are all with resolution
$1024^2$ and Galerkin truncation wavenumber $\kg =342 =(1024+2)/3$.  The
particular realization used as initial condition in the present calculation can be retrieved
from \url{http://www.kyoryu.scphys.kyoto-u.ac.jp/}\url{~takeshi/populated/}.
Fig.~\ref{f:notyger} shows the vorticity and its Laplacian at $t=0$ and
$t=0.66$, the latest time at which no tyger is seen (at least in the
fields displayed). Although for the untruncated solution real singularities are ruled out at any
finite time, there is strong enhancement of spatial derivatives
of the vorticity \cite{yudovich}. The highest values of the Laplacian is found in the
straight cigar-like structure seen near the center of the figure
which---as we shall see---will play an important role in tyger generation
\footnote{All the simulations at resolution up to $8192 ^2$ (not shown
here) indicate that, just before being affected by truncation, the
strongest small-scale activity is in such cigars; they are very thin
in the transverse direction and their centerlines have a large radius
of curvature; they are located near hyperbolic critical points of the
vorticity where there is strong compression in the transverse
direction and strong extension in the longitudinal direction.}.
Furthermore this cigar moves very little because there is a velocity
stagnation point near its center \footnote{We also observed in a
number of simulations that, before truncation becomes important,
critical hyperbolic points of the stream function (stagnation points)
and of the vorticity are close to each other; this may be due to having most of the
enstrophy concentrated within a relatively narrow wavenumber band.}.
Fig.~\ref{f:2dtygerdevelopment}, which is centered on the strongest
cigar, shows the development of tygers. In terms of the Laplacian of
the vorticity  they become visible around $t=0.71$ and then become
much 
stronger. 

\begin{figure}[htbp]
\iffigs
\begin{center}
\includegraphics[height=6cm]{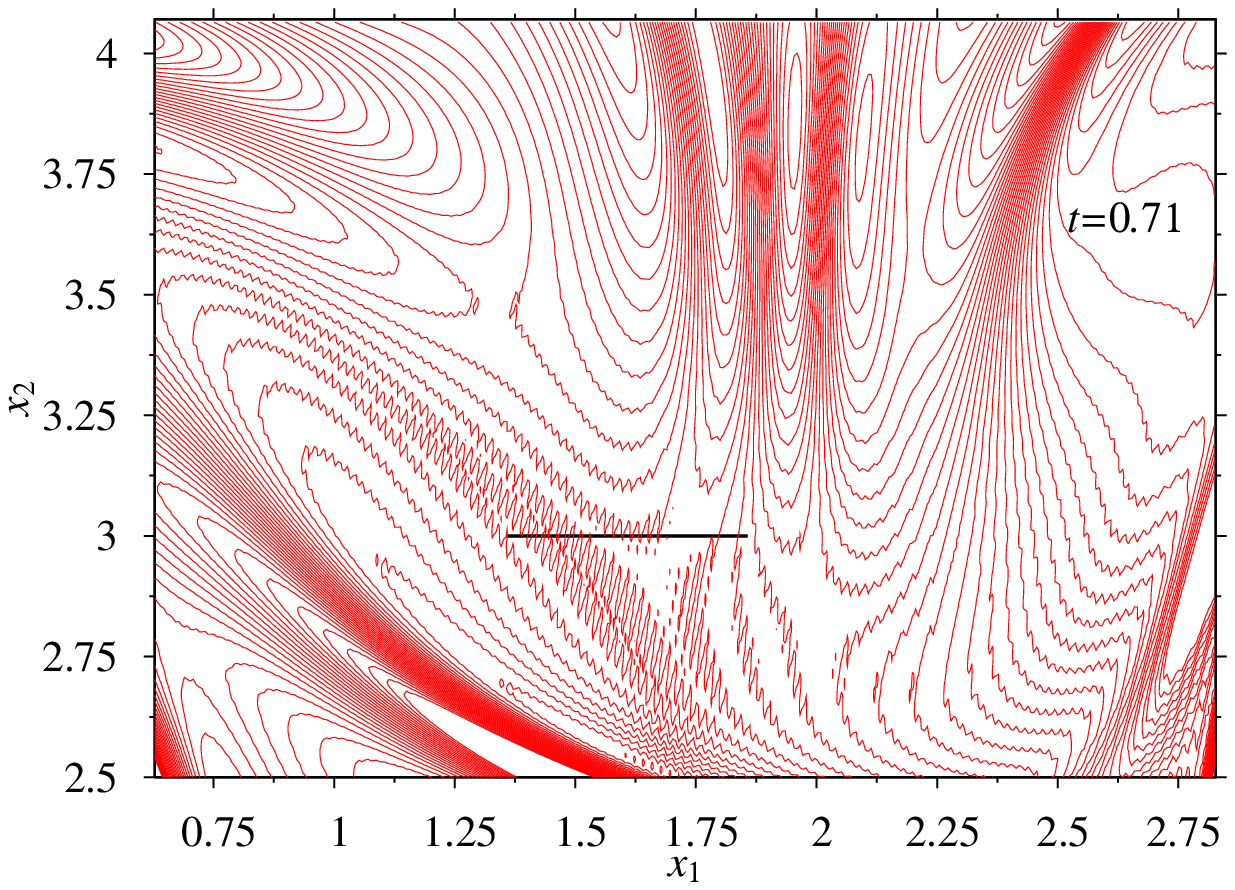}
\includegraphics[height=4cm]{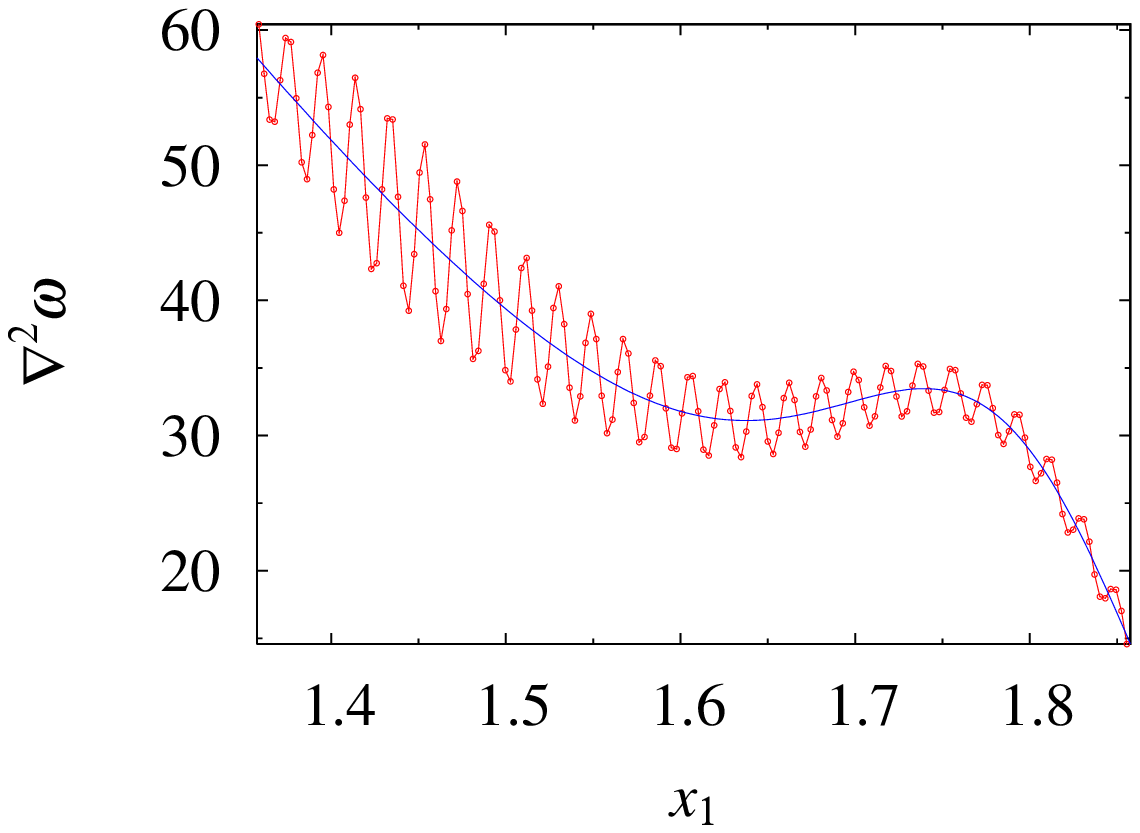}
\end{center}
\fi
\caption{(Color online) Upper: zoomed version of contours of the Laplacian of
  vorticity at $t=0.71$. Lower: plot of the Laplacian of vorticity
  along the horizontal segment near $x_2 =3$, shown in the upper panel.}
\label{f:tygersuperzoomed}
\end{figure}

A further look at a tyger is provided in Fig.~\ref{f:tygersuperzoomed} which
zooms into one of the tygers and also shows the Laplacian of
the vorticity along a cut. 
\begin{figure}[htbp]
\iffigs
\includegraphics[height=5cm]{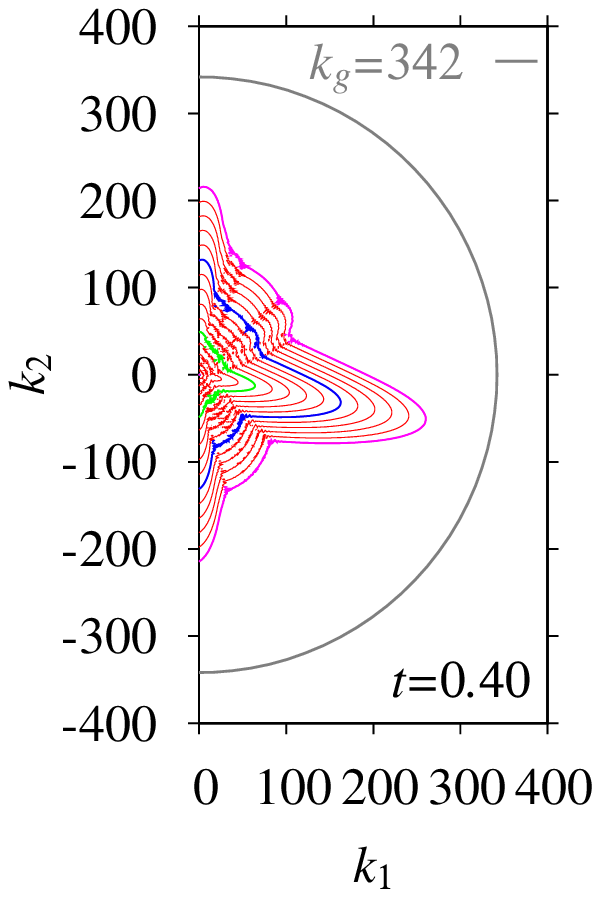}
\includegraphics[height=5cm]{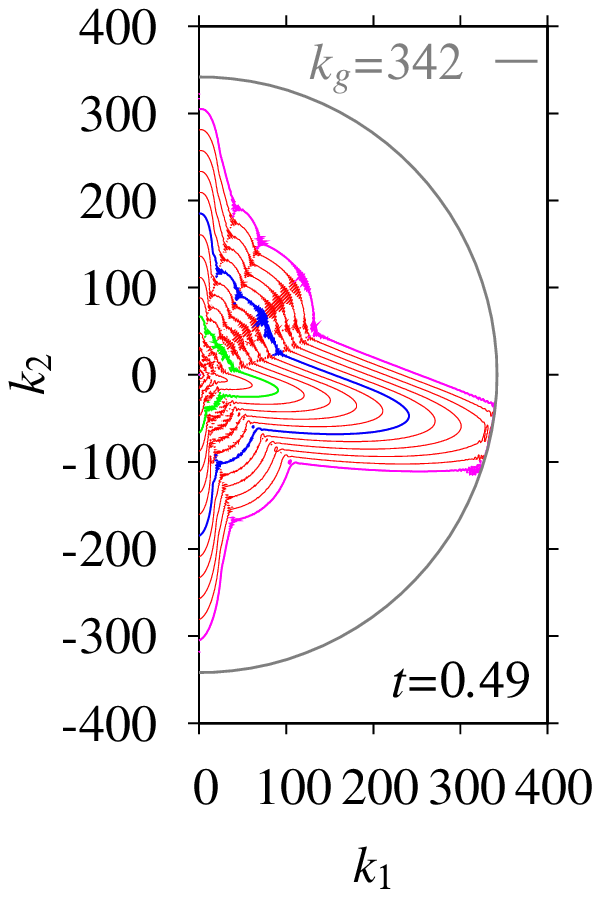}
\includegraphics[height=5cm]{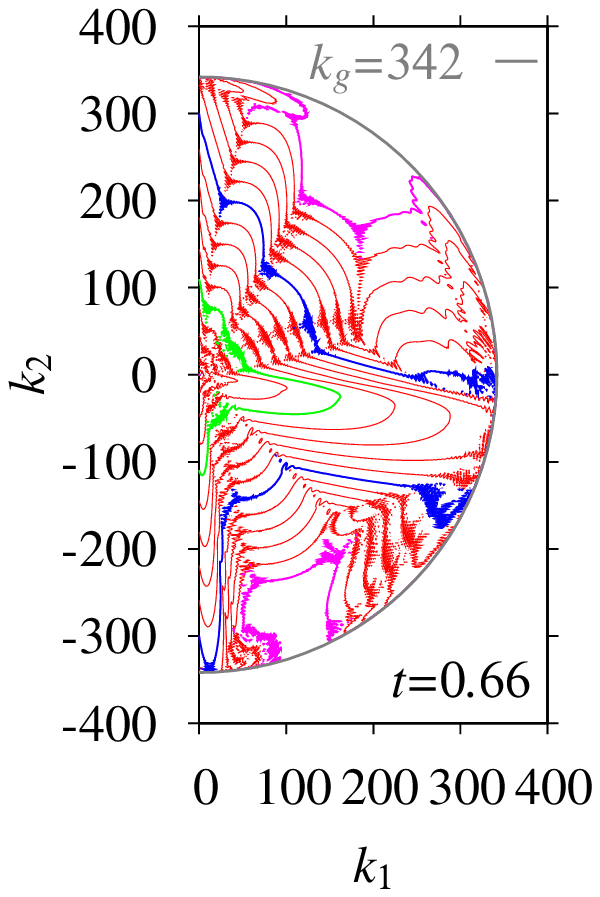}
\includegraphics[height=5cm]{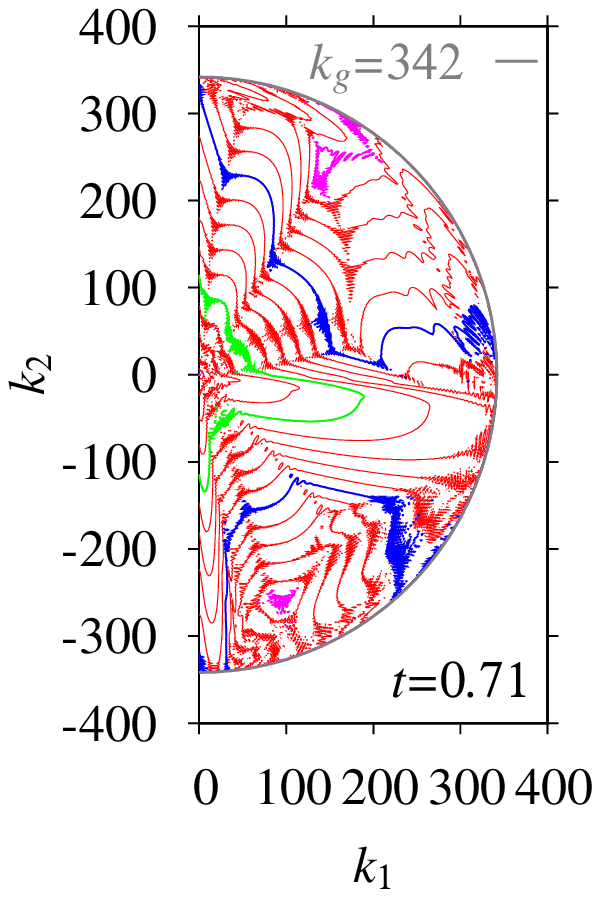}
\fi
\caption{(Color online) Contours of the modulus of the vorticity Fourier coefficients 
at various times. Negative $k_1$ values not shown because of Hermitian
symmetry. Contour values are $10^{-1},\, 10^{-2}, \ldots,  10^{-15}$
from inner to outer (in color version, green, blue and pink highlight
the values $10^{-5}, 10^{-10}$,
and $10^{-15}$, respectively). Galerkin truncation effects are visible
above the rounding level already at $t=0.49$ and become more and more invasive.}
\label{f:fouriervorticity}
\end{figure}
\begin{figure}[htbp]
\iffigs
\begin{center}
\includegraphics[height=5cm]{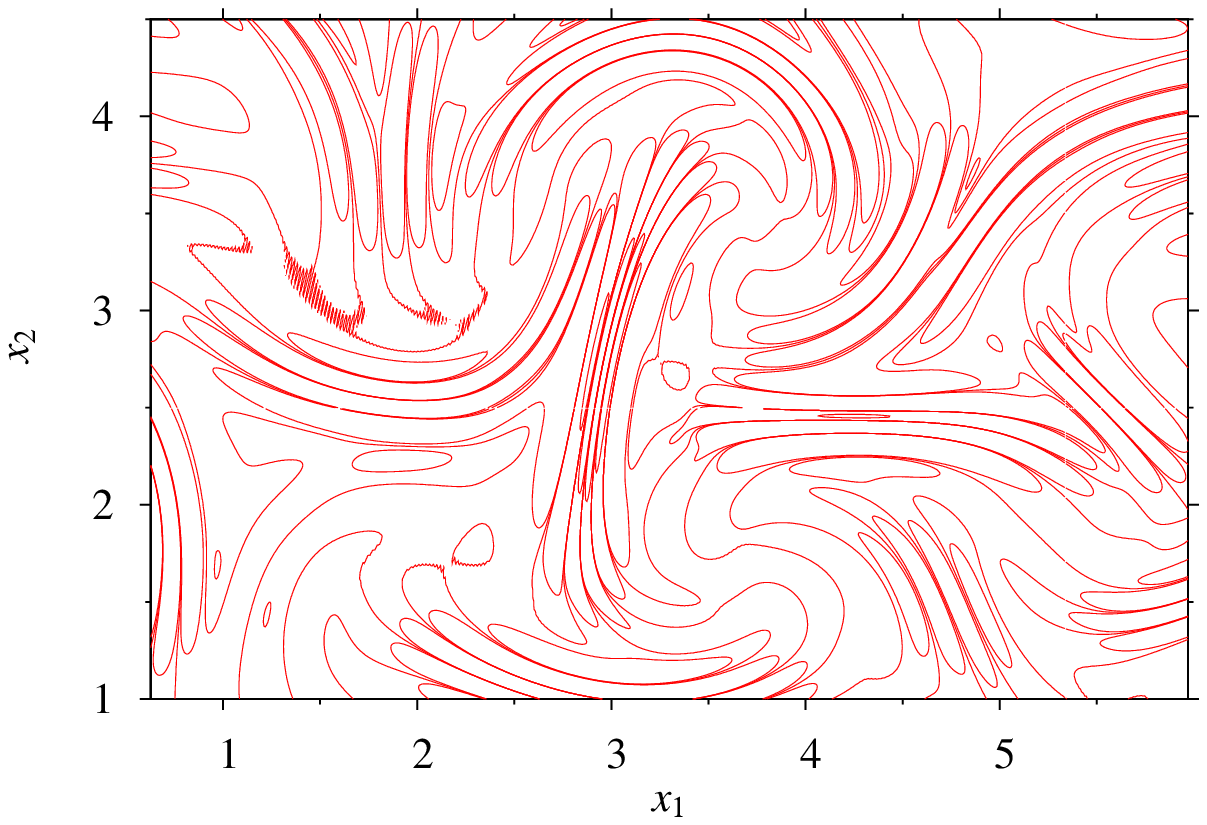}
\includegraphics[height=3cm]{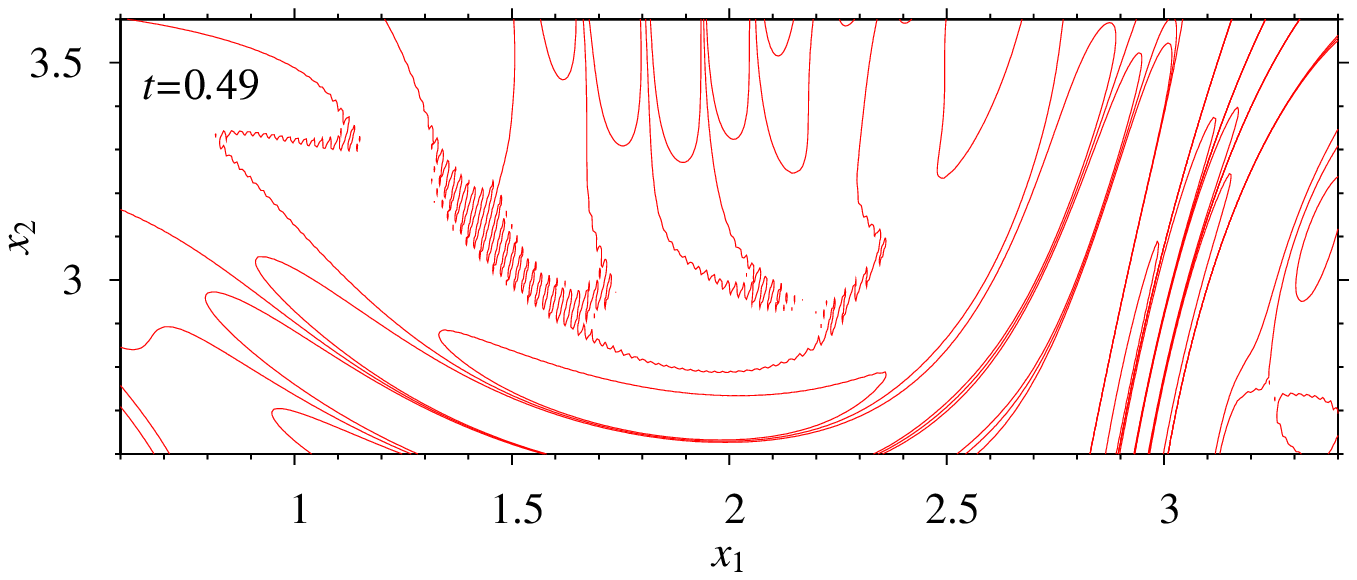}
\end{center}
\fi
\caption{(Color online) Contours of tri-Laplacian of the vorticity (upper: zoomed;
lower: superzoomed) showing  a tyger already at $t = 0.49$.}
\label{f:trilaplacian}
\end{figure}

 We immediately see
that, as for the 1D Burgers equation, most of these tygers have come ``out of the blue,'' namely
appearing at places which had no preexisting small-scale activity;
more precisely, they appear when complex-space singularities come within
one Galerkin wavelength $(2\pi)/\kg$ of the real domain.
The streamlines shown in Fig.~\ref{f:2dtygerdevelopment} indicate that tyger
activity appears at places where the velocity is roughly parallel to the
central cigar. As already pointed out, the cigar hardly moves; this condition is thus
equivalent to having fluid particles whose distance to the cigar
remains roughly constant. In so far as the cigar may be considered as
a one-dimensional straight object, the
truncation waves generated by the cigar will have crests  parallel
to the cigar and those fluid particles which move parallel to the crest keep
a constant phase and thus have resonant interactions with the truncation
waves. So far, this is basically  the same mechanism as discussed in 
Sec.~\ref{ss:tygerresonance} for 1D Burgers dynamics except, of course,
that the flow being now incompressible, the velocity within the tyger
is mostly perpendicular to the direction of fastest variation. If we now consider the one-parameter family
of straight lines perpendicular to a given
cigar, each such line will have some number (possibly zero) of  resonance points;
altogether they form the tygers. Since the flow
outside cigars is fully two-dimensional, these tygers have no reason to be parallel to
the cigars.

Observe that there are some points where this kind of resonance
condition holds but no tyger is seen, for example in
Fig.~\ref{f:2dtygerdevelopment} at $t=0.71$ near $x_1=3.8$ and $x_2=
3.4$.  This can be
interpreted in terms of strain: an incompressible flow has at each point a 
strain matrix with two perpendicular eigendirections, one for positive
strain and
the other one for negative strain. Fig.~\ref{f:2dtygerdevelopment} has 
little pink (light gray)
segments indicating the direction of positive strain. Tyger activity is
found only at resonance points where the (positive) strain direction
is not far from being 
perpendicular to the cigar. More precisely, it is easily shown that it has to
be  within less than $\pi/4$ of 
this direction. Otherwise the near-truncation activity generated
by resonance  is sheared quickly beyond the truncation  horizon.

\begin{figure}[htbp]
\iffigs
\begin{center}
\includegraphics[height=5cm]{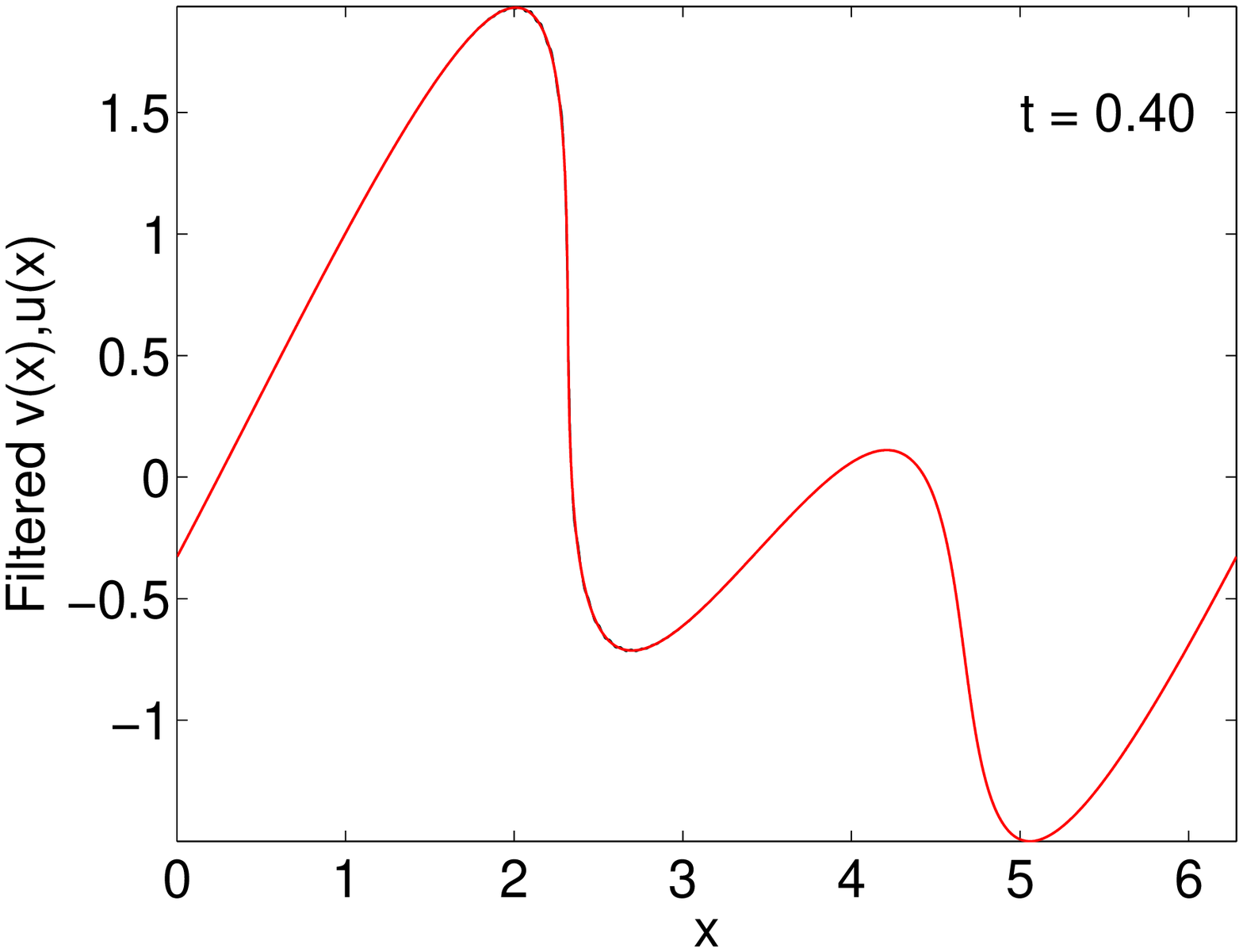}
\includegraphics[height=5cm]{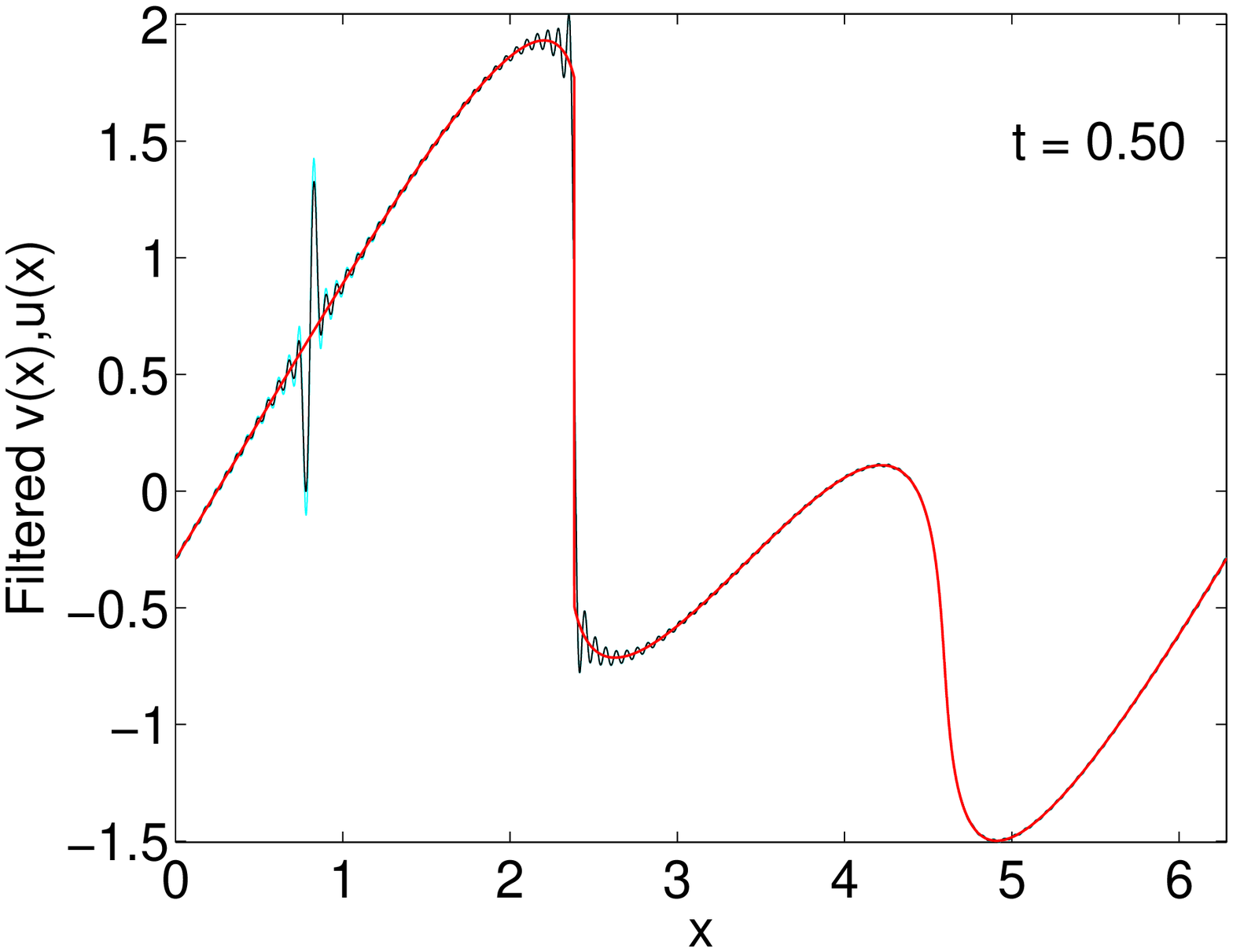}
\includegraphics[height=5cm]{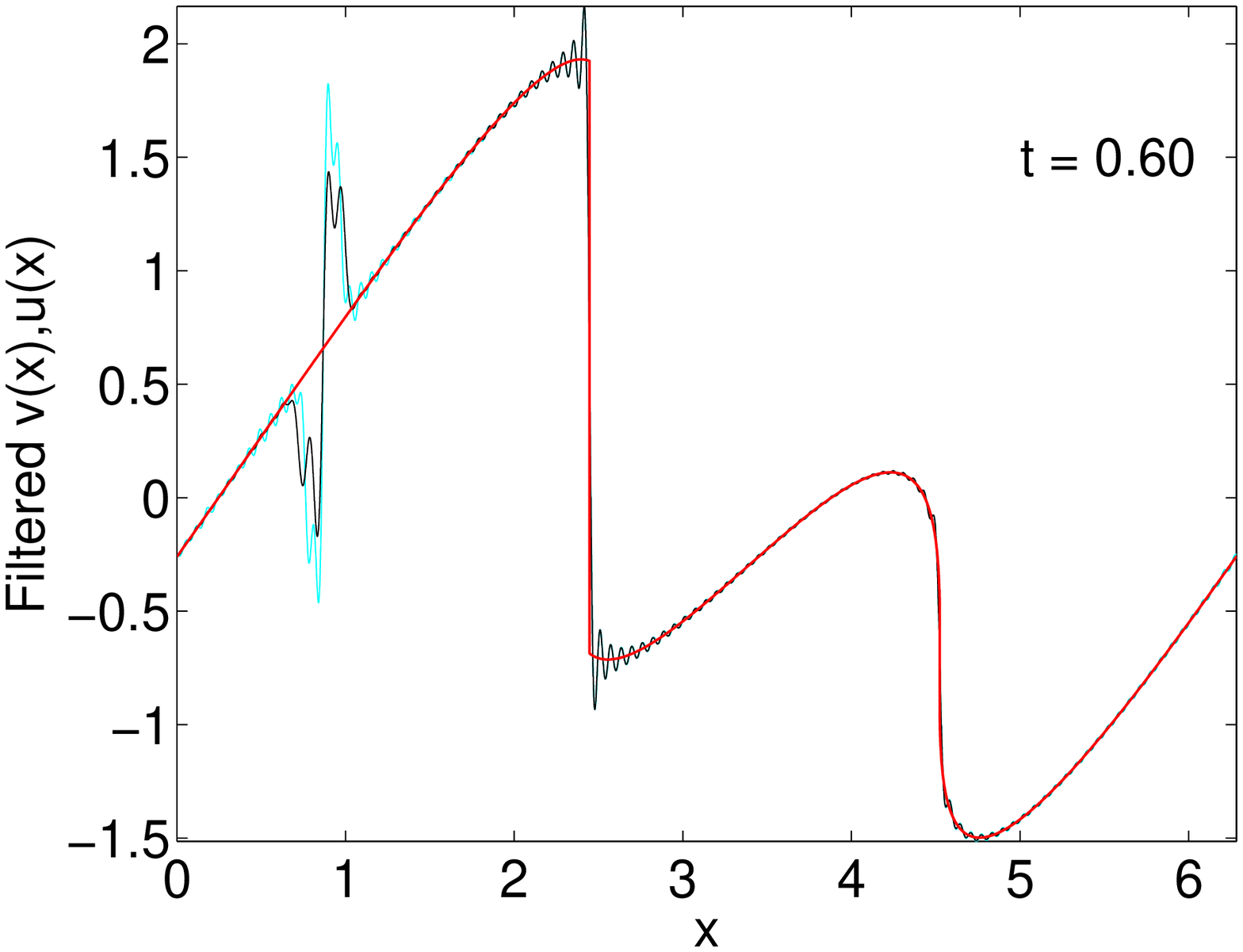}
\includegraphics[height=5cm]{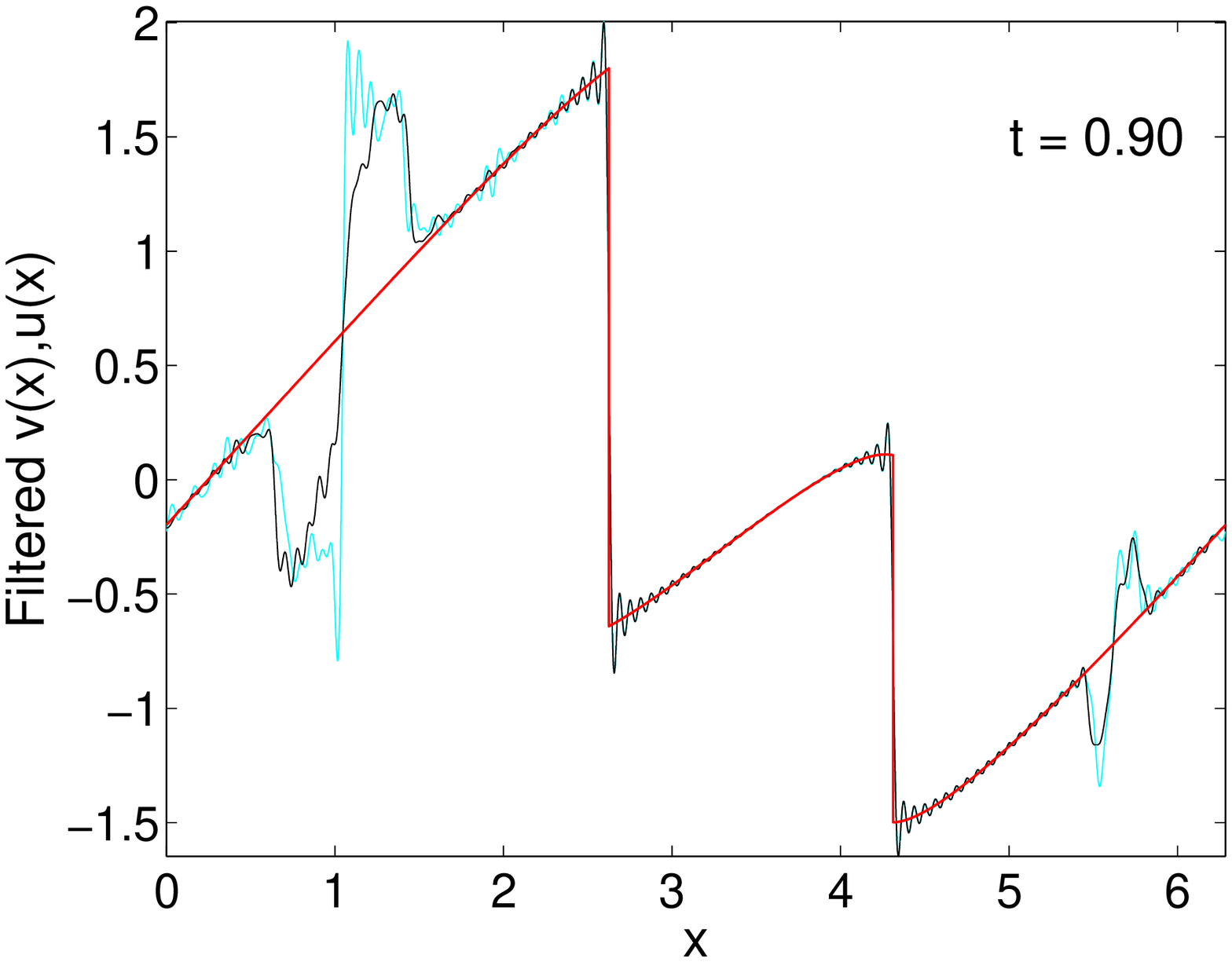}
\end{center}
\fi
\caption{(Color online) Plots of solution of the Galerkin-truncated Burgers
equation, with $K_G = 5,461$ in cyan (light grey) and $K_G = 21,845$ in black, low-pass filtered at
wavenumber $K = 100$, at various times. Initial condition
$v_0(x) = \sin(x) + \sin(2x - 0.741)$. The untruncated solution is shown in red (black, mostly straight line).}
\label{f:noweak}
\end{figure}

It is also of interest to show truncation effects and tygers in
Fourier space. Fig.~\ref{f:fouriervorticity} shows, at various times,  
the moduli of the
Fourier coefficients in the $(k_1,\,k_2)$ plane on a logarithmic
scale. The lowest value contours are at the $10^{-15}$
level, while rounding errors are about $10^{-16}$.  Note that the
Fourier-space picture is organized in the form of one main lobe,
perpendicular to the physical-space central cigar and secondary lobes
associated to less intensive small-scale structures. At the earliest
time $t=0.4$ no truncation effect near $|\kb| =\kg = 342$ is
visible \footnote{With much higher precision, truncation effects would
become visible.}. At $t=0.49$, long before tygers become visible on the
Laplacian of the vorticity in physical space, Fourier-space truncation
becomes visible. This truncation at first affects the wavevectors in the
direction of the main lobe that is perpendicular to the central
cigar. Truncation effects then  spread progressively to other
angular directions but appear to do so continuously, in contrast to
physical space where tygers are born ``out of the blue.'' We can
actually see such early truncation effects in physical space by taking
more spatial derivatives and thus putting more weight on high
wavenumbers. At $t=0.49$, Fig.~\ref{f:trilaplacian} shows contours of the tri-Laplacian $(\nabla ^2)^3
\omega$ with wiggly tygers. By performing various cuts (not shown),
we checked that the spatial variation is mostly perpendicular to the
cigar.

Recently we checked that many features observed for 2D incompressible
Euler tygers are also present in the 3D case. The details will be
reported elsewhere.

\subsection{The dissipative anomaly  and the lack of weak limit}
\label{ss:energy}


Earlier in this paper, we have seen that for large values
of the truncation wavenumber $\kg$, the Galerkin-truncated
solution remarkably preserves many features of the inviscid
limit such as shocks and their dynamics. So we ask: could
it be that the Galerkin-truncated Burgers equation converges in a
suitable sense
to the inviscid limit solution as $\kg \to \infty$?
This question was actually the main motivation of the present work. We
shall see that the answer is ``no'', but a qualified no.

First, for the kind of analytic initial conditions considered
here that go singular at some finite time $\ts$, the answer
to the above questions is actually ``yes'' for times
$0 \le t < \ts$. At such times, the solution of the Burgers
equation in Fourier space for large wavenumbers is bounded  by
$Ce^{-\delta k}$ (see Appendix~\ref{a:sing}). The
effect of truncation is thus exponentially small in $\kg$ and should go
away when $\kg \to \infty$. For the two-dimensional incompressible
Euler equation with periodic analytic initial data, analyticity holds
for arbitrary large times and thus the Galerkin-truncated solution 
is expected to converge to the untruncated solution \footnote{Evidence
  for this may be found in \cite{nguyenetal}. This result was
  proven recently by Bardos and Tadmor \cite{BardosTadmor10}.}.
For three-dimensional
flow the situation may depend on whether or not there is finite-time
blow up, a question which is very much open (cf. \cite{blue,gibbon}
and references therein.)

Returning to the one-dimensional Burgers case, what about $t>\ts$, when shocks
are present and the solution is \textit{dissipative}, whereas the
Galerkin-truncated solution is \textit{conservative}? How can such a
conservative system mimic the dissipative anomaly? One could imagine that
the small-scale tyger activity plays the role of molecular motion and
that motion on scales much larger than the Galerkin wavelength is governed
by the inviscid-limit Burgers equation. 

This is however not the case. Tadmor investigated the limit---in a suitable
sense to be defined below---of the truncated solution and found that it cannot
be dissipative (Ref.~\cite{tadmor1989} p.~31). The limit considered by Tadmor 
and in other papers studying conservative modifications of the Burgers
equation with strong oscillations \cite{laxlevermore,goodmanlax,houlax} 
is a distributional \textit{weak} limit. Assuming that the
Galerkin-truncated
solution $v$ has a weak limit $\overline{v}$ satisfying the Burgers equation, and using the basic
dynamical equations \eqref{inviscidburgers}-\eqref{gtburgers}, Tadmor
shows that $v^2$ has the weak limit $\left(\overline{v}\right)^2$,
from
which he infers that the limit is actually a strong one which implies
energy conservation and contradicts the dissipative character of the
solution
to the Burgers equation. Recently numerical
simulations of the inviscid Galerkin-truncated Burgers equation with $\kg$ up
to about $10^4$ showed indeed that such solutions do not converge to the
inviscid limit of the untruncated solution \cite{nguyenetal}.

The simplest instance of such a weak limit is just to apply
a low-pass filter to the solution with a  fixed threshold  $K$ for the
modulus of the wavenumber, while
letting $\kg\to\infty$. It is then easy to show that one can find a 
subsequence $n_1,n_2,\ldots$ of the sequence of integers $\kg$ such that 
the low-pass filtered solution has a limit \cite{bardostadmorprivate}.
We have obtained evidence that, without taking subsequences,
there is no weak limit. Indeed Fig.~\ref{f:noweak} shows the low-passed
solution with the same threshold $K=100$ for the same initial condition
and the same output time, but for two  very large and well-separated values
of the truncation: $\kg = 5,461$ (denoted 5K) and $\kg = 21,845$ (denoted 
21K). The solutions agree very well with the untruncated solution at  shocks
and nearby but the 5K and 21K tygers
differ significantly, even after application of the low-pass filter. 
It may thus be that there is no weak limit as $\kg \to
\infty$ \footnote{In contrast to what happens for the Lax--Levermore
  study of the KdV equation \cite{Bardos10}.}.

We can supplement this by a more physical and fluid mechanical explanation
of why the  truncated solution cannot converge (weakly) to the
inviscid-limit solution.  We have seen in Sec.~\ref{ss:temporal} that shocks behave just
as predicted by the inviscid limit for a substantial length of time
(until tyger spreading on the ramp reaches the shocks). Because of this, the hypothetical
limit of the truncated solution would be losing energy at the shocks,
just as the ordinary Burgers equation. The energy lost has to 
be found in the tygers in the form of high-wavenumber
oscillations. Decomposing
the truncated solution $v =u +\ut$ (where $u$ is the inviscid-limit solution)
we obtain tyger Reynolds stresses $\overline{ \ut ^2/2}$, where the overline
means a \textit{mesoscopic} spatial averaging over a distance large compared to the
Galerkin wavelength and small compared to any macroscopic
scale (for example using a low pass filter with threshold $K=100$ for
the case $\kg = 5,461$). If the mesoscopic tyger energy and thus the  Reynolds
stress is not spatially uniform, its gradient will drive the flow away
from the inviscid limit. This 
is the same mechanism that makes the tyger asymmetrical, as already 
mentioned.  Is there a way to obtain the correct inviscid limit
by eliminating the undesirable Reynolds stresses through some kind
of \textit{tyger purging}? We shall come back to this important
practical issue in the last section.

Finally, let us remark on our choice of the word \textit{tyger} for the
oscillations which are a result of Galerkin truncation.
Historically, the distinction between conservative and dissipative
systems has played a crucial role in not only man's scientific
pursuits, but also in a deeper cultural context. For centuries
there was a certain sanctity associated with things conservative
as opposed to being dissipative. Hence, before Galileo's
telescope revealed the ``transient'' nature of celestial
occurrences (e.g., Sun spots), man had always ascribed heavenly
objects as conservative and the more mundane,
transient, and earthly occurrences as dissipative. The fine
balance of the two seemed essential for all existence.

In the backdrop of this and in our investigations of
truncated systems, William Blake's poem ``The Tyger'' assumes special
significance \url{http://en.wikipedia.org/wiki/The_Tyger}.
In most interpretations, Blake's tyger is of course not the
animal itself \footnote{Actually, at the time of William Blake, the
spelling of the animal with a ``y'' was already obsolete.}; it is 
a metaphor for the symmetry of seemingly
different, even opposite, processes which nonetheless combine
to make a coherent whole. Thus apparent oppositions such as
life and death, light and darkness are seamlessly unified.
In our present study, we explore the interplay of conservative
and dissipative dynamics: how one, surprisingly,
might be embedded into another. As a result, it was quite natural for
us to call this phenomenon a tyger.

\section{Detailed analysis of the birth of the Burgers tyger}
\label{s:birth}

So far our point of view has been that of the (numerical)
experimentalist, with
some amount of phenomenological theory used to interpret the results 
whenever possible.  In this section we shall use a lot more ``systematic
theory'', namely expansions for large values of the Galerkin
truncation wavenumber $\kg$. In principle, such expansions can be
carried out beyond leading order; we shall however not attempt this,
let alone obtain rigorous bounds for errors. Indeed
some of our approximations used below are akin to what one calls ``patching''
(as opposed to ``matching'') in boundary layer theory, that is
approximations which can give the correct exponent of a power-law
leading term but cannot predict the correct constant in front.

Let us now give a general overview of how we intend to proceed.
We shall concentrate on a single problem, that of the birth of the
tyger for the one-dimensional Burgers equation with single-mode initial
condition. The birth, which takes place around the ``Galerkin
time'' $\tg$  when complex singularities come within one Galerkin
wavelength from the real domain, is part of an early phase which
extends from $t=0$ to
the time of the first singularity $t=\ts =1$. In Sec.~\ref{ss:tstardata} we shall 
see that the early tyger at $t=\ts$   has remarkable scaling 
properties with $\kg$. Our intention here is to understand analytically
how this comes about. For this, our strategy will be to devise various
models/approximations which make the problem simpler while hopefully
keeping the leading-order behavior unaffected. There will be three
levels of modelling: (i) linearization around the untruncated
solution, (ii) ignoring the ``exponentially small'' phase up to the
Galerkin time $\tg$, and (iii) ``freezing'', i.e. replacing the
untruncated solution $u(x,t)$ by $\us(x) \equiv u(x,\ts)$ for
$t\in [\tg,\,\ts]$. In Sec.~\ref{ss:3approx} we explain how to do
this and why it is justified. The problem is then reduced to studying
a linear first-order differential equation with
constant coefficients in a  finite dimensional space (Sec.~\ref{ss:minimal}).
It was brought to our attention recently by J.~Goodman that this
differential equation has features in common with that studied by 
Goodman, Hou and Tadmor (subsequently cited as GHT) in connection with the
stability of the pseudospectral method in the presence of aliasing
\cite{GHT}.

\subsection{Scaling properties of the early tyger}
\label{ss:tstardata}

\begin{figure}[htbp]
\iffigs
\centerline{\includegraphics[height=5cm]{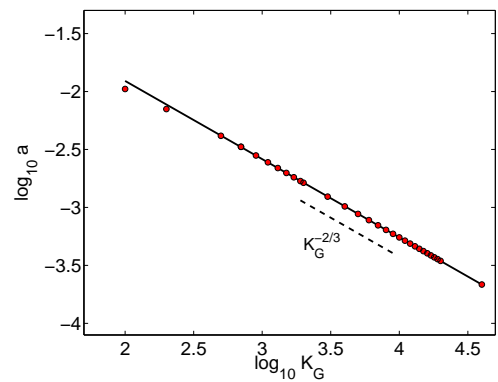}}
\fi
\caption{(Color online) A log-log plot of the discrepancy-based amplitude $a$ of the tyger  
at time $t=\ts$ as a
  function of $\kg$. Initial condition $u_0(x) =\sin x$. The data points are
shown by filled red circles (black circles) and the thick black line is
the best power-law fit $\propto K_G^{-2/3}$, which holds over two decades.}
\label{f:minus23rdlaw}
\end{figure}

\begin{figure}[htbp]
\iffigs
\centerline{\includegraphics[height=5cm]{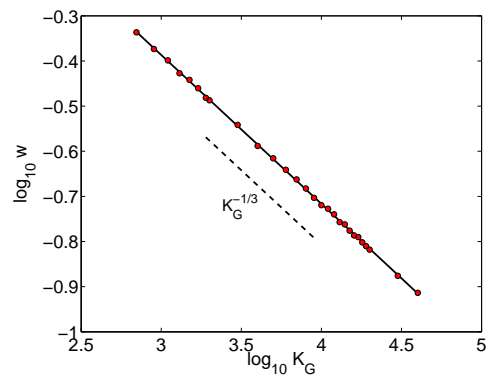}}
\fi
\caption{(Color online) Log-log plot of the width (at half amplitude) of the
  tyger. Otherwise, same as in Fig.~\ref{f:minus23rdlaw}. The best power-law 
fit is now $\propto K_G^{-1/3}$, which holds over nearly two decades.}
\label{f:minus13rdlaw}
\end{figure}

In Sec.~\ref{s:simulpheno} we have seen that tygers are born at
suitable resonance points in the form of bulges made of oscillation at
the Galerkin wavelength  with a very symmetrical
envelope. Eventually, Reynolds stresses will distort this envelope. At
the time $\ts$ of the first singularity this is not yet the case. Now
we
concentrate on the scaling properties for high $\kg$ of the amplitude $a$
and
width $w$ of such early tygers, in terms of the discrepancy $\ut =v-u$. Here we limit ourselves to the
single-mode
initial condition. Figs.~\ref{f:minus23rdlaw} and \ref{f:minus13rdlaw}
show respectively the amplitude and width as a function of $\kg$ for
values
ranging from 100 to 40,000. The amplitude is measured at the maximum closest to the
center of the tyger, located a distance $\pi/(2\kg)$ away. The width is
measured at half of this amplitude. It is seen that both have clean
scaling laws for sufficiently large $\kg$: subdominant corrections
are perceptible on the log-log plots only for $\kg < 200$. The
leading-order behavior is
 \begin{equation}  
  a \propto \kg ^{-2/3}, \qquad w \propto \kg ^{-1/3}.
\label{amplitudeandwidth}
\end{equation}
The following sections are mostly devoted to explaining these scaling laws.

\subsection{Three successive approximations}
\label{ss:3approx}

As already observed in Sec.~\ref{ss:energy}, initial conditions with
a finite number of Fourier harmonics, such as the single-mode case, 
are for a while analytic in the complexified  space variable within a strip around
the real domain of width $\delta(t)$. As long as $\delta(t)\kg \gg 1$ the
effect of truncation is exponentially small. This is why, the kind of tygers
reported in Sec.~\ref{s:simulpheno} are not seen before $\tg$, when complex 
singularities come within roughly one Galerkin
wavelength from the real domain. For large $\kg$, this happens only a short time 
$O(\kg ^{-2/3})$ before the time $\ts$  of the first singularity
(cf. Appendix~\ref{a:sing}). Hence, by the time $\ts$, truncation has
been felt significantly only for a lapse of time $O(\kg ^{-2/3})$. The phase
mixing argument given in Sec.~\ref{ss:tygerresonance} and in particular
\eqref{width} tell us that the coherent build up of a tyger will affect only
those locations whose velocity differs from that at resonance by an
amount $\Delta v$ such that
\begin{equation}  
  \Delta v \lesssim \frac{2\pi}{\kg ^{-2/3} \kg} \propto \kg ^{-1/3}.
\label{tswidth}
\end{equation}
Since at such times, the velocity $v$ of the truncated solution is expected to stay
close to the velocity $u$ of the untruncated solution and the latter varies
linearly with $x$ near the resonance point, the width of the $\ts$ tyger is
itself proportional  to  $\kg ^{-1/3}$, as indicated by the simulations.

Is there an equally simple argument to understand why the amplitude 
scales as $\kg ^{-2/3}$? One possibility would be to observe that at
$\ts$ the untruncated solution has a cubic root behavior near the preshock
location (cf. Appendix~\ref{a:sing}). If we cut out a small interval of
length one Galerkin wavelength $\lambdag$ , we will be  ``missing'' 
an energy $\sim \int_0^{\lambdag} x^{2/3} dx\sim \kg ^{-5/3}$. Remembering
that the Galerkin-truncated equation conserves energy, if we assume that this
missing energy is transferred entirely to a tyger  of width $\kg^{-1/3}$, we
obtain precisely an amplitude $\propto \kg ^{-2/3}$. There is however no
reason to assume that near the preshock the effect of truncation can
be reduced to carving out a little interval of one Galerkin wavelength.
It is doubtful that this energy argument can be turned into something
rigorous.

We turn now to more systematic arguments. Before $\ts$, when $\delta(t)>0$, it 
is easily shown that the Galerkin-truncated solution converges strongly
to the untruncated solution. The simulations reported in Sec.~\ref{ss:tstardata}
suggest that this still holds at $t=\ts$, since the amplitude of the
discrepancy goes to zero. Furthermore, the nonlinear effect of Reynolds
stresses, as discussed in Sec.~\ref{ss:energy} is of even higher order.
Indeed, within the bulge the Reynolds stresses will be $\sim \kg
^{-4/3}$; since they change spatially on a scale $\sim \kg ^{-1/3}$, the
gradient of Reynolds stresses is $\sim \kg ^{-1}$; over a time interval
$\sim \kg ^ {-2/3}$ this will change the bulge amplitude by $ \sim \kg
^{-5/3}$, which is small compared to the amplitude of the bulge itself.

All this suggests that the early tyger development can be captured
by somehow linearizing the Galerkin truncated solution around the untruncated
one. Let us rewrite the basic dynamical equations 
\eqref{inviscidburgers}-\eqref{gtburgers} in terms of the discrepancy
\begin{equation}  
\ut \equiv v -u.  
\label{defdiscrepancy}
\end{equation}
We obtain
\begin{equation}  
\partial_t \ut +\pkg \partial_x\left(u \ut +\frac{\ut ^2}{2}\right) =
\left({\rm I} -\pkg
\right) \partial_x\frac{u ^2}{2}, \quad \ut(0) =0,  
\label{utildequation}
\end{equation}
where ${\rm I}$ stands for the identity operator and the zero initial
condition follows from $u_0 =v_0$, a consequence of having a finite
number of modes initially.

Observe that the r.h.s. of \eqref{utildequation} provides no input to
wavenumber
below the truncation. Actually this input is hidden in the l.h.s. To make
this clear, we need to decompose the various fields into their
Galerkin-truncated part and the remainder. We set
\begin{eqnarray}  
 u &=& \ul + \ug \\
\ul &\equiv& \pkg u \qquad \ug \equiv ({\rm I} -\pkg)u.
\label{defulug}
\end{eqnarray}
Next, to similarly decompose $\ut$, we use  $({\rm I} -\pkg)\ut = -\ug$, which follows from the fact that
$v= u+\ut$ has no harmonics beyond the truncation. Hence   we have
\begin{eqnarray}  
\ut &=& \up -\ug, \\ 
\up &\equiv& \pkg \ut.
\label{defup}
\end{eqnarray}
In what follows we shall work mostly with $\up$ which has no harmonics
beyond the truncation and which we call the
\textit{perturbation}. As we shall see, around $\ts$, the perturbation
is small. In contrast, beyond the truncation, the discrepancy is just equal to minus
the untruncated flow; it is thus known but in no way small.
Now, we apply $\pkg$ to \eqref{utildequation} and use the various
decompositions to obtain
\begin{equation}  
\partial_t \up  +\pkg \partial_x\left(u\up+\frac{(\up)^2}{2}\right)
=\pkg \partial_x\left(\ul\ug +\frac{(\ug)^2}{2}\right).
\label{hassource}
\end{equation}
The r.h.s. of \eqref{hassource} is a known function which we shall
call the \textit{beating input} and denote  $f$ because it describes how harmonics
of the untruncated solution, located beyond the truncation, interact
with themselves or with subtruncation harmonics to give a
subtruncation input. This beating input, which is shown  in the lower
panel
of Fig.~\ref{f:beating}, consists basically of spatial oscillations
at the Galerkin wavelength, modulated  by an envelope that peaks
at the preshock. This is the precise content of what we called
``truncation waves'' in the phenomenological approach of Sec.~\ref{ss:tygerresonance}

We are now in a position to define the three approximations made for 
large $\kg$, which we call respectively \textit{linearization}, 
\textit{reinitialisation} and \textit{freezing}:
\begin{itemize}
\item 1. The term $(\up)^2$ in the l.h.s. of
  \eqref{hassource} is discarded;
\item 2. The perturbation $\up$ is set to zero at time $\tg$;
\item 3. The untruncated solution is frozen to its $\ts$ value.
\end{itemize}

Concerning linearization, we already observed that nonlinear effects
will be weak if the scaling laws for the $\ts$ tyger are indeed given
by \eqref{amplitudeandwidth}. Reinitialization is justified because, prior
to the Galerkin time  $\tg$, only exponentially small perturbation are
present and, here, we are only trying to capture algebraically
small terms. It is then convenient to introduce a new shifted time variable
\begin{equation}  
\tau \equiv t - \tg.  
\label{deftau}
\end{equation}
To avoid unnecessary constants,  we choose, $\tg = \ts -\kg ^{-2/3}$. Hence,
in the $\tau$ variable, the first real (preshock) singularity is at
\begin{equation}  
\taus \equiv  \kg ^{-2/3}. 
\label{deftaus}
\end{equation}
We shall also take the liberty to still denote the perturbation by $\up$ when
it is expressed in terms of the shifted time.
As to the freezing, replacing $u(t)$ by $\us\equiv u(\ts)$,  it appears justified since the untruncated solution
hardly changes between $\tg$ and $\ts$, except at 
wavenumbers much larger than $\kg$ which do not contribute to \eqref{hassource}.
Freezing changes \eqref{hassource} into an equation with constant
coefficients. In particular the beating input becomes time-independent.

\begin{figure}[htbp]
\iffigs
\centerline{\includegraphics[height=5cm]{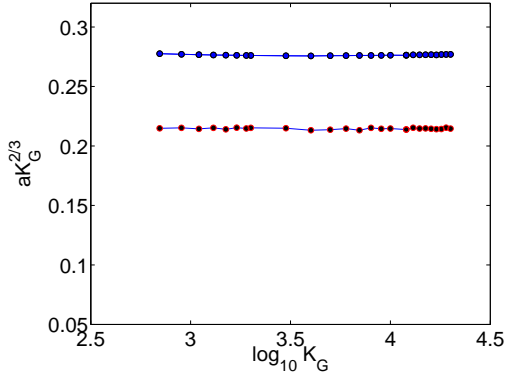}}
\fi
\caption{(Color online) Log-linear plot of the compensated amplitude of the tyger
$K_G^{2/3}\,a(\kg)$, calculated (i) from $\ut$, (ii) from
the linearised approximation for $\up$, and (iii) from the freezing
plus reinitialization approximation, all versus
$\kg$. The data corresponding to (i) and (ii) are
indistinguishable and shown as blue filled circles (black upper curve). The
data corresponding to (iii), shown by black filled circles with
a red border (black lower curve), have the same scaling but a multiplicative constant about
twenty per cent lower.
}
\label{f:3approx}
\end{figure}

Although there are good theoretical reasons to make these three
approximations,
we also tested them numerically. Fig.~\ref{f:3approx} shows the  scaling
law for the tyger amplitude at $\ts$ in compensated form (after multiplication
by $\kg ^{+2/3}$) for (i) the full problem, (ii) with only linearization 
assumed and (iii) with in addition reinitialization and freezing assumed.
All three cases have the same scaling law. Linearization brings about only
a minuscule change, as expected. The other two approximations make a difference
of about twenty percent, an indication that they can be refined.

\subsection{Tyger birth: the minimal model}
\label{ss:minimal}

With all three approximations formulated in the previous section, the
temporal dynamics of the perturbation near $\ts$ is 
simply given by
\begin{eqnarray}
&&\frac{d}{d\tau} \up = A \up + f, \qquad \up(0) = 0, \label{simple}\\
&& A \equiv -\pkg \partial_x\left(u_{\star} \,\bullet\right),
\label{defA}\\
&& f \equiv \pkg \partial_x\left(u_{\star} ^{\scriptscriptstyle <} \,u_{\star}
^{\scriptscriptstyle >}+\frac{(u_{\star}
^{\scriptscriptstyle >})^2}{2}\right).
\label{deff}
\end{eqnarray}
It is important to stress that, because of the freezing approximation,
$A$ and $f$ are evaluated at $\ts$ and thus time-independent.

We shall now write this equation as a finite-dimensional linear
differential equation by working in Fourier space and see that $A$ is actually
a matrix. Some of the quantities
that we shall look at are more conveniently represented by taking the 
initial condition $u_0 =\sin x$, which has the tyger born near $x=0$ and
the preshock at $x=\pi$. For other quantities the  choice $u_0 =-\sin x$
is better. We shall refer to the former as ``origin at the tyger''
and to the latter as ``origin at the preshock''. Note that in both
cases, the initial condition being real and odd, so are the untruncated
solution $u$, the truncated one $v$, the perturbation $\up$ and the
beating input $f$. Hence their Fourier coefficients $\uhk$, $\vhk$, $\uhpk$ 
and $\fhk$ are pure imaginary and odd functions of $k$. With this notation, it
is easy to rewrite \eqref{simple}-\eqref{deff} as a system of $2\kg+1$ equations, indexed by $k\in [-\kg,\,\kg]$:
\begin{eqnarray}
&&\frac{d}{d\tau} \uhpk = \sum_{k'=-\kg}^{\kg} A_{kk'}\, \uhpkp + \fhk\,,
  \qquad \uhpk(0) = 0, \label{simplek}\\[1ex]
&& A_{kk'} \equiv -\ui k\,\hat u_{\star,\, k-k'}\,,
\label{defAk}\\[1ex]
&& \fhk \equiv \ui k\sum_{p+q=k}(\uhlp\,\uhgq +\frac{1}{2}\uhgp\,\uhgq).
\label{deffk}
\end{eqnarray}
It is important to observe that, $\uhlkmkp$ being pure imaginary, the 
entries of the matrix $A$ are all real.

Here we observe that GHT were led, at the technical level, to studying a
\textit{homogeneous} version of \eqref{simplek} without the beating input
$f$ (and thus without resonant wave interactions). In their work, the velocity
$u$ is also prescribed but mostly taken to be  $\sin(x)$ or $\sin(px)$. The
operator/matrix $ A_{kk'}$ differs only marginally from ours, due to the
deliberate presence of aliasing.

\begin{figure*}[htbp]
\iffigs
\begin{center}
\includegraphics[height=5cm]{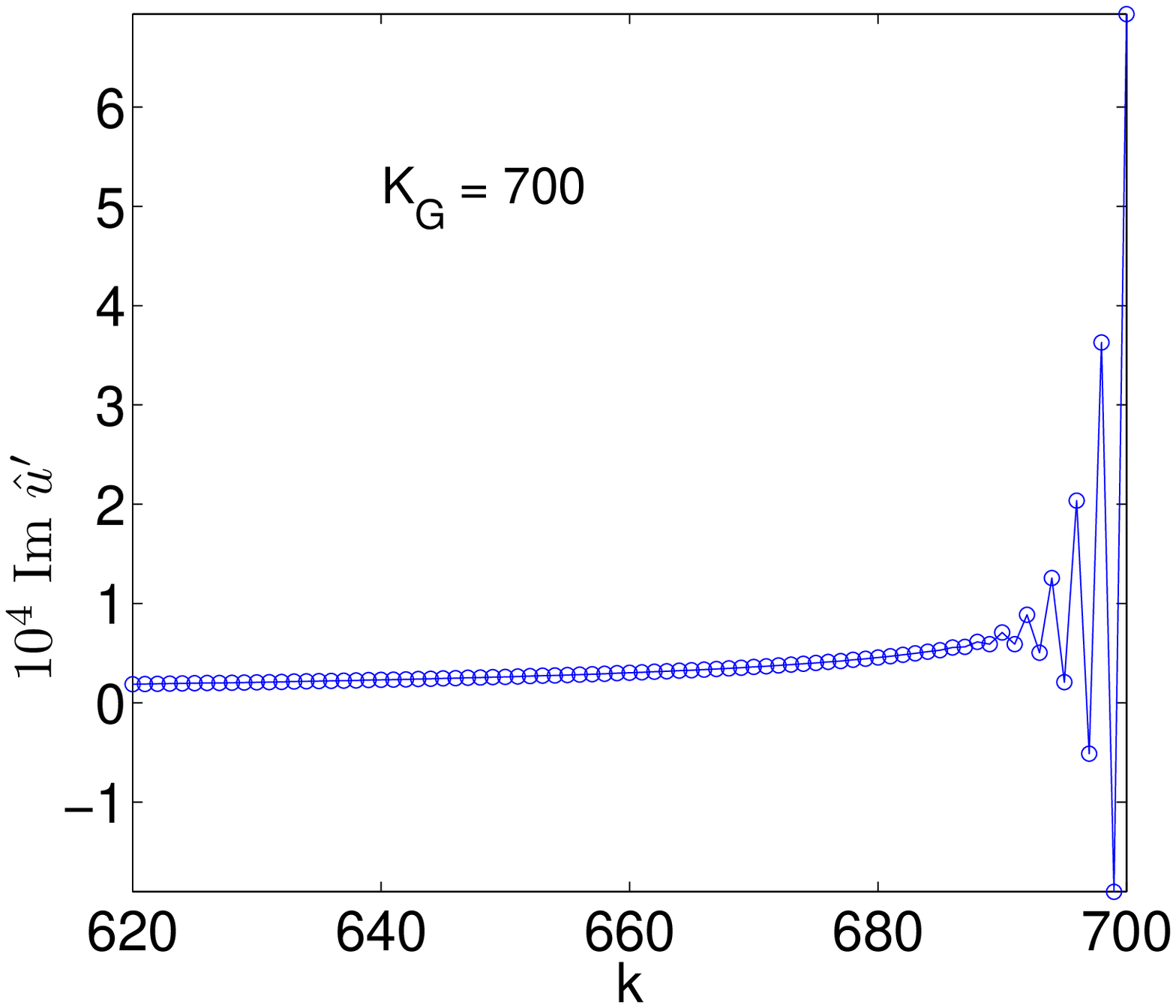}
\includegraphics[height=5cm]{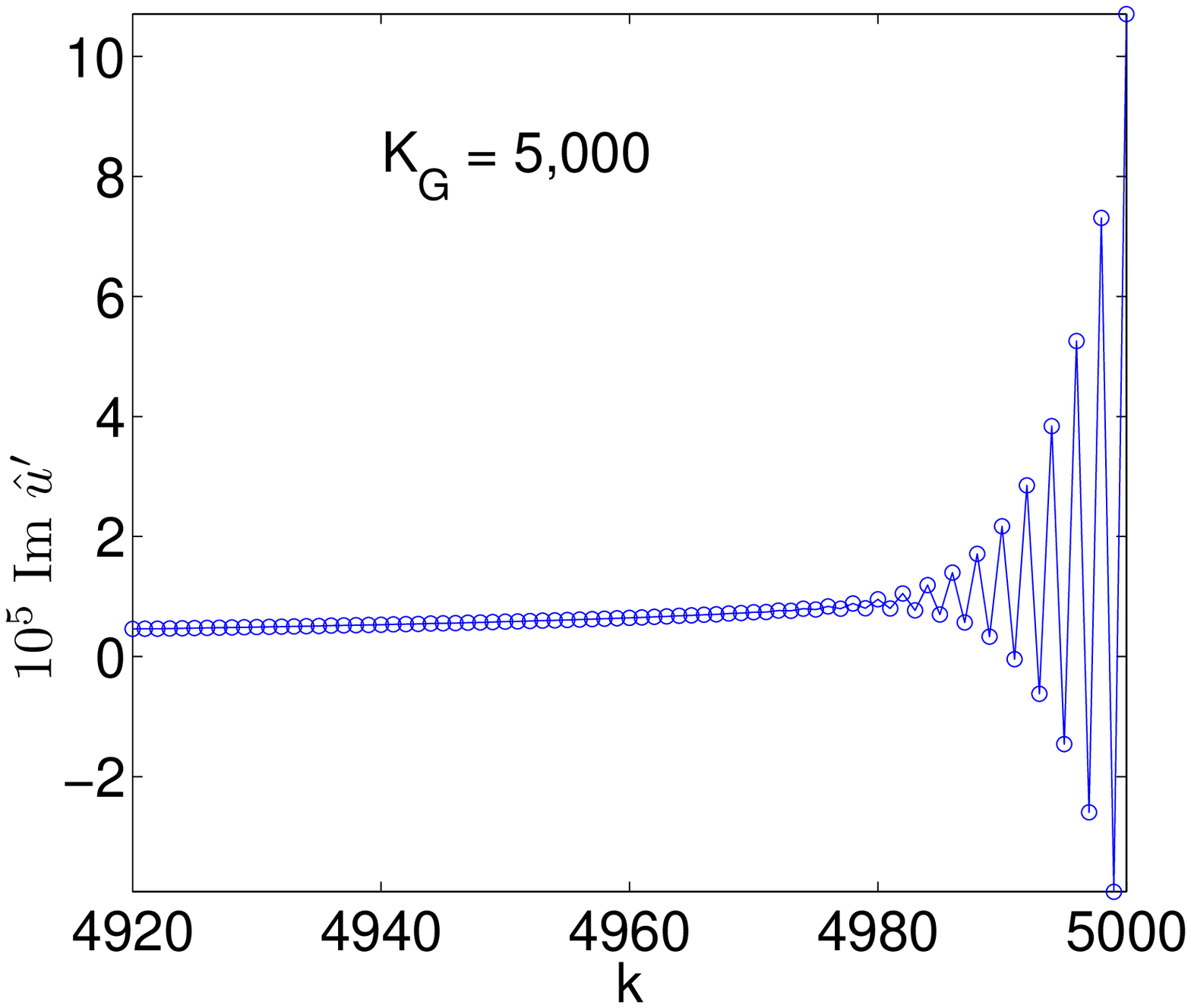}
\includegraphics[height=5cm]{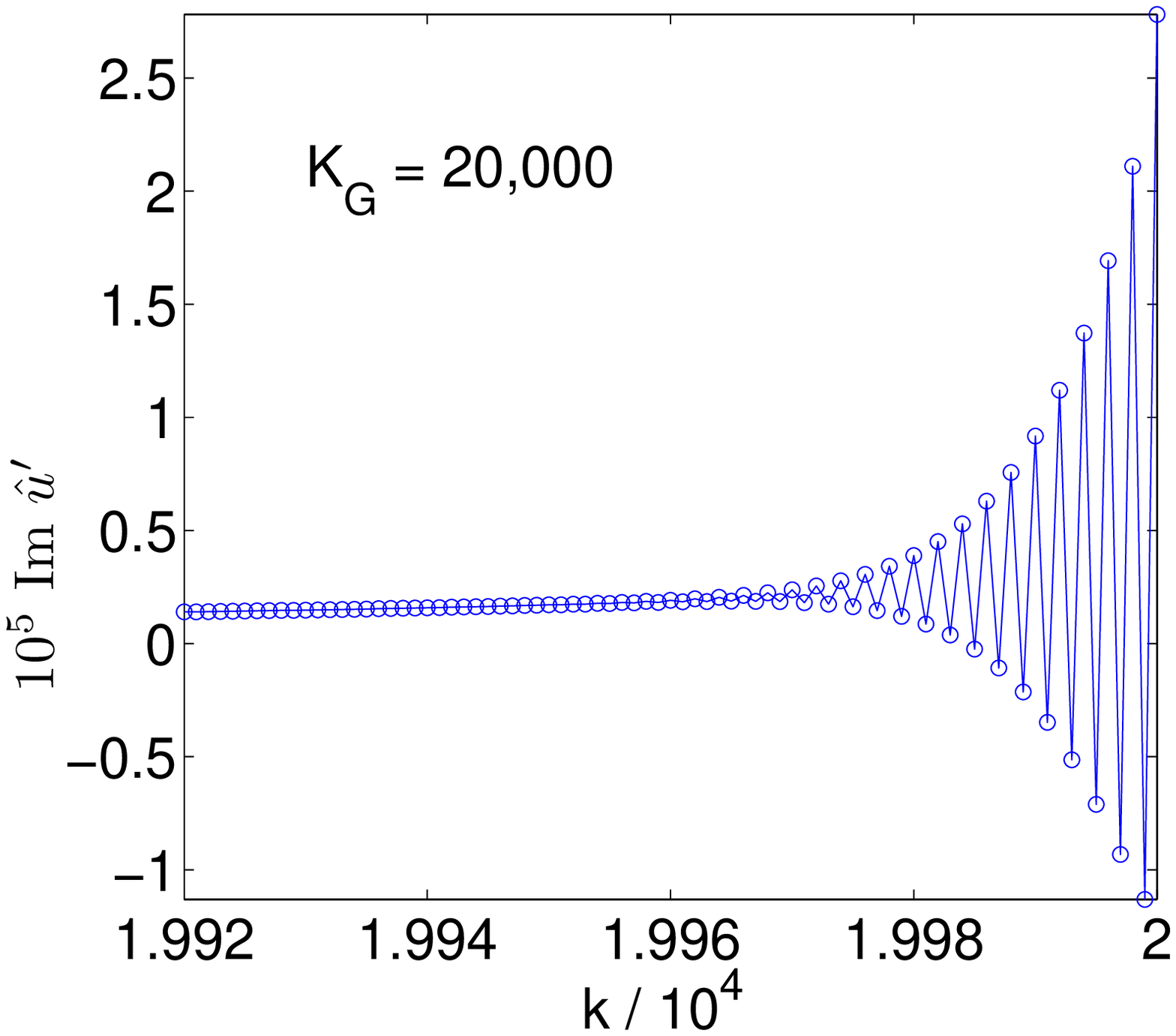}
\end{center}
\fi
\caption{(Color online) The boundary layer in Fourier space near $\kg$. Shown are
the imaginary parts of $\uhp(\ts)$ for three values of $\kg$. The origin is at
the preshock. The even-odd oscillations indicate that most of the activity
is at the tyger, a distance $\pi$ away.}
\label{f:blayer}
\end{figure*}

\begin{figure}[htbp]
\iffigs
\centerline{\includegraphics[height=5cm]{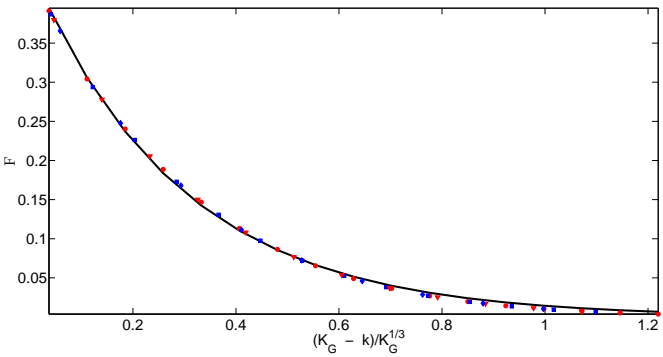}}
\fi
\caption{(Color online) The envelopes of the various boundary layers shown in
  Fig.~\ref{f:blayer} (with preshock contributions subtracted out),
  collapsed into a single curve after rescaling. Red (black) circles: $\kg =
  20,000$, blue (black) squares: $\kg = 15,000$, red (black) triangles: $\kg = 10,000$,
  blue (black) diamonds: $\kg = 5,000$. The thick black line is the exponential
  fit \eqref{scaling}.}
\label{f:blcollapse}
\end{figure}

It is of course quite easy to solve \eqref{simplek}-\eqref{deffk}
numerically for the perturbation $\uhp(\taus)$.  Fig.~\ref{f:blayer} 
shows its imaginary part for three large  values of $\kg$. Most of the activity is
concentrated in boundary layers near $\pm\kg$ (we are only showing the
right boundary layer because the function is odd). In the ``tail''
outside of this
boundary layer the amplitude of the perturbation is very small.  The
Fourier space boundary layer contains all the information about
spatial oscillations on scales close to the Galerkin wavelength. If a
tyger is localized in physical space near $X$, the
Fourier amplitude will have a phase factor $\ue ^{-\ui k X}$.  For
this particular set of figures we have taken the origin at the preshock. In the
boundary layer the most conspicuous features are even-odd oscillations which
are a signature of a tyger located at $X= \pi$. The oscillations are not
completely symmetrical between positive and negative values,  an indication that there is also some small-scale
activity near the preshock at $X=0$. The low-amplitude tail to the left of the
boundary layer has no oscillations, indicating that it comes mostly from the
neighborhood of the preshock. The width of the boundary layer is found to scale
approximately as $\kg ^ {1/3}$, that is the inverse of the tyger width, as
expected from Heisenberg's ``uncertainty principle'' \footnote{Here we mean of
course only a property of the Fourier transformation which underlies
Heisenberg's proof of the quantum-mechanical uncertainty principle.}.
As to the peak
amplitude 
of the perturbation in the boundary layer, it  is found to scale   
as $\kg ^ {-1}$. After subtraction of the high-$k$ contribution
stemming from the neighborhood of the preshock \footnote{This can be
  done by replacing pairs of successive even-odd Fourier amplitudes by
  their half differences and their opposites, respectively.}, we found
that the upper part of the envelope of the boundary layer has the following
scaling representation:
\begin{equation}  
 {\rm Im}\, \uhpk = \kg ^{-1}F\left(\frac{\kg -k}{\kg ^{1/3}}\right),
 \qquad F(x) = 0.448 \, \ue ^{-3.45x} ,
\label{scaling}
\end{equation}
which implies a collapse of all the boundary layer data after suitable
rescaling, as illustrated in Fig.~\ref{f:blcollapse}.

Note that by Fourier transformation, \eqref{scaling} immediately
implies the basic scaling laws \eqref{amplitudeandwidth} for the width
and amplitude of the tyger. As we have seen, the  $\kg ^{-1/3}$ dependence of the width
is just a consequence of the phase mixing condition
\eqref{width}. In the next section we shall try to understand how
the amplitude factor comes about.

\subsection{Tyger birth reduced to a simple (?) linear algebra problem}
\label{ss:matrix}

Obviously, \eqref{simple} can be solved for the perturbation at time $\taus =
\ts-\tg$:
\begin{equation}  
 \up(\taus) = A ^{-1}\left(\ue ^{\taus A} - {\rm I}\right)f =
\left(\sum_{n=0}^\infty A ^n\frac{\taus ^{n+1}}{(n+1)!}\right)f.
\label{incredible}
\end{equation}
If the matrix $A$ is singular (as it actually is), the middle equation of
\eqref{incredible} is not directly meaningful, but the r.h.s. remains meaningful.

From this it becomes clear that much is controlled by the
\textit{spectral} properties of the operator $A$. We are thus led
to consider the associated eigenvalue/eigenvector equation
\begin{equation}  
A \psi = \lambda \psi,  
\label{orrsommerfeld}
\end{equation}
which plays for the Galerkin-truncated problem the role of the standard
Orr-Sommerfeld equation \cite{drazinreid} and will be thus called.
Detailed spectral properties of the Orr--Sommerfeld operator $A$, which
is neither Hermitian nor antihermitian, are discussed in Appendix~\ref{a:OS}. The
eigenvalues, most of which are complex, come in opposite pairs; the
associated complex eigenvectors being either  even or odd functions of
$k$. In addition there is a zero mode, i.e. an eigenvector with
eigenvalue zero. We denote the eigenvalues by $\lambda_j$ where $j$ is a signed
integer varying from $-\kg$ to $\kg$. For positive $j$, $\lambda_j$ is the 
$j$th eigenvalue with positive imaginary part, eigenvalues being ordered by increasing
moduli; $\lambda_{-j} = - \lambda_j$ and $\lambda_0 =0$. A complete set of
complex eigenmodes is denoted $\psij$ and their  Fourier coefficients by
$\psijk$. When $\kg$ is large, the eigenvalues with large indices $j$
are
almost pure imaginary and the largest (in moduli) eigenvalues
are very close to $\pm\ui \kg$ \footnote{For the case $u_{\star}
  = \sin x$, which is not of great relevance here,
it may be shown that the eigenvalues of $A$ are pure imaginary. In contrast,
for the same velocity but aliased boundary conditions, GHT find a real part
equal to $\pm 1/2$.}. The typical spacing between the moduli
of successive eigenvalues is order unity for large $\kg$. 

Also of  importance are the strength and scaling properties of the
beating input, which are discussed in Appendix~\ref{a:beating}. Its
Fourier coefficients $\fhk$ are pure imaginary and odd functions of
$k$. We saw that the tyger has  the unexpected feature that it appears
away from the preshock. However the  beating input, which is caused directly by
truncation, is mostly localized where the untruncated solution displays its
highest small-scale activity, namely near the preshock. With the origin 
taken at the preshock, the imaginary part of the beating input 
${\rm Im}\,\fhk$   peaks at truncation $k=\kg$ with $|\fhkg|= O\left(\kg
^{-1/3}\right)$.
When moving down from $\kg$ to lower $k$, it   falls off rather
slowly as $(\kg -k )^{-1/3}$ while keeping a constant sign. Thus in Fourier space we have a rather
broadband beating input. In the physical space the beating input is
mostly
an oscillation at the Galerkin wavelength, whose envelope falls
off as $|x|^{-2/3}$, where $|x|$ is the distance from the preshock location.

Let us now decompose the beating input $f$ and the solution
$\up(\taus)$ in terms of the eigenmodes:
\begin{equation}  
\up(\taus) = \sum_j\upj(\taus) \psij, \qquad f= \sum_j f^{(j)}\psij . 
\label{jdecompose}
\end{equation}
It then follows from \eqref{incredible} that
\begin{equation}  
\upj(\taus)  = \frac{\ue ^{\lambda_j \taus} -1}{\lambda_j} f^{(j)},
\label{solution}
\end{equation}
in which it is understood that the fraction takes the value $\taus$ when
$\lambda =0$. An important role is played by those eigenvalues for
which 
\begin{equation}  
|\lambda_j| \taus \sim 1;  
\label{defthreshold}
\end{equation}
these will be called \textit{threshold eigenvalues}. Since $\taus
\propto \kg^{-2/3}$ there is a whole range of eigenvalues well below
and well above the threshold. The corresponding eigenmodes will be
called
``low-lying'' and ``high-lying'' modes, respectively. Well below threshold the fraction in
\eqref{solution} can be Taylor expanded, yielding to leading order
$\taus f^{(j)}$: those modes are essentially unaffected by their
interaction with the flow through the Orr--Sommerfeld operator $A$. We
have
checked that such modes are responsible for the low-amplitude
oscillatory
tail to the left of the boundary layer seen in Fig.~\ref{f:blayer}.
Well above threshold, the modulus of the fraction is much smaller than
$\taus$ because $\ue ^{ \lambda_j \taus}$ is essentially a phase
factor \footnote{The real part of the eigenvalues becomes negligibly small
  for all $j$ 
when multiplied by $\taus$.} and one can thus suspect that high-lying
modes do not contribute much to the boundary layer. Making this
argument solid requires a better control over the phases than we have
been able to achieve analytically. We have thus carried out
numerically
partial summations of the r.h.s. of  \eqref{jdecompose}, on
the one hand starting from low-lying eigenmodes and adding progressively
higher-lying eigenmodes and on the other hand doing it in reverse. In
both instances we found that the boundary layer of
Fig.~\ref{f:blayer} emerges mostly from modes near the
threshold \footnote{Films of such partial summations are available 
at \cite{partialsummations}.}.

When \eqref{solution} is applied near threshold, the $1/ \lambda_j$
yields a factor $\taus \propto \kg ^{-2/3}$, while the beating input
has
an overall factor $\propto \kg ^{-1/3}$. Together, this produces a
$\kg ^{-1}$ amplitude factor in a boundary layer of thickness $\propto \kg ^{1/3}$, needed to explain
the $\kg ^{-2/3}$ law for the amplitude of the $t =\ts$ tyger in physical
space.  What we just explained is however far from a proof since (i)
we
did not  show analytically that the dominant contribution to the
boundary layer  comes from threshold modes and  (ii) a $\kg ^{-1}$
amplitude factor for the beating input $f$ does not necessarily imply
the same factor for its threshold components $f^{(j)}$.

\section{Open problems and conclusion}
\label{s:conclusion}

We must now conclude our adventures in Tygerland.  Although this
project has been unfolding over three years, we have the
feeling that we only indented the subject, as far as true mathematical
understanding is concerned.  For example in Sec.~\ref{s:birth},
devoted just to the birth of tygers, we have not identified
analytically the important mechanism which allows threshold modes to
populate the boundary layer seen in Fig.~\ref{f:blayer}.  

As to the after-birth events, they have so far only been the subject
of numerical experimentation and occasional phenomenological theory. 
The collapse of the tyger, shortly after the time of appearance of the
first shock, seen in Fig.~\ref{f:collapse} is strongly reminiscent
of  the collapse phenomenon in plasmas \cite{zakharov1972}.
The immediately subsequent development, as we have seen in Sec.~\ref{ss:temporal}, involves at
least two phenomena. One is the spreading out on the ramp, the moving of the tyger along
the ramps of the untruncated Burgers solution, which perhaps can be explained
by advection effects; the other one is that the small-scale motion
looses the highly organized structure seen around $\ts$; in other words
the Fourier  spectrum broadens away from the Galerkin wavenumber. This may
signal the onset of the thermalization of the solution which, eventually,
becomes a Gaussian noise in the space variable with a flat
spectrum.  Here a digression is in order.
The ordinary untruncated Burgers equation---with or without viscosity---has 
played a major role, not only as testing ground for numerical schemes, but also
for helping us to find mistakes in excessively naive
ideas intended for Navier--Stokes turbulence. The Galerkin-truncated
Burgers equation may take us a step further, being paradoxically closer
to Euler--Navier--Stokes: it is nonlocal (in a way consistent with
energy conservation) and its solutions display spatio-temporal chaos,
as documented, e.g., in \cite{majdatimofeyev2000}.

Of course all this would be quite academic if we did not already have
good evidence that the key phenomena associated to tygers are also
present in the two-dimensional incompressible Euler equation, as discussed
in Sec.~\ref{ss:twod}.  Our understanding of this, so far,  based on
what we know about the analytic structure of 2D flow, is far from
complete. It seems important to find out how 
the more systematic theory of the birth can be carried over to
2D---and perhaps 3D---Euler. So far, it is clear that complex-space
singularities approaching the real domain within one Galerkin wavelength
are the triggering factor, as in the 1D Burgers case. 

Now, a few remarks about tyger purging, which is definitely not
the central issue of the present investigation. We have seen in Sec.~\ref{s:simulpheno} that
tygers, being born far from shocks, do not modify shock dynamics but do modify
the flow elsewhere because the tygers induce Reynolds stresses on scales much
larger than the Galerkin wavelength; hence the weak limit of the
Galerkin-truncated solution as $\kg \to \infty$ is definitely not the inviscid
limit of the untruncated solution. Can we ``purge tygers away'' and thereby
obtain a subgrid-scale method which describes the inviscid-limit solution
right down to the Galerkin wavelength?  

How to do this practically? Several ideas come to mind. One is simply
to apply some amount of viscosity. If the viscosity is sufficiently
large, the truncation becomes irrelevant but the solution thus obtained
will not coincide with the inviscid limit down to the Galerkin
wavelength or anywhere close to it. A second idea is to look for
embryonic tygers in physical space and selectively abort them. This can perhaps
be done by a suitable wavelet or filtering technique but may be
tricky \footnote{See however an attempt in this direction in Refs.~\cite{fargeetal,nguyenetal}.}.
A simpler idea is to purge the boundary layer near $\kg$ at each
time step. However this amounts to applying a Galerkin truncation
with a slightly smaller $\kg$ and will produce more tygers. A more subtle
way worth exploring is to wait until a low amplitude tyger has appeared
that is concentrated in a sufficiently narrow    boundary layer near $\kg$
\footnote{This may take some time.} and then to perform the purging,
an operation 
which clearly should not take place too often. Such ideas will of course
have to be tested carefully in future work. One may also wonder to what 
extent such a purging technique can be carried over to 2D and 3D
incompressible flow. We have already checked that for the case of 2D
and 3D
incompressible flow, the birth of the tyger takes
place in a narrow boundary layer near the Galerkin truncation.\\

\centerline{ACKNOWLEDGMENTS}

We are most grateful to the late Steven A. Orszag who introduced us
to spectral methods.
We had many useful discussions, in particular with C.~Bardos, J.~Bec,
S.~Bhattacharjee, M.~Blank, M.E.~Brachet, P.~Constantin, H.~Frisch, J.~Goodman,
J.-L.~Guermond, K.~Khanin, A.~Majda, R.~Nguyen Van Yen, R.~Pandit,
R.~Pasquetti, W.~Pauls and J.-Z.~Zhu.. Computations used the M\'esocentre de calcul of the Observatoire de
la C\^ote d'Azur and the SX-8 at the Yukawa Institute of Kyoto University and
SERC (IISc). UF's and SSR's work was supported in part by COST Action MP0806
and by ANR ``OTARIE'' BLAN07-2\_183172.  TM's work was supported by the GCOE
Program ``The Next Generation of Physics, Spun from Universality and
Emergence'' from the MEXT of Japan.  SSR  acknowledges DST (including its coming Indo-French IFCAM program) and UGC (India) for
support. SN and TM thank the French Ministry of
Education for funding several extended visits to the Observatoire de la C\^ote
d'Azur. 
\section*{Appendices} 
\appendix
\setcounter{equation}{0}
\section{Numerics and graphical representation}
\label{a:num}

With the exception of the inviscid-limit solutions of the untruncated
Burgers equation which were obtained by the Fast Legendre Transform method
\cite{noullezvergassola}, all the numerical simulations presented in this paper used the
pseudo-spectral method \cite{orszag}, in combination with a fourth-order
Runge--Kutta time marching with a time step $10^{-4}$ for $\kg \le
5,000$ and $10^{-5}$ for $\kg >5,000$. Up to $2^{16}$ collocation points were used. In order to implement Galerkin truncation 
it was essential to remove aliasing. In principle this can be done
by the so-called two-thirds rule: $\kg \le (2/3) K_{\rm max}= N/3$, where 
$N$ is the number of collocation points. This allows a calculation of
the solution at collocation points which suffers only from  double-precision
rounding errors and temporal truncation. 

\begin{figure}[htbp]
\iffigs
\centerline{\includegraphics[height=5cm]{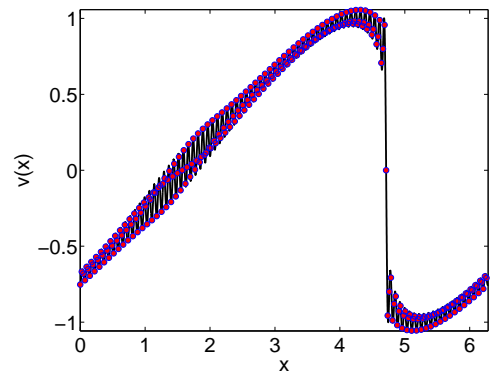}}
\fi
\caption{(Color online) Plots of the solution $v(x)$ of the
Galerkin-truncated Burgers equation, with
$\kg = 85$,  using two different numbers of collocation points,
namely, $N = 16,384$ (black continuous line) and with red-filled circles (black circles) 
 for $N = 256$ for the single-mode initial condition 
$v_0(x) = \sin(x - \pi/2)$
shown at time $t = 1.10$. This plot illustrates the
stroboscopic effect, a purely graphical interpolation artefact, caused by insufficient
resolution for a given $\kg$.}
\label{f:stroboscopic}
\end{figure}

However, when it comes to representing tygers graphically, using the
two-thirds rule can produce a stroboscopic graphical artefact, illustrated
in Fig.~\ref{f:stroboscopic}. Since in the presence of tygers there
is a strong excitation at and near the truncation wavenumber $\kg$, the
velocity is very close to a sine wave with Galerkin wavelength.
Unless proper interpolation is used, such a sine wave cannot be correctly
represented using only three points per wavelength. Otherwise the
rapid oscillations disappear in favor of an illusory triple valuedness.
One easy way to do the interpolation is to use a much higher number
of collocation points whenever graphical output is needed.

\section{Real and complex singularities for the Burgers equation}
\label{a:sing}

As is well known, the solutions of hydrodynamical equations such as the Burgers
or Euler equations in any dimension with space-periodic and analytic initial
data, remain so for at least a finite time (cf., e.g., Refs.~\cite{blue,gibbon} and references therein). In the 1D Burgers
case, a real singularity appears after a finite time (finite-time blow up). For 2D Euler, 
analyticity is preserved for all times but complex-space singularities
will typically approach the real domain arbitrarily close at long times.
For 3D Euler it is not clear if there is or not finite-time blow up.
The signature of complex-space singularities---more precisely of those
closest to the real domain---is an exponential fall-off at high wavenumbers $k$ of the spatial
Fourier transform $\propto \ue ^{-\delta(t) k}$ (within algebraic
prefactors), where $\delta(t)$ is the width of the analyticity
strip, that is the distance from the real domain of the nearest complex-space
singularity. As long as $\delta(t)\kg \gg 1$ the effect of Galerkin
truncation is exponentially small and can be ignored for all the purposes
of the present study. For convenience we define the threshold as the time
$\tg$ when $\delta(\tg) \kg=1$.

Let us here recall how $\delta(t)$ can be obtained explicitly for the case of 
the untruncated Burgers equation with the initial condition $u_0(x) = -\sin x$
(for details, cf. Ref.~\cite{fournierfrisch}). In Lagrangian coordinates,
the motion of an inviscid Burgers fluid particle is given by
\begin{equation}  
  a \mapsto x =a+tu_0(a)= a -t \sin a.  
\label{lagrangian}
\end{equation}
A singularity is obtained when the Jacobian $\partial x/\partial a$ of this map vanishes.
For $t$ slightly less than $\ts =1$, the Lagrangian and Eulerian 
locations of the singularity nearest to the
real domain is close to the origin and can be obtained by expanding
$\sin a$ to cubic order. We thus have, up to higher-order terms:
\begin{equation}  
x =  (\ts -t)a +\frac{a ^3}{6}; \quad \partial x/\partial a = \ts -t +\frac{a ^2}{2}.  
\label{expand}
\end{equation}
The Jacobian is seen to vanish at $a_\star(t) = \pm \ui \sqrt{2(\ts -t)}$.
The corresponding Eulerian singularities are at
\begin{equation}  
  x_\star(t)=\pm\frac{2  \sqrt 2}{3} \ui (\ts -t)^{3/2}.
\label{xstar}
\end{equation}
Equating the modulus $\delta(t)$ of the imaginary part of  $x_\star(t)$ to
$1/\kg$, we obtain:
\begin{equation}  
\ts -\tg = \left(\frac{3}{2\sqrt 2}\right)^ {2/3} \kg ^{-2/3} \approx 1.04 \kg ^{-2/3}.
\label{obtaintg}
\end{equation}
For convenience we have  replaced $1.04$ by unity and thus used in Sec.~\ref{s:birth}
\begin{equation}  
  \tg = \ts -\kg ^{-2/3}.
\label{tgts}
\end{equation}

\section{The beating input}
\label{a:beating}

\begin{figure}[htbp]
\iffigs
\begin{center}
\includegraphics[height=5cm]{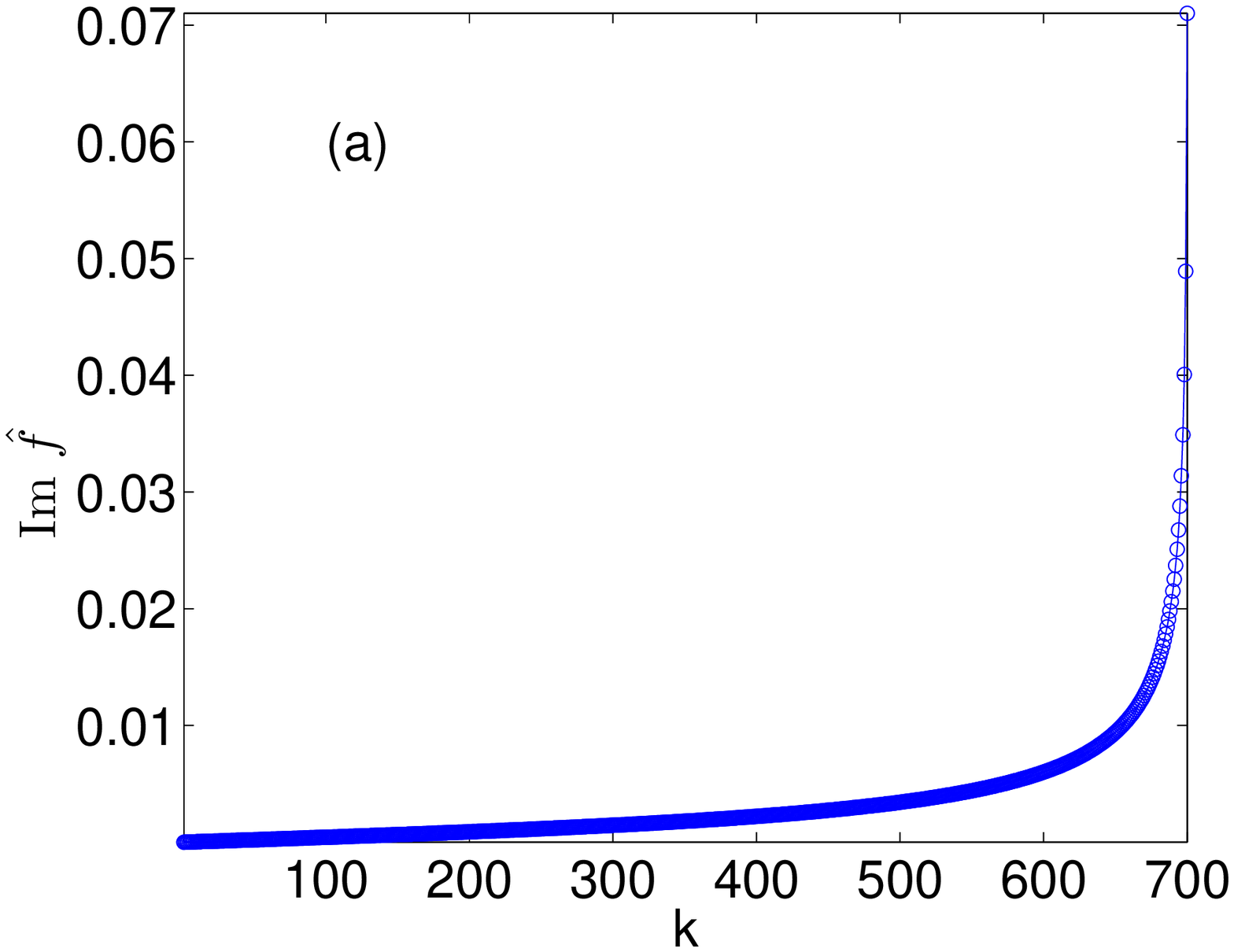}
\includegraphics[height=5cm]{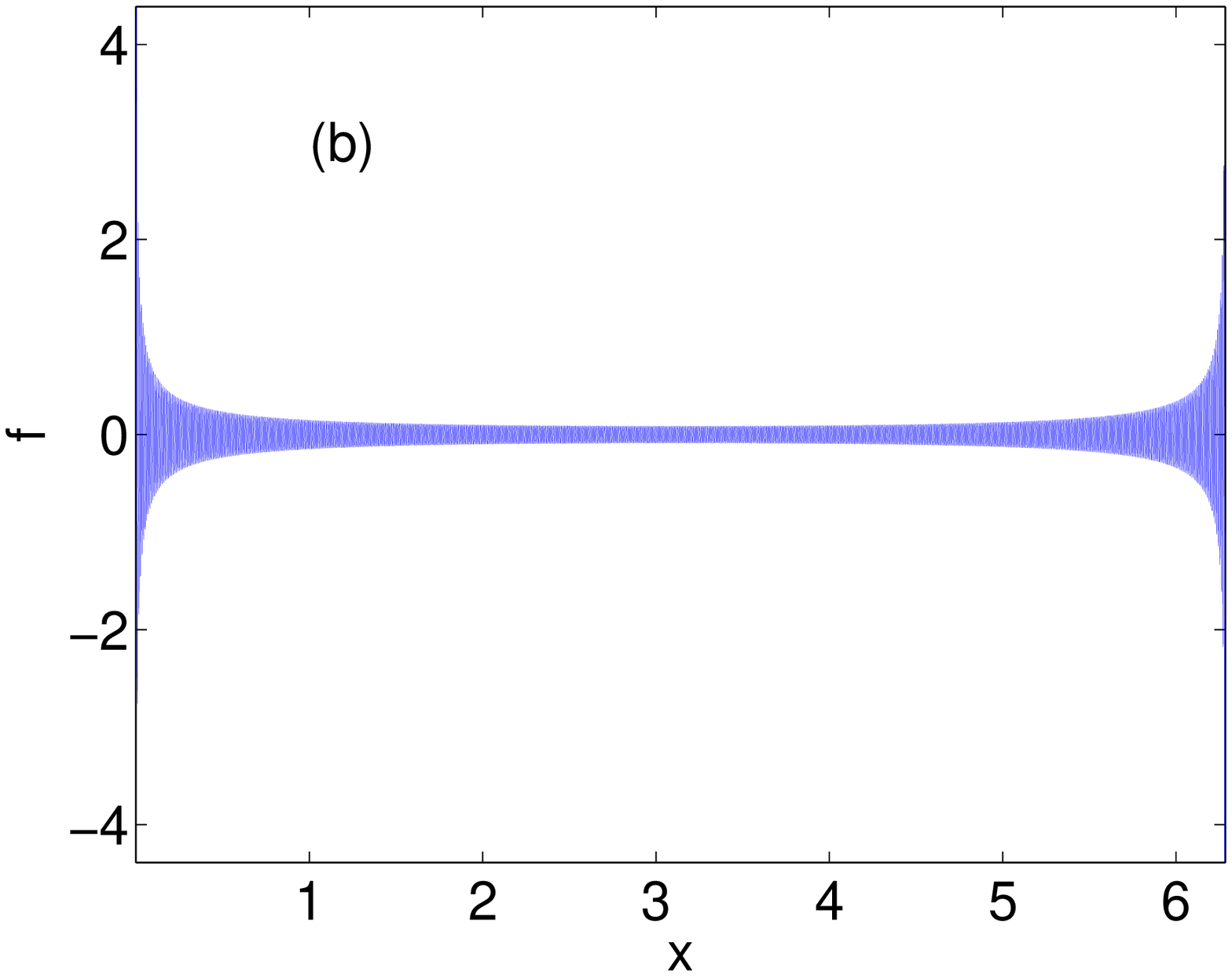}
\end{center}
\fi
\caption{(Color online) The beating input term, calculated at $\ts$ and used
together with the freezing approximation, evaluated for the
single mode
initial condition $v_0(x) = u_0(x) = -\sin (x)$ (origin at
preshock) for
$K_G = 700$. (a): imaginary part in Fourier space; (b): physical
space.}
\label{f:beating}
\end{figure}

Our purpose it to find the large-$\kg$ asymptotic behavior of the beating input
\eqref{deffk}, repeated here for convenience,
\begin{equation}  
  \fhk \equiv \ui k\sum_{p+q=k}(\uhlp\,\uhgq +\frac{1}{2}\uhgp\,\uhgq),
\label{deffkrep}
\end{equation}
in the range $1\ll \kg -k \ll \kg$. Although it is this input which
eventually
permits the birth of the tyger, the beating input is strongest
at the preshock and it is best to place the origin there. The function
$\us$, whose high- and low-passed filtered Fourier transforms appear
in \eqref{deffkrep}, is odd and has a cubic root singularity at the
origin. Hence its Fourier transform is pure imaginary, odd and given
to leading order at large wavenumbers by
\begin{equation}  
\uhk = \ui C k^{-4/3},  
\label{asuhk}
\end{equation}
where $C$ is a real constant and it is understood that  $k^{-4/3}$ has
the same sign as $k$. Actually, \eqref{asuhk} is a very good
representation of the Fourier transform of $\uhk$ at all but a few
low-lying wavenumbers. Observe that the r.h.s. of \eqref{deffkrep} has
two terms, the first has $|p|\le \kg$ and $|q|>\kg$ while the
second has both $|p|$ and $|q|$ greater than $\kg$. The latter is
easily bounded in modulus by $D |k| \kg ^{-5/3}$ where $D$ is a
  positive constant and thus will be seen to
contribute negligibly to the asymptotics. Hence, from \eqref{deffkrep}
and \eqref{asuhk}, we have
\begin{eqnarray}  
  \fhk & \simeq& -C^2 \ui k g_k \label{fkft}\\
g_k &\equiv& \sum_{p+q=k,\, |p|\le \kg,\, |q|>\kg} p^{-4/3} q^{-4/3}.
\label{defgk}
\end{eqnarray}
We introduce dimensionless variables:
\begin{equation}  
\tilde p \equiv -\frac{p}{\kg}, \quad \tilde q \equiv \frac{q}{\kg},\quad
\tilde k \equiv \frac{\kg -k}{\kg}, 
\label{defdiml}
\end{equation}
and rewrite \eqref{defgk} as
\begin{equation}  
g_k = -\kg ^{-8/3} \sum_{\tilde k<\tilde p\le 1} \tilde p ^{-4/3}
(1+\tilde p -\tilde k) ^ {-4/3},  
\label{gkdiml}
\end{equation}
where the summation on $\tilde p$ is over integer multiples of
$1/\kg$. Now we set 
\begin{equation}  
y\equiv \frac{\tilde p -\tilde k}{\tilde k}  
\label{defy}
\end{equation} 
and approximate the summation by an integral to obtain, to leading
order,
\begin{equation}  
g_k \simeq -\frac{1}{3}\kg ^{-4/3} (\kg -k)^ {-1/3},
\label{gklo}
\end{equation}
whence 
\begin{equation}  
 \fhk  \simeq \ui \frac{1}{3}C^2\kg ^{-1/3} (\kg -k)^ {^{-1/3}}, \quad 1\ll \kg -k \ll \kg.
\label{fhklo}
\end{equation}
There are also subleading corrections involving 
$(\kg -k)^0$, \,\,$(\kg -k)^ {^{+1/3}}$,
etc  \cite{helene}. Because the exponent gap is only $1/3$, it is difficult in 
simulations to see a clean leading order term.
Fig.~\ref{f:beating} shows the beating input in Fourier space and in
physical space.

\section{The Orr--Sommerfeld problem for Galerkin truncation}
\label{a:OS}

\begin{figure}[htbp]
\iffigs
\begin{center}
\includegraphics[height=5cm]{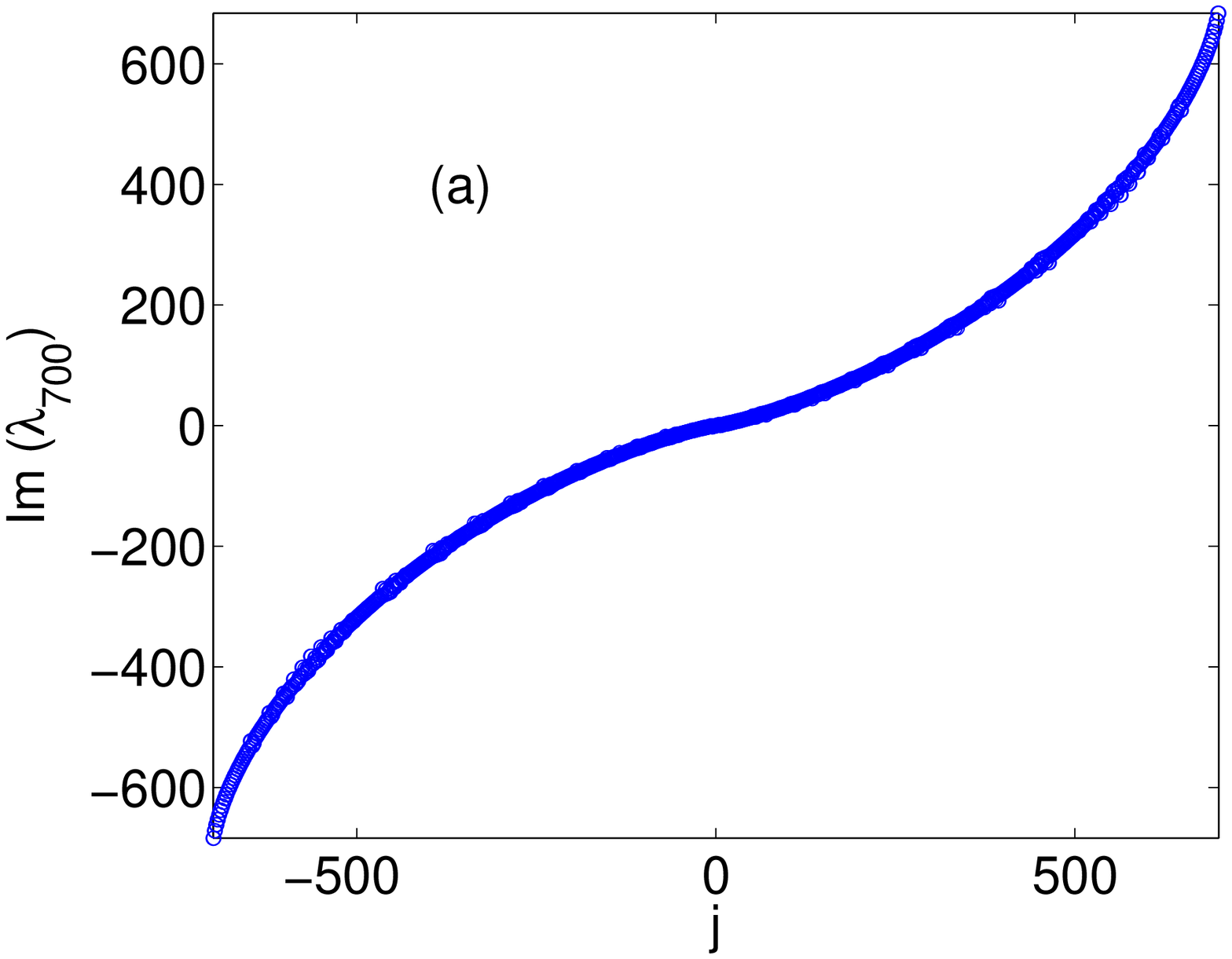}
\includegraphics[height=5cm]{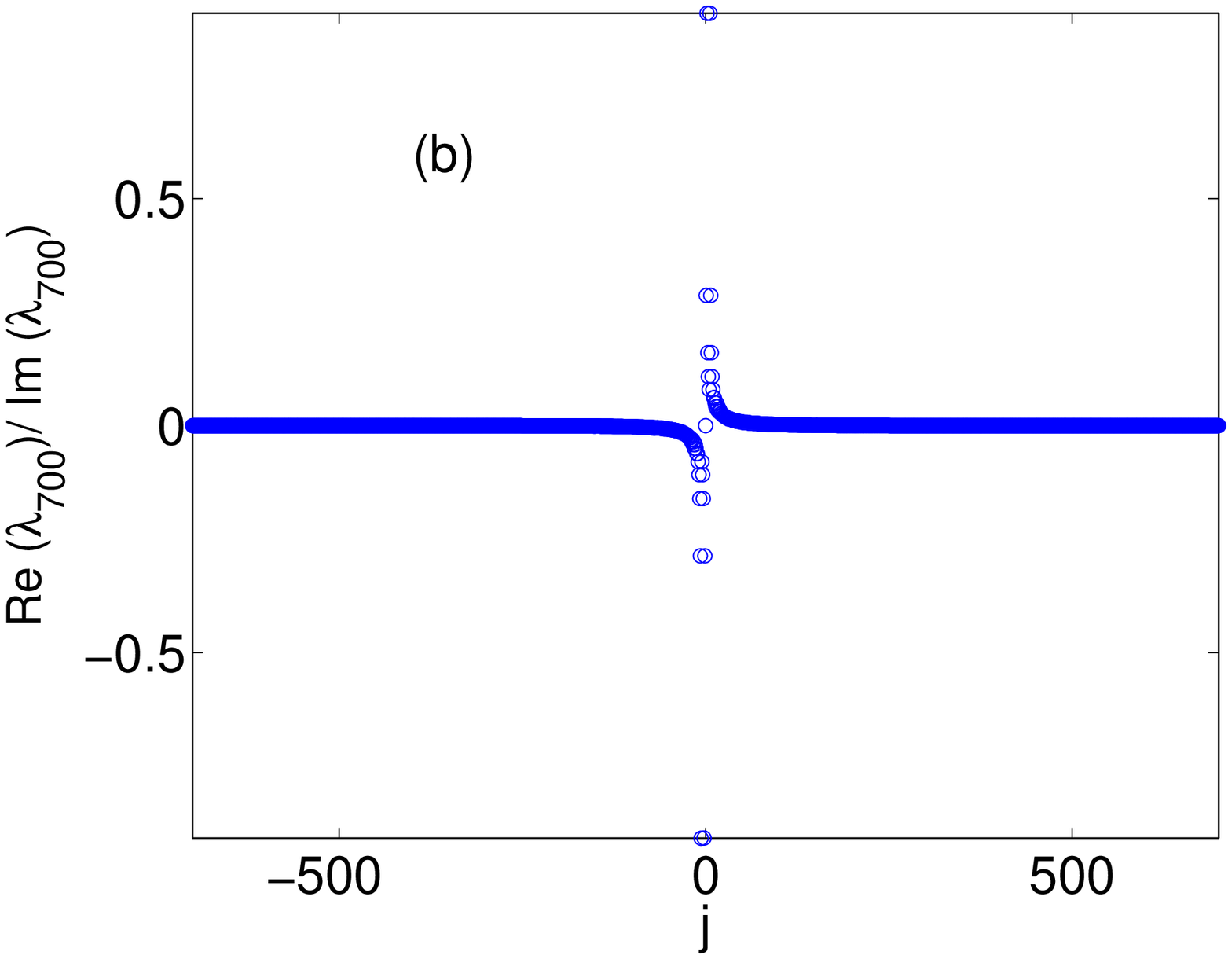}
\includegraphics[height=5cm]{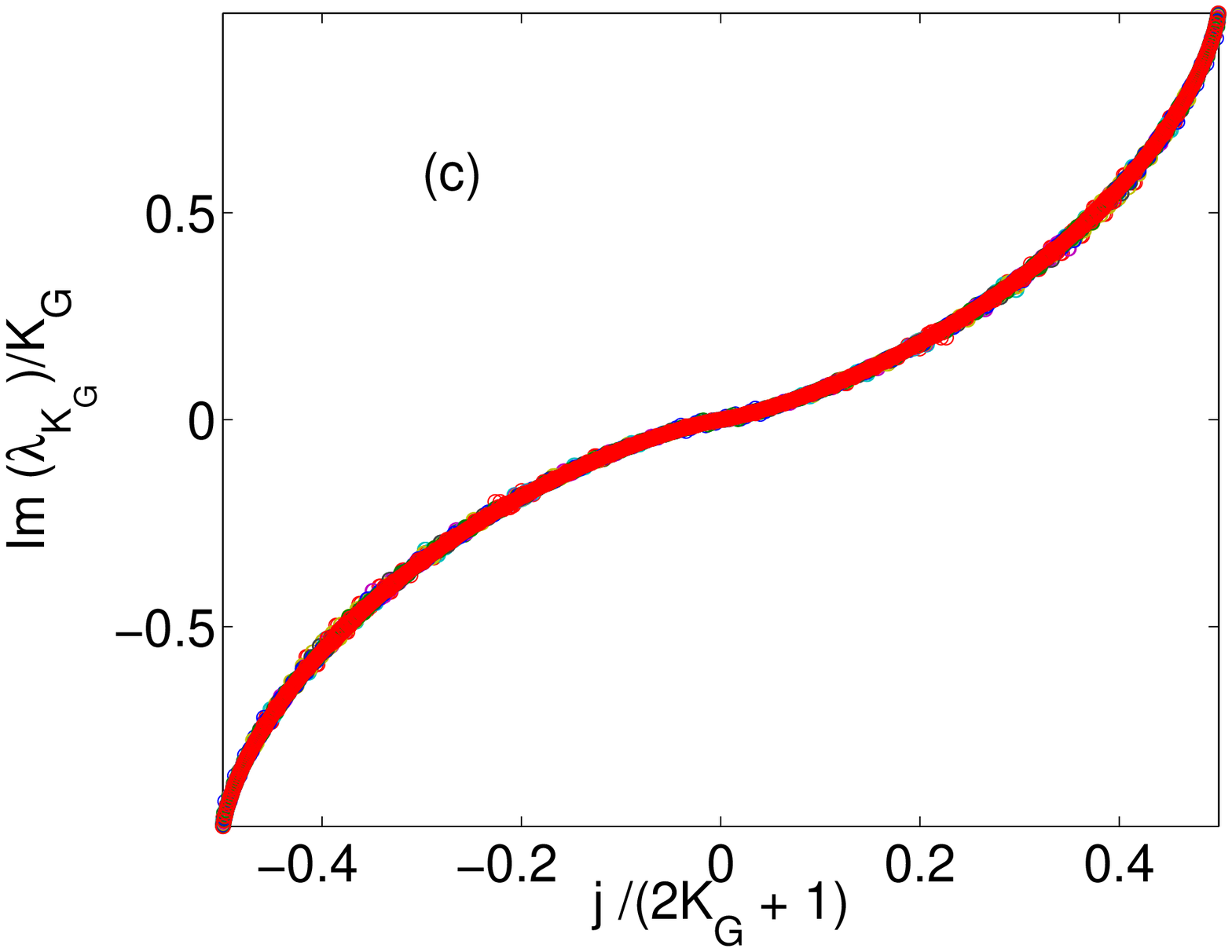}
\end{center}
\fi
\caption{(Color online) (a): imaginary part of the eigenvalues for
$\kg$ = 700; (b):  ratio of imaginary to real part of the
eigenvalues for
$\kg$ = 700; (c): rescaled imaginary parts of the eigenvalues
for $\kg = 100,\, 200,\, 300,\ldots,\, 1000$ showing collapse.}
\label{f:eigenvalues}
\end{figure}
\begin{figure}[htbp]
\iffigs
\begin{center}
\includegraphics[height=3.5cm]{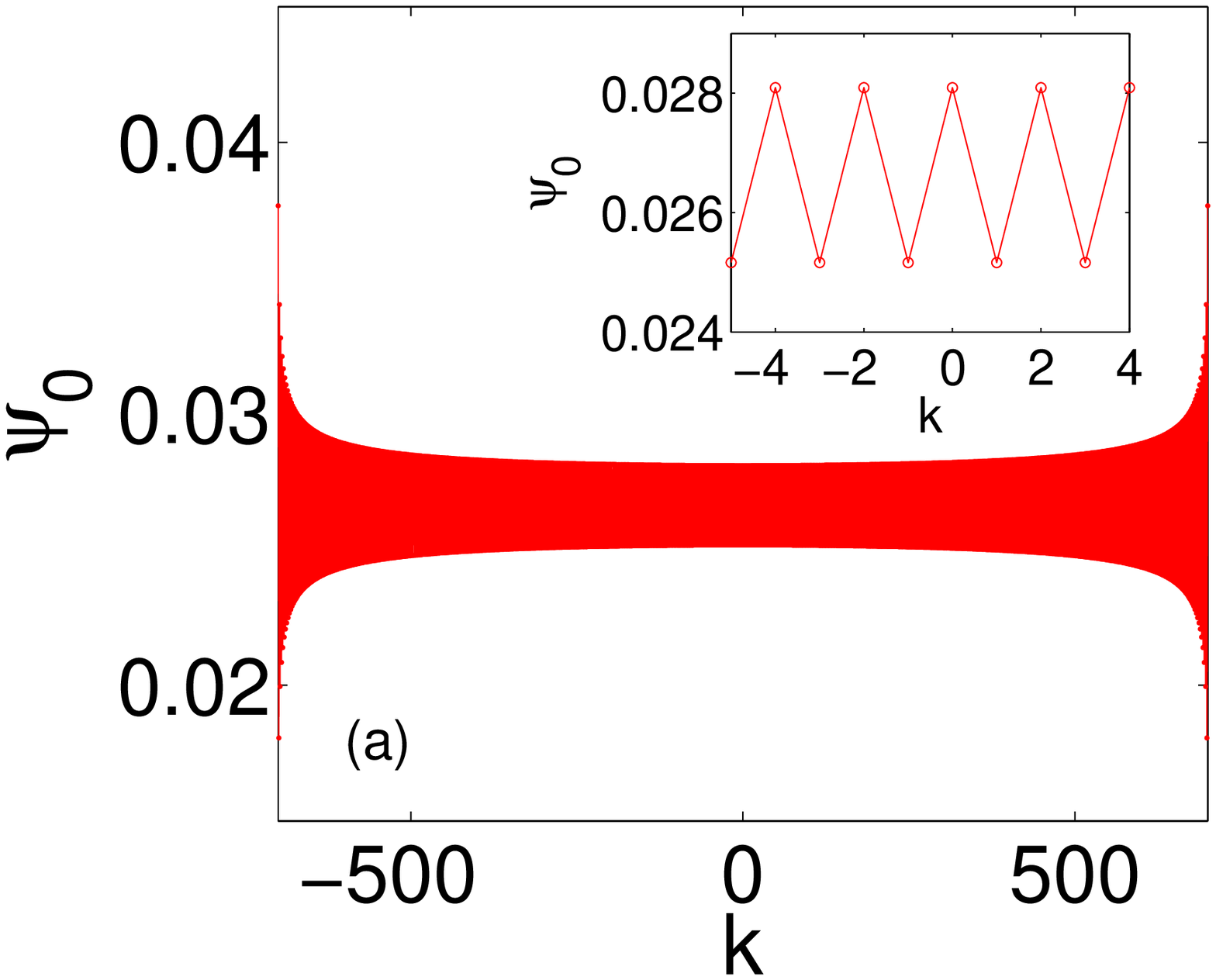}
\includegraphics[height=3.5cm]{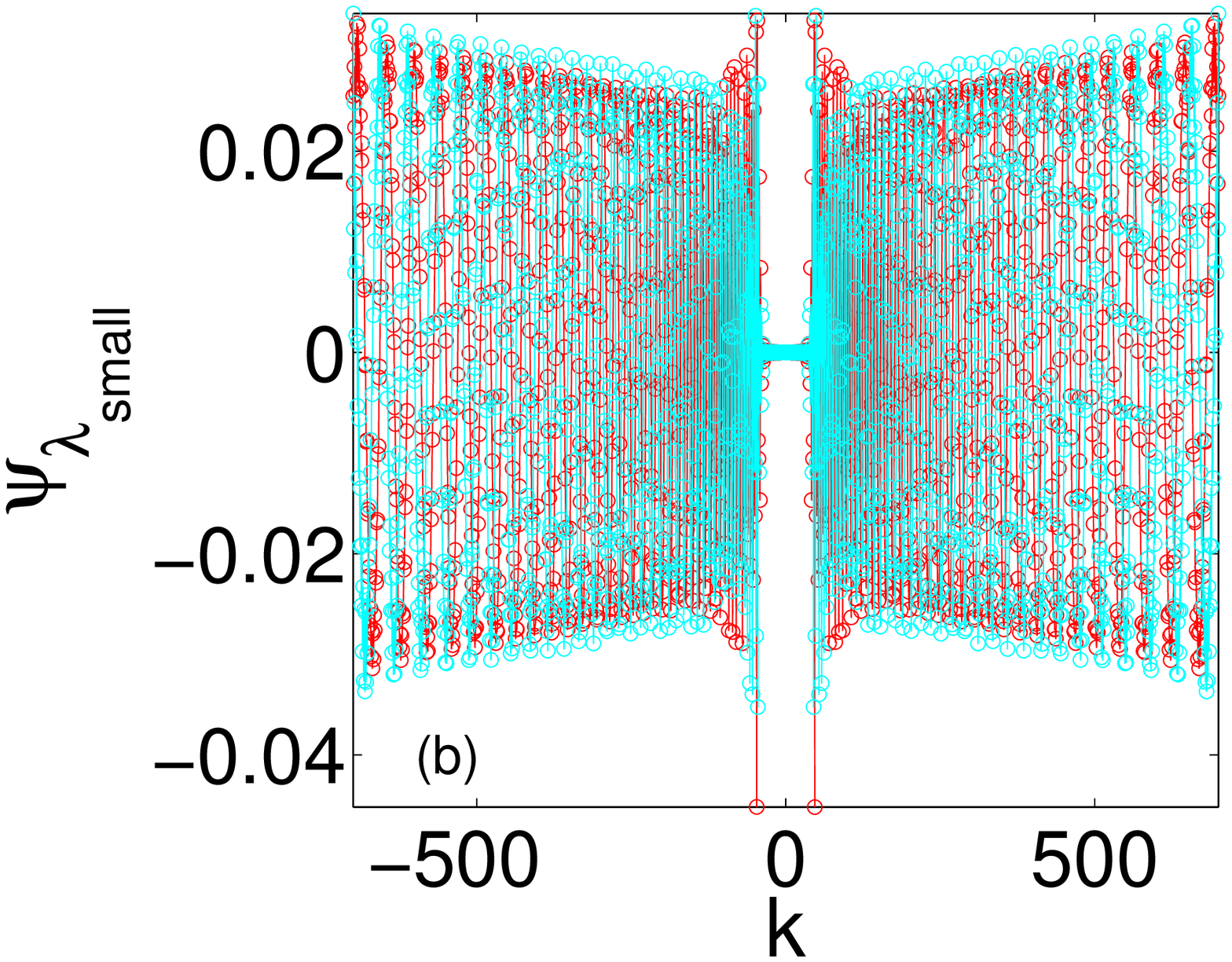}
\includegraphics[height=3.5cm]{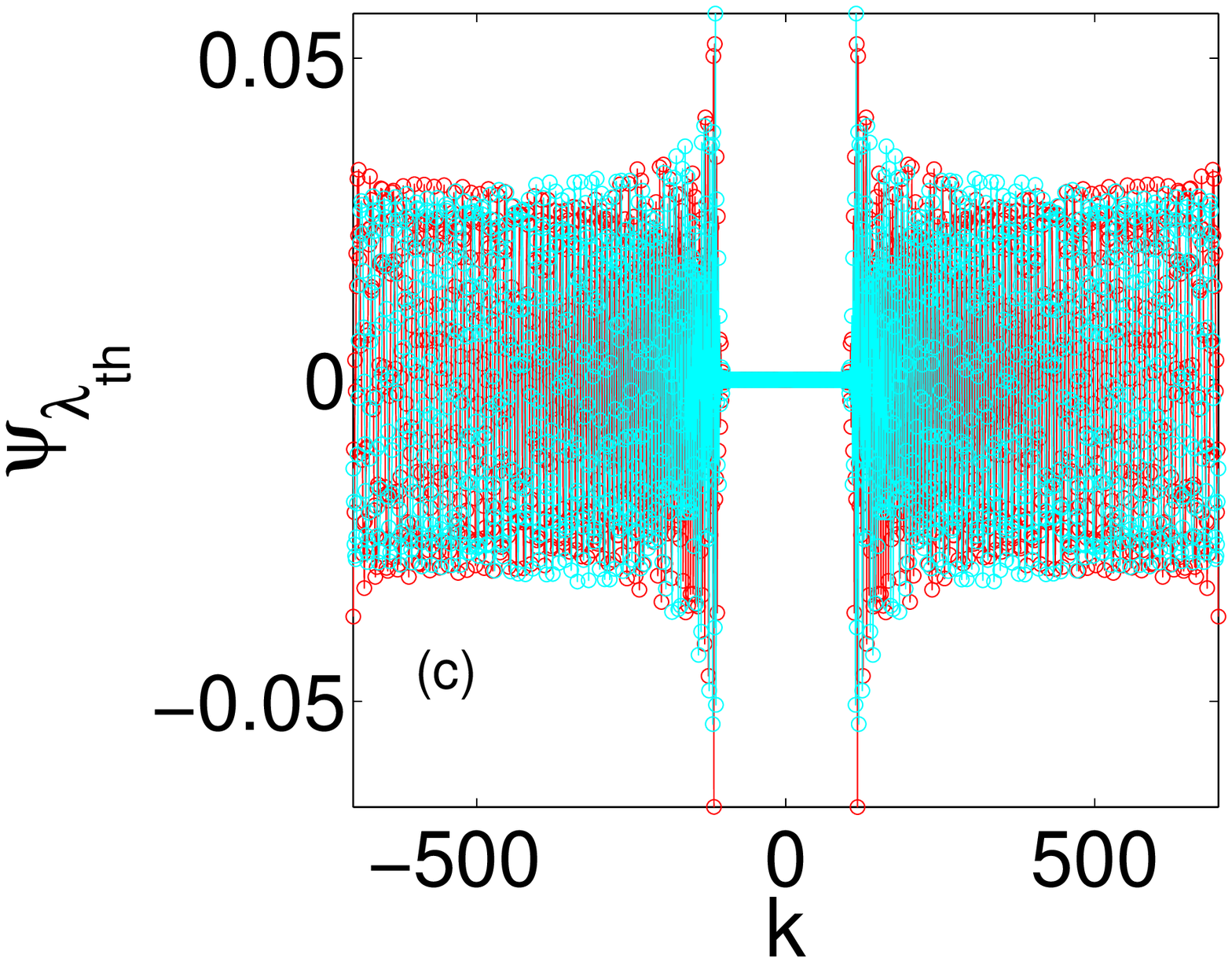}
\includegraphics[height=3.5cm]{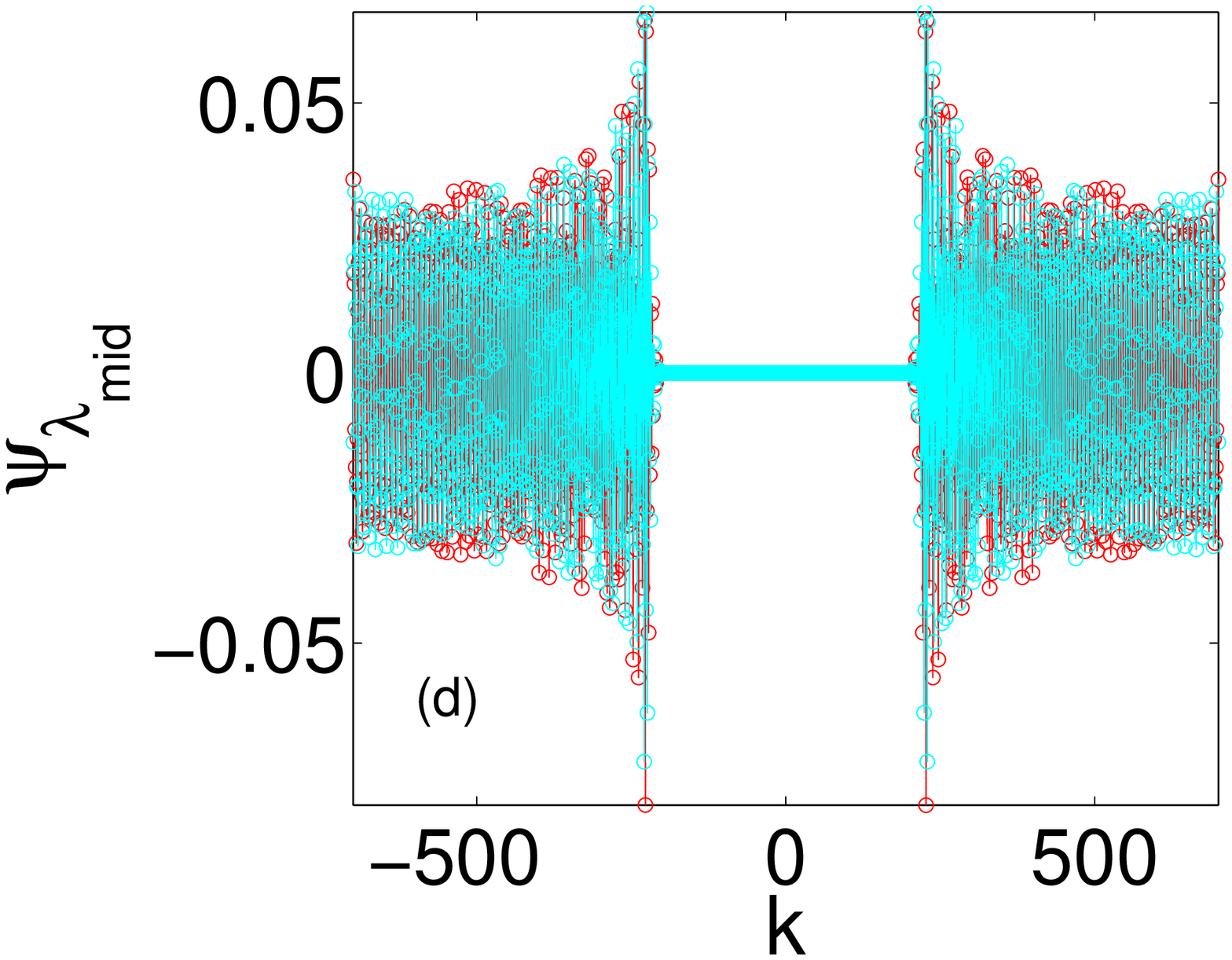}
\includegraphics[height=3.5cm]{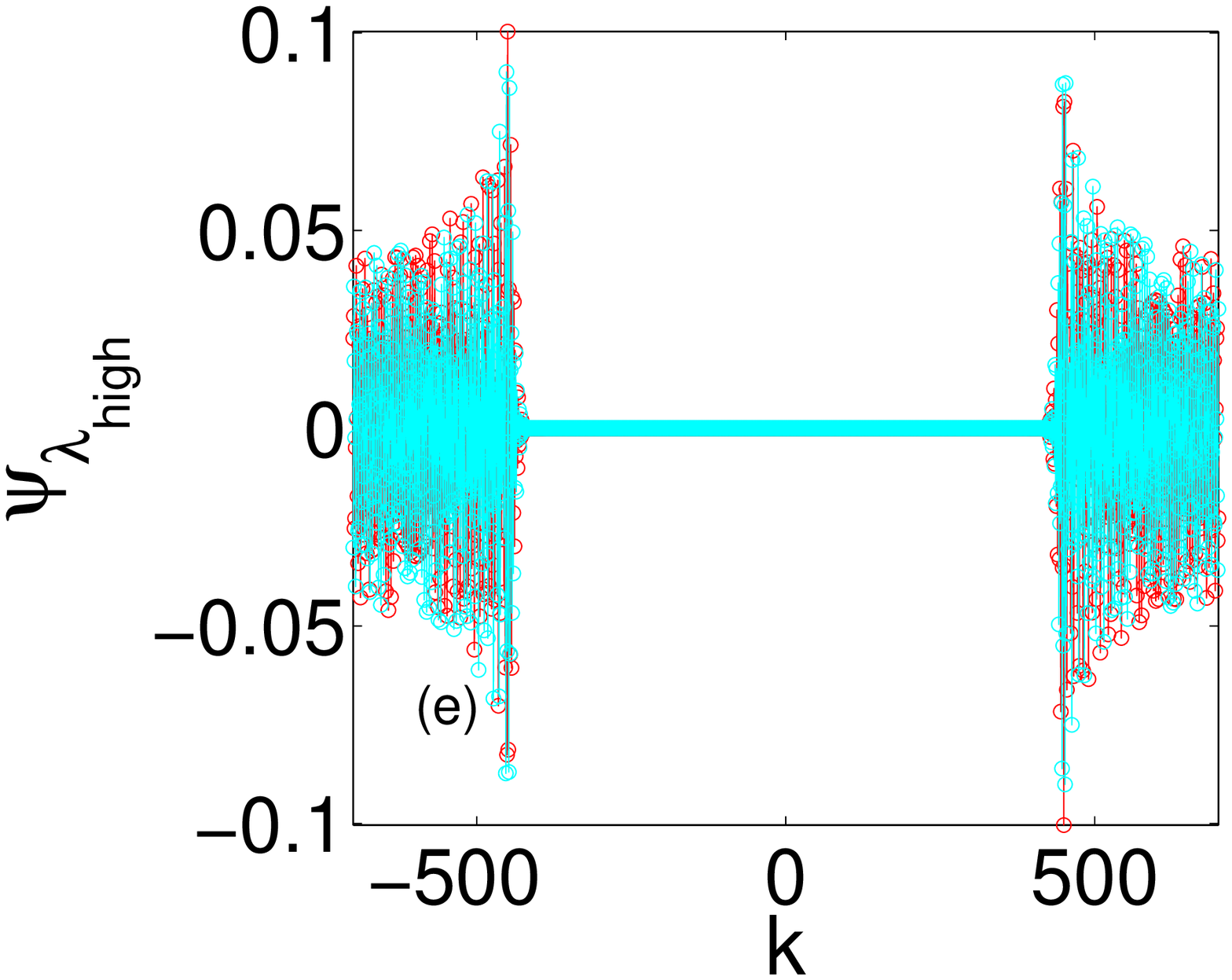}
\includegraphics[height=3.5cm]{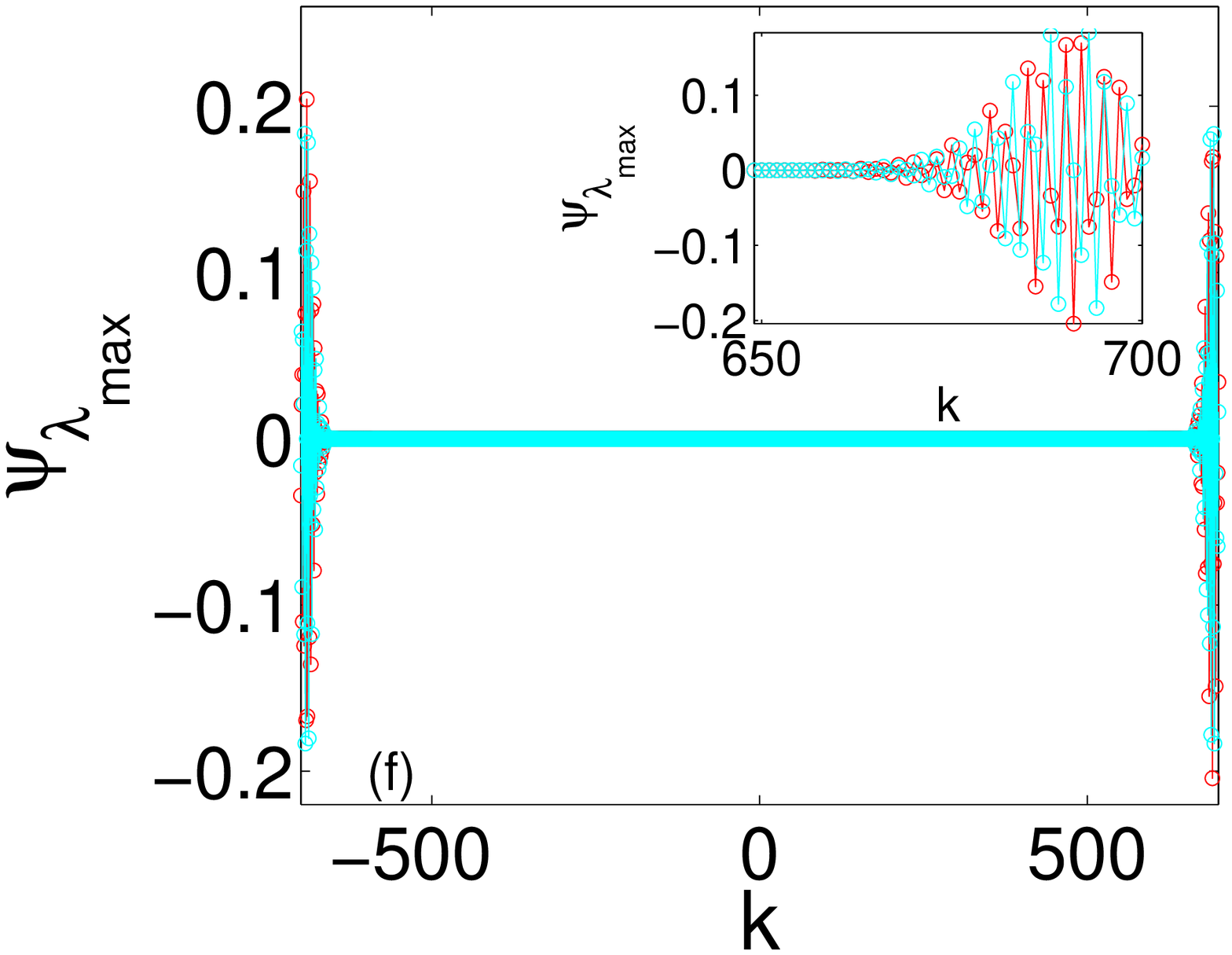}
\end{center}
\fi
\caption{(Color online) Eigenvectors of the Orr--Sommerfeld operator in Fourier space (origin
  at the tyger) for $\kg=700$; real parts are shown in red (black) and the imaginary parts in cyan 
 (light grey).  
 (a) modes corresponding to an imaginary part of the eigenvalue
which is zero (zero mode, which can be taken purely real), (b) halfway between
zero and threshold, (c) at threshold, (d) halfway between threshold and the largest 
value,  (e) close to the largest value, (f) largest 
value (the inset zooms on the largest wavenumbers).}
\label{f:eigenmodes}
\end{figure}

Here we study the spectral properties of the Orr--Sommerfeld operator
governing weak perturbations near the time $\ts=1$ of the first singularity
of the untruncated solution for the initial
condition $u_0 =\sin x$, as introduced in Sec.~\ref{s:birth}. This 
operator involves the Fourier transform  $\hat u_{\star,\,k}$ of the solution of the
untruncated Burgers equation at 
time $\ts$. Because this is an odd pure imaginary function
of $k$, we set $\hat u_{\star,\,k} = \ui \uvk$, where $\uvk$ is a real odd function.
With this notation the operator becomes the following real $(2\kg +1) \times
(2\kg +1)$ matrix 
\begin{equation}  
   A_{kk'} \equiv  k\,\uvkmkp, \quad -\kg \le k\le \kg, \quad -\kg \le k'\le \kg.
\label{redefA}
\end{equation}
Indices are integers but not restricted to nonnegative values, unless
otherwise stated. Henceforth, unless otherwise stated, all summations
are from $-\kg$ to $\kg$.

We studied the spectral properties of $A$ by standard numerical techniques
using MATLAB$^{\hbox{\textregistered}}$, but we also have some
analytical results of an algebraic nature.
It is appropriate to begin  with the latter.

\textit{The matrix $A$ is singular, that is it has a vanishing
  determinant.} This is equivalent to stating that there exists a
  non-vanishing vector (here called a \textit{zero mode}) $\psi$ such
that $\sum_{k'} A_{kk'} \psi_{k'}=0$ for all $k$. If we can find a
  $\psi$
such that 
\begin{equation}  
  \sum_{k'} \uvkmkp  \psi_{k'}=0,\quad \hbox{for all}~k,
\label{utchzeromode}
\end{equation}
then it follows from \eqref{redefA} that $\psi$ is also a zero mode of $A$.
The matrix $\uvkmkp$ is skew-symmetric and of odd dimension. By an elementary
theorem of Jacobi its determinant vanishes and thus it has a zero mode. Since
the entries $\uvkmkp$ are real, the zero mode can also be taken real.

\textit{The non-vanishing (complex) eigenvalues come in opposite
  pairs; the associated eigenvectors which are even or odd in $k$.}
The eigenvalue/eigenvector equation for the Orr--Sommerfeld operator $A$ reads
\begin{equation}  
  \sum_{k'}   k\,\uvkmkp \psi_{k'} = \lambda \psi_{k}
\label{eigenequation}
\end{equation}
 Observe that the operator $A$ \eqref {redefA}
  is neither Hermitian nor antihermitian and that, for non-vanishing
  $\lambda$,
we have \hbox{$\psi_0 =0$.} We now exploit the oddness
  of  $\uvk$, to look for even and odd eigenvectors. First assume
that $\psi_k = \psi_{-k}$. In \eqref{eigenequation}, limiting
  ourselves to $k>0$ we separate the $k'$ contributions into positive
and negative ones and obtain, using the oddness of $\uvk$:
\begin{equation}  
  \sum_{k'=1}^{k'=\kg} k\, \left(\uvkmkp+\uvkpkp\right)\psi_{k'}
  =\lambda \psi_k, \quad k>0,\, k'>0.
\label{eigeneven}
\end{equation}
Now we rescale our eigenvectors by a factor of $1/\sqrt{k}$ \footnote{Observe
  that the $1/\sqrt{k}$ rescaling is the $k$-space analog of the trick used
on p.~96  by
  GHT (in $x$-space)  to prove stability for nonnegative velocities.}:
\begin{equation}  
  \phi_k \equiv \frac{\psi_k}{\sqrt{k}},
\label{defphi}
\end{equation}
so as to rewrite the eigenvalue/eigenvector equation as
\begin{equation}  
  \sum_{k'=1}^{k'=\kg} \sqrt{kk'} \left(\uvkmkp+\uvkpkp\right)\phi_{k'}
  =\lambda \phi_k, \quad k>0,\, k'>0.
\label{niceeven}
\end{equation}
Proceeding similarly under the asumption of an odd eigenvector, we obtain 
instead of \eqref{niceeven}
\begin{equation}  
   \sum_{k'=1}^{k'=\kg} \sqrt{kk'} \left(\uvkmkp-\uvkpkp\right)\phi_{k'}
  =\lambda \phi_k, \quad k>0,\, k'>0.
\label{niceodd}
\end{equation}
We now observe that \eqref{niceeven} and \eqref{niceodd} are two
eigenvalue/eigenvector equations involving two $\kg \times\kg$ ``reduced'' matrices which
are negative transposed of each other. Hence their eigenvalues are opposite.

If all eigenvalues of $A$ and of the two reduced matrices are distinct
(something for which we have so far only numerical evidence), then the even
eigenvectors, the odd ones and the zero mode exhaust the list of $2\kg+1$
eigenvectors of $A$.

The other results on the spectral properties of the Orr--Sommerfeld operator
are obtained numerically (mostly for the case $\kg =700$)
and described now, with occasional soft phenomenological interpretations
of the findings. 

The eigenvalues other than zero are all complex but very close to being
pure imaginary. Fig.~\ref{f:eigenvalues}, for the case $\kg =700$, 
shows the imaginary parts of all 1401 eigenvalues: they range  almost
exactly  from  $-\kg$ to $\kg$. Probably, this is related to the fact
that the Orr--Sommerfeld operator \eqref{defA} is a modified
advection operator with an advecting  velocity that ranges from $-1$
to $+1$  and that the $k$ factor which stems from the space derivative
cannot exceed $\kg$ because of the truncation operator.

The second panel in Fig.~\ref{f:eigenvalues} shows the ratio of the
real to the imaginary part, which is quite small, even for the
eigenvalues close to zero. Actually, the real parts can be neglected
altogether when used in \eqref{solution}. The explanation may lie in
the decomposition of the matrix \eqref{niceeven} into the
antihermitian matrix $\sqrt{kk'} \uvkmkp$ and the Hermitian matrix
$-\sqrt{kk'}\uvkpkp$. Remember that in this reduced formulation both
$k$ and $k'$ are positive.  In the Hermitian part $k$, $k'$ and $k+k'$
must be less or equal to $\kg$. This prevents $k$ and $k'$
from being simultaneously close to $\kg$. But we also found that most
of the eigenmodes are confined essentially in a relatively narrow
boundary layer near $\kg$. Hence the Hermitian part cannot contribute
much \footnote{For the aliased case of GHT, it seems also that 
high-lying eigenmodes are confined to such a boundary layer and this may
explain why some of their $x$-space figures are tyger-like.}.

The third panel in Fig.~\ref{f:eigenvalues} shows that by simply
rescaling by factors $\propto\kg$ the horizontal and vertical axes of
the distributions of the imaginary parts of the eigenvalues, the
curves for different $\kg$ all nicely collapse on top of each other,
for large enough $\kg$, suggesting that some limiting distribution
exists
as $\kg \to \infty$
This we interpret---in highly speculative mode---as follows. Because of the confinement near $\kg$ of
most of the eigenmodes, we can approximate the reduced matrix
$\sqrt{kk'} \uvkmkp$ by $\kg \uvkmkp$, which is $-\ui \kg$ times the
convolution with the Fourier transform of the velocity $\us$. The
eigenfunctions of this operator are Dirac measures and the
corresponding eigenvalues are $-\ui \kg$ times the values the
velocity takes at the supports of these Dirac measures.

We now turn to the eigenmodes (Fig.~\ref{f:eigenmodes}). It is seen
that the zero mode (first panel: eigenvalue zero) is nearly
constant, around $0.0265$ with additional small-amplitude even-odd
oscillations and some edge effects near $\pm \kg$. This structure is not
surprising: if it was not for truncation effects, \eqref{utchzeromode} would
be the Fourier transform of the equation $\us(x)\psi(x) =0$ whose solution is
an arbitrary linear combination of Dirac measures at the two zeros of $\us$,
one at the tyger ($x=0$) and one at the preshock ($x=\pi$). By Fourier
transformation, these go over into a constant vector and a vector proportional
to $(-1)^k$.  As we move to higher eigenvalues, we find that the
eigenmodes are localized at higher and higher wavenumbers. This is
unexplained but consistent with the almost purely imaginary character
of the eigenvalues, as given above.


\end{document}